\documentclass{aa}

\usepackage[utf8]{inputenc}

\usepackage{txfonts}
\usepackage{stmaryrd}
\usepackage{soul}

\usepackage{natbib}

\usepackage{amsmath,amsfonts,amssymb}
\usepackage{mathtools}
\usepackage{xspace}
\usepackage{textcomp,mathrsfs}

\usepackage[squaren,Gray]{SIunits}
\newcommand{\micron}{\micro\meter}

\usepackage[normalem]{ulem}

\usepackage[algoruled,vlined]{algorithm2e}
\DontPrintSemicolon
\usepackage{algpseudocode}
\usepackage{comment}

\newcommand*{\Iter}[1]{^{[#1]}}

\SetKwComment{Comment}{\ensuremath{\vartriangleleft} }{}

\SetCommentSty{CommentFormat}

\SetKw{And}{and}
\SetKw{Or}{or}
\SetKw{Break}{break}
\SetKw{True}{true}
\SetKw{False}{false}

\usepackage{wrapfig}
\usepackage{graphicx}
\graphicspath{{figures/}}
\usepackage{tikz}

\definecolor{DarkBlue}{RGB}{0,69,134}      
\definecolor{Orange}{RGB}{255,66,14}       
\definecolor{Yellow}{RGB}{255,211,32}      
\definecolor{Green}{RGB}{87,157,28}        
\definecolor{DarkViolet}{RGB}{126,0,33}    
\definecolor{LightBlue}{RGB}{131,202,255}  
\definecolor{DarkGreen}{RGB}{49,64,4}      
\definecolor{LightGreen}{RGB}{174,207,0}   
\definecolor{Violet}{RGB}{75,31,111}       
\definecolor{Golden}{RGB}{255,149,14}      
\definecolor{Red}{RGB}{197,0,11}           
\definecolor{Blue}{RGB}{0,132,209}         

\usepackage{multirow}
\usepackage{threeparttable}

\usepackage[colorlinks=true,allcolors=DarkBlue]{hyperref}

\newcommand*{\eg}{\emph{e.g.}\xspace}
\newcommand*{\ie}{\emph{i.e.}\xspace}

\DeclarePairedDelimiterX{\Paren}[1]{(}{)}{#1}
\DeclarePairedDelimiterX{\Brace}[1]{\{}{\}}{#1}
\DeclarePairedDelimiterX{\Brack}[1]{[}{]}{#1}
\DeclarePairedDelimiterX{\Abs}[1]{\rvert}{\lvert}{#1}
\DeclarePairedDelimiterX{\Norm}[1]{\lVert}{\rVert}{#1}
\DeclarePairedDelimiterX{\Avg}[1]{\langle}{\rangle}{#1}
\DeclarePairedDelimiterX{\Round}[1]{\lfloor}{\rceil}{#1}
\DeclarePairedDelimiterX{\Floor}[1]{\lfloor}{\rfloor}{#1}
\DeclarePairedDelimiterX{\Ceil}[1]{\lceil}{\rceil}{#1}
\DeclarePairedDelimiterX{\IntRange}[1]{\llbracket}{\rrbracket}{#1}
\newcommand*{\delimsize}{}
\newcommand*{\SuchThat}{\:\delimsize\vert\:} 

\DeclareMathOperator*{\argmin}{arg\,min}
\DeclareMathOperator*{\argmax}{arg\,max}

\DeclareMathOperator{\Diag}{diag}

\newcommand*{\from}{{:}\:}

\DeclarePairedDelimiterXPP{\Var}[1]{\mathrm{Var}}(){}{#1}
\DeclarePairedDelimiterXPP{\Cov}[1]{\mathrm{Cov}}(){}{#1}
\DeclarePairedDelimiterXPP{\Expect}[1]{\mathrm{E}}(){}{#1}
\DeclarePairedDelimiterXPP{\LogPr}[1]{\ell}(){}{#1}
\newcommand*{\estim}[1]{\hat{#1}}
\newcommand*{\proxy}[1]{\hat{#1}}

\newcommand*{\Tag}[1]{\mathrm{#1}}

\newcommand*{\Set}[1]{\mathbb{#1}}
\newcommand*{\Reals}{\Set{R}}

\newcommand*{\V}[1]{\boldsymbol{#1}}
\newcommand*{\M}[1]{\mathbf{#1}}
\newcommand*{\TransposeLetter}{\top}
\newcommand*{\T}{^{\TransposeLetter}}

\newcommand{\nth}{\text{-th}\xspace}

\newcommand*{\planet}{\oplus} 
\newcommand*{\Criterion}{\mathcal{C}}
\newcommand*{\Regul}{\mathcal{R}}
\newcommand*{\RegulXYZ}{\Regul_{\V x\,\V y\,\V z}} 
\newcommand*{\RegulX}{\Regul_{\V x}}
\newcommand*{\RegulY}{\Regul_{\V y}}
\newcommand*{\RegulZ}{\Regul_{\V z}}

\newcommand{\mask}{\Tag{msk}}

\newcommand{\calib}{\Tag{cal}}

\newcommand{\HadamardProduct}{\odot} 

\newcommand*{\mas}{\mathrm{mas}}

\newcommand{\VioletBlue}[1]{\textcolor{violet!70!blue}{#1}}
\newcommand{\DarkYellow}[1]{\textcolor{yellow!70!black}{#1}}
\newcommand{\DarkPink}[1]{\textcolor{pink!80!black}{#1}}

\newcommand{\IDEA}[1]{}
\newcommand{\HIDE}[1]{}
\newcommand{\Exospeco}{\textsc{Exospeco}\xspace}
\newcommand{\noun}[1]{\textsc{#1}}

\begin{document}

\title{Characterization of stellar companion\\ from high-contrast long-slit
  spectroscopy data}

\subtitle{The EXtraction Of SPEctrum of COmpanion (\Exospeco) algorithm}

\author{Samuel Thé\inst{\ref{inst1}}
  \and Éric Thiébaut\inst{\ref{inst1}}
  \and Loïc Denis\inst{\ref{inst2}}
  \and Thibault Wanner\inst{\ref{inst1}}
  \and Rémi Thiébaut\inst{\ref{inst1}}
  \and Maud Langlois\inst{\ref{inst1}}
  \and Ferréol Soulez\inst{\ref{inst1}}}

\institute{Université de Lyon, Université Lyon1, ENS de Lyon, CNRS, Centre de
  Recherche Astrophysique de Lyon UMR 5574, F-69230, Saint-Genis-Laval, France
  \label{inst1} \\ \email{surname.name@univ-lyon1.fr} \and
  Université Jean Monnet Saint-Etienne, CNRS, Institut d Optique Graduate
  School, Laboratoire Hubert Curien UMR 5516, F-42023, SAINT-ETIENNE, France
  \label{inst2} \\
  \email{surname.name@univ-st-etienne.fr}}

\idline{Pre-print version, article under review}
\date{June 5, 2023}

\abstract
{}
{High-contrast long-slit spectrographs can be used to characterize exoplanets.
  High-contrast long-slit spectroscopic data are however corrupted by stellar
  leakages which largely dominate other signals and make the process of
  extracting the companion spectrum very challenging. This paper presents a
  complete method to calibrate the spectrograph and extract the signal of
  interest.}
{The proposed method is based on a flexible direct model of the high-contrast
  long-slit spectroscopic data. This model explicitly accounts for the
  instrumental response and for the contributions of both the star and the
  companion. The contributions of these two components and the calibration
  parameters are jointly estimated by solving a regularized inverse problem.
  This problem having no closed-form solution, we propose an alternating
  minimization strategy to effectively find the solution.}
{We have tested our method on empirical long-slit spectroscopic data and by
  injecting synthetic companion signals in these data. The proposed
  initialization and the alternating strategy effectively avoid the
  \emph{self-subtraction} bias, even for companions observed very close to the
  coronagraphic mask. Careful modeling and calibration of the angular and
  spectral dispersion laws of the instrument clearly reduce the contamination
  by the stellar leakages. In practice, the outputs of the method are mostly
  driven by a single hyper-parameter which tunes the level of regularization of
  the companion SED.}
{}


\maketitle

%
%

\section{Introduction}
\label{sec:Introduction}

High-contrast extreme adaptive optics (AO) systems such as SPHERE
\citep[Spectro-Polarimetry High-contrast Exoplanet REsearch,][]{SPHERE}, GPI
\citep[Gemini Planet Imager,][]{macintosh-2006-gpi, macintosh-2014-gpi}, or
SCExAO \citep{Jovanovic_etal-2015-SExCAO} have been developed to directly
observe the close environment of stars in the visible and the near infrared.
The study of exoplanets and their formation is one of the main scientific
objective of these instruments. One of the advantages of high-contrast extreme
AO systems is that they can provide direct access to the light from the
exoplanet which is crucial to perform spectral characterizations. Substantial
contamination by the light from the host star occurs, though: in the visible
and the near infrared, in spite of the real-time correction by the AO system
and of the masking of the host star by a coronagraph, the residual stellar
light diffracted by the instrument is much brighter than that received from
most exoplanets of interest. For this reason, dedicated post-processing methods
have been developed to track evidences of exoplanet presence in data corrupted
by strong stellar leakages. The number of published detection algorithms,
\noun{Loci} \citep{Lafreniere_etal-2007-LOCI}, \noun{Tloci}
\citep{Marois_etal-2013-TLOCI}, \noun{Klip} \citep{Soummer_et_al-2012-KLIP},
\noun{Moods} \citep{Smith_et_al-2009-exoplanet_detection}, \noun{Andromeda}
\citep{Mugnier_et_al-2009-Andromeda}, \noun{PeX} \citep{devaney_pex_2017}, and
\noun{Paco} \citep{Flasseur-2018-PACO, Flasseur-2020-robust_PACO,
  Flasseur-2020-PACO_ASDI} to name a few, reflects the scientific interest but
also the intrinsic difficulty of trustfully detecting an exoplanet from
sequences of high-contrast images. The most successful of these methods are the
ones that take into account the statistics of the stellar leakages (notably
their correlations) whether they consist in sequences of images
\citep{Smith_et_al-2009-exoplanet_detection, Flasseur-2018-PACO,
  Flasseur-2020-robust_PACO}, in sequences of multi-spectral images from
Integral field spectrographs (IFS) \citep{Flasseur-2020-PACO_ASDI}, or even in
multi-epoch sequences of images \citep{Dallant_etal-2022-PACOME-SPIE}.

After its detection, the direct characterization of an exoplanet is possible
with high-contrast extreme AO systems equipped with a spectrograph. Both SPHERE
and GPI are equipped with low resolution IFS. In addition SPHERE/IRDIS is
equipped with a medium (MRS) resolution long-slit spectrograph (LSS) in J, H,
and K bands, the latter being also available at low (LRS)
resolution\footnote{$\lambda/\Delta\lambda = 35$ for the LRS mode or
  $\lambda/\Delta\lambda = 400$ for the MRS mode} \citep{IRDIS}. With
SPHERE/IRDIS/LSS, the spectrum of a detected companion can then be measured by
aligning the slit of the spectrograph with the host star and the companion
while the host star is occulted by an opaque mask combined with the slit. In an
LSS image, the stellar leakages take the form of speckles spectrally dispersed
along oblique lines generally brighter than the companion spectrum (see
Fig.~\ref{fig:LSS_data_HR3549} for an example). In order to get rid of these
stellar leakages, \citet{IRDIS_LSS, Vigan_etal-2012-high_contrast_spectroscopy}
have developed a \emph{spectral deconvolution} (SD) method following the work
of \citet{Sparks_Ford-2002-spectral_deconvolution}. The SD method consists in a
filtering of the LSS image after a geometrical transform to align the speckles
along a given direction. In practice, the SD method is quite sensitive to the
alignment of the instrument, requires to fix defective pixels, and suffers from
a \emph{self-subtraction} bias. The latter is due to an overestimation of the
stellar leakages caused by the presence of the companion. To improve on the SD
method and reduce the \emph{self-subtraction} bias, \cite{MesaHR3549B} have
adapted the strategy implemented in \noun{Tloci} \citep{Marois_etal-2013-TLOCI}
to the case of LSS data. In spite of these improvements, existing extraction
methods suffer from a number of defaults, most of them steming from the
requirement to geometrically transform the LSS data to align the dispersed
speckles. In particular, they provide, at best, a least squares estimation of
the stellar leakages which is sub-optimal as the noise is not independent and
identically distributed (i.i.d.) in the geometrically transformed images
\citep{Thiebaut_et_al-2016-physical_constraints}. To overcome the drawbacks of
existing methods, we propose to formulate the extraction of the spectrum of a
companion as an inverse problem. The inverse problem corresponds to the joint
estimation of the contributions of the star and of its companion from the LSS
data. Not only does this approach require no transform of the LSS data (thus
avoiding the introduction of correlations) but it also yields statistically
optimal estimators. To cope both with possible instrumental misalignment and
the lack of a closed-form solution for the inverse problem, we implemented an
alternating optimization strategy with optional self-calibration stages to
solve the problem.

The outline of the paper is as follows. In Section~\ref{sec:state-of-the-art},
we present a model of the distribution of the light on the detector of a LSS
instrument. This model is used to illustrate how, after a geometrical
transform, stellar leakages can be partially removed by a truncated singular
value decomposition (TSVD) before extracting the companion signal. Such an
approach is representative of the performance that can be reached, at best, by
standard methods. We then present in Section~\ref{sec:inverse-problems} our
approach to jointly estimate the stellar leakages and the contribution of the
companion without transforming the data. For the processing of LSS data,
  knowing the spectro-angular coordinates of each detector pixel is mandatory
  and we describe in Section~\ref{sec:calibration} a numerical method to
  estimate the spatio-spectral dispersion laws from given calibration data. In
Section~\ref{sec:validation}, we validate the proposed method on both real data
from SPHERE/IRDIS/MRS and on injections of synthetic companions in real data.
We show the importance of the calibration (described in
Appendix~\ref{sec:calibration_details}) and compare our method with more standard
approaches.

\section{State of the art processing}
\label{sec:state-of-the-art}

In spite of the coronagraphic mask in high-contrast data, the stellar leakages,
in the form of quasi-static dispersed speckles, largely dominate the signal of
interest, \ie, the spectrum of the companion. These speckles, whose
contribution cannot be precisely determined by using other stars \citep[\ie by
reference differential imaging,][]{Xie_2022} or by rotating the slit to hide
the planet \citep{2016ascl.soft03001V} are a major source of nuisance for
extracting the companion spectrum. This section introduces a modeling of the
LSS data that is used in Section \ref{sec:inverse-problems} to design our
spectrum extraction method. This model is also useful to explain how previous
approaches perform the suppression of the stellar leakages \citep{IRDIS_LSS,
  MesaHR3549B}.

\subsection{Image formation}
\label{sec:image-formation}

\begin{figure}[t]
  \begin{tikzpicture}
    \draw (0,0) node{\includegraphics[scale=0.58]{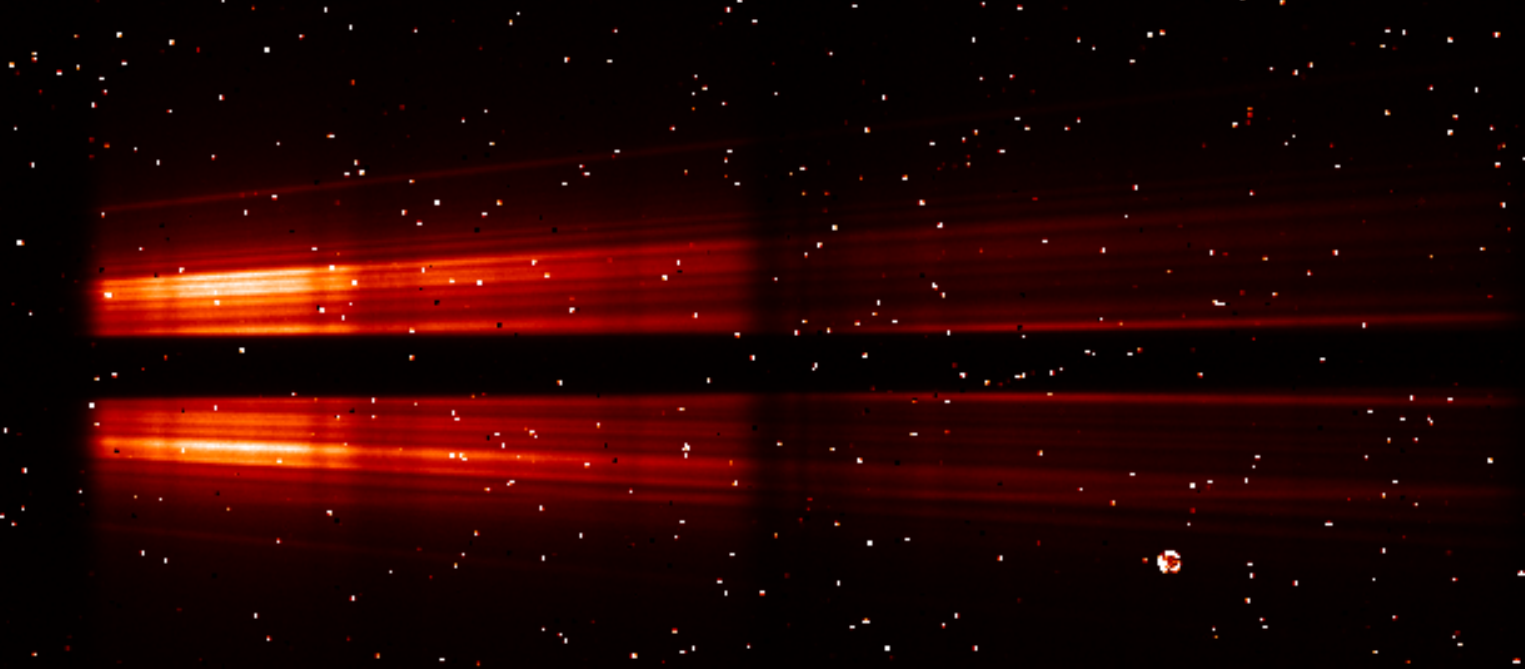}};
    \draw[>=latex,->,color=white] (-4.30,-0.20) -- (+4.00,-0.20) node[right]{$\lambda$};
    \draw[>=latex,->,color=white] (-4.20,-1.60) -- (-4.20,+1.40) node[above]{$\rho$};
    \draw[>=latex,->,color=cyan] (+4.30,-0.95) -- (+3.80,-0.95);
    \draw[>=latex,->,color=cyan] (-4.30,-0.95) -- (-3.80,-0.95);
  \end{tikzpicture}
  \caption{Long-slit medium resolution spectroscopy data of HR\,3549 taken by
    IRDIS, with horizontally the spectral axis $\lambda$ and vertically the
    angular separation axis $\rho$. The blue arrows indicate the position of
    the companion.}
  \label{fig:LSS_data_HR3549}
\end{figure}
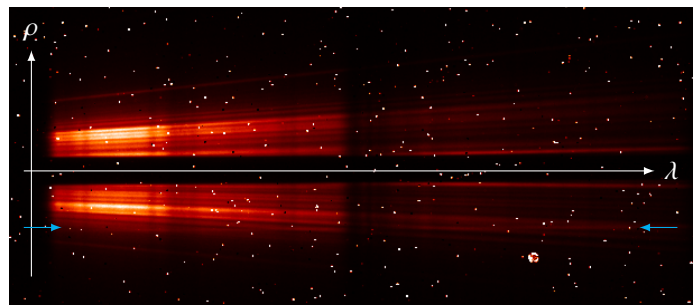

Figure \ref{fig:LSS_data_HR3549} shows a single exposure captured by the LSS of
SPHERE/IRDIS. The vector $\V d \in \Reals^{N}$, with $N$ the number of pixels,
can be modeled by:
\begin{equation}
  \label{eq:data-model}
  d_{n} = m(\rho_{n},\lambda_{n}) + \varepsilon_{n}
\end{equation}
with $m(\rho,\lambda)$ the distribution of light in the detector plane at
angular coordinate $\rho$ along the slit and wavelength $\lambda$, $\rho_{n}$
and $\lambda_{n}$ the angular and spectral coordinates at the $n$-th pixel, and
$\varepsilon_{n}$ the contribution of the noise. Our notations are
  summarized in Table~\ref{tab:notations}. The light distribution in the
detector plane is the sum of the contributions by the star and by the
companion:
\begin{equation}
  \label{eq:light-distribution}
  m(\rho,\lambda) = f_\star(\lambda)\,h_\star(\rho,\lambda) +
  f_\planet(\lambda)\,h_\planet(\rho,\lambda)
\end{equation}
with $f_\star$ and $f_\planet$ the spectral energy distributions (SEDs) of the
star and of the companion as seen by the detector\footnote{these SEDs are
  affected by the chromatic transmission of the atmosphere and of the
  instrument}, $h_\star$ and $h_\planet$ the point spread functions (PSFs) for
a source at the respective angular positions of the star and of the companion,
the so-called on-axis and off-axis PSFs. The on-axis PSF,
$h_\star(\rho,\lambda)$ explains the oblique bright lines due to stellar
leakages in Fig.~\ref{fig:LSS_data_HR3549} while the stellar SED
$f_\star(\lambda)$ explains the variations of intensity along these lines. As
can be seen in Fig.~\ref{fig:LSS_data_HR3549}, the companion signal, that is
$f_\planet(\lambda)\,h_\planet(\rho,\lambda)$, is barely distinguishable in the
LSS data and it is mandatory to get rid of the stellar leakages
$f_\star(\lambda)\,h_\star(\rho,\lambda)$.

\begin{table}
  \caption{Notations. Lowercase letters are for continuous functions and
      scalars (\eg $f_\star$), boldface lowercase letters for \emph{vectors}
      (\eg $\V x$), and boldface uppercase letters for linear mappings, a.k.a.
      \emph{matrices}, (\eg $\V F_\star$). Vectors with a hat (\eg
      $\estim{\V x}$) are estimators. The main unknowns of the problem are
      $\V x$, the sampled star SED, $\V y$, the sampled on-axis PSF, and $\V z$,
      the sampled companion SED. \label{tab:notations}}
  \begin{tabular}{ll}
    \hline
    \textbf{Notation} & \textbf{Description} \\
    \hline
    subscript $\star$ & Parameters of the stellar model\\
    subscript $\planet$ & Parameters of the companion model\\
    \hline
    $\V d \in \Reals^N$ & Science data\\
    $\V m \in \Reals^N$ & Sampled model of $\V d$\\
    $\V w \in\Reals^N$ & Diagonal of the precision matrix of $\V d$\\
    $\V\lambda \in \Reals^N$ & Pixel-wise wavelengths\\
    $\V\rho \in \Reals^N$ & Pixel-wise angular positions\\
    \hline
    $f_\star$ & Continuous star SED \\
    $\V x\in\Reals^{N_{\V x}}$
                      & Sampled star SED $f_\star$\\
    $\V\lambda^\Tag{grd}_\star \in \Reals^{N_{\V x}}$
                      & Sampling wavelengths for $\V x$\\
    $\M F_\star\in\Reals^{N\times N_{\V x}}$
                      & Interpolation operator: $\V x$ to pixel-wise $f_\star$\\
    \hline
    $h_\star$ & Continuous on-axis PSF \\
    $\V y\in\Reals^{N_{\V y}}$
                      & Sampled on-axis PSF $h_\star$ at $\lambda^\Tag{ref}$\\
    $\V\rho^\Tag{grd}_\star \in \Reals^{N_{\V y}}$
                      & Sampling angles for $\V y$\\
    $\rho_\star$      & Angular position of the star\\
    $\V \nu_\star\in\Omega_\star$
                      & Calibration parameters of $h_\star$\\
    $\Omega_\star$ & Feasible set $h_\star$ parameters\\
    $\M H_\star\in\Reals^{N\times N_{\V y}}$
                      & Interpolation operator: $\V y$ to pixel-wise $h_\star$ \\
    & at $\lambda^\Tag{ref}$\\
    \hline
    $f_\planet$ & Continuous companion SED \\
    $\V z\in\Reals^{N_{\V z}}$
                      & Sampled companion SED $f_\planet$\\
    $\V\lambda^\Tag{grd}_\planet \in \Reals^{N_{\V x}}$
                      & Sampling wavelengths for $\V z$\\
    $\M F_\planet\in\Reals^{N\times N_{\V z}}$
                      & Interpolation operator: $\V z$ to pixel-wise $f_\planet$\\
    \hline
    $h_\planet$ & Continuous off-axis PSF \\
    $\V h_\planet\in\Reals^N$
                      & Sampled off-axis PSF $h_\planet$ at $\lambda^\Tag{ref}$\\
    $\V \nu_\planet\in\Omega_\planet$
                      & Calibration parameters of $h_\planet$\\
    $\Omega_\planet$ & Feasible set of $h_\planet$ parameters\\
    $\rho_\planet\in\Reals$
                      & Angular position of the companion\\
    \hline
    $\V\mu = (\mu_{\V x}, \mu_{\V y}, \mu_{\V z})$&Hyper-parameters\\
    \hline
    $\V \gamma\in\Reals^N$
                      & Pixel-wise chromatic scaling factors\\
    $\V \nu = (\V\nu_\star,\V\nu_\planet) \in \Omega$
                      & Calibration parameters\\
$\Omega = \Omega_{\star}\times\Omega_{\planet}$   & Feasible set of calibration parameters\\ \hline
     $\Lambda$ & Spectral dispersion law\\
     $\V a$ & Parameters of the spectral dispersion law\\
    $\varrho$ &Angular dispersion law\\
     $\V s$ & Parameters of the angular dispersion law\\
    $\Delta\rho$ & Width of the coronagraphic mask\\

    \hline
  \end{tabular}
\end{table}

\subsection{Low rank approximation of the stellar leakages}
\label{sec:TSVD}

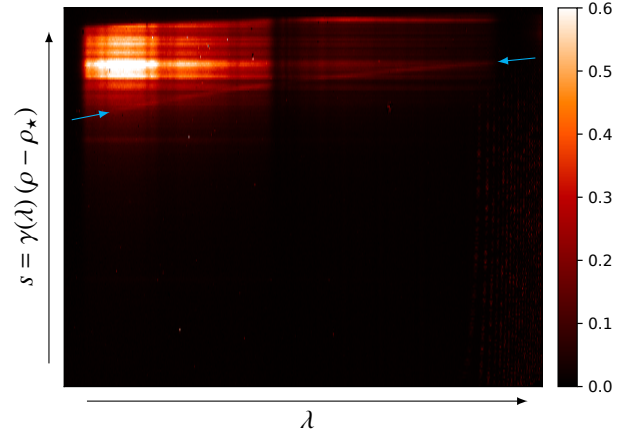
\begin{figure}[t]
  \centering
  \begin{tikzpicture}
    \draw (0,0) node{\includegraphics[scale=0.58]{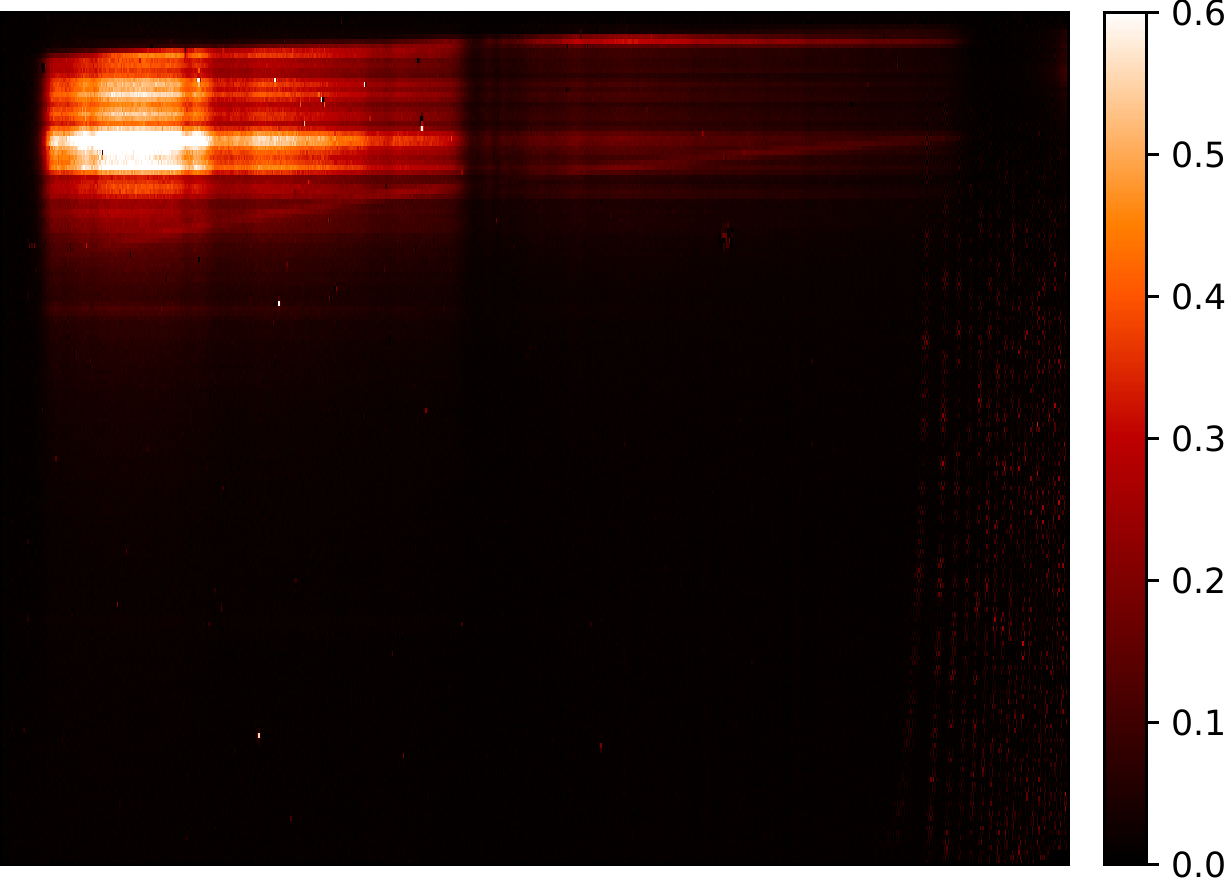}};
    \draw[>=latex,->] (-3.3,-2.7) -- (2.5,-2.7) node[midway,below]{$\lambda$};
    \draw[>=latex,->] (-3.8,-2.2) -- (-3.8,2.2);
    \draw (-4.1,0) node[rotate=90]{$s = \gamma(\lambda)\,(\rho - \rho_{\star})$};
    \draw[>=latex,->,color=cyan] (+2.60,+1.85) -- (+2.10,+1.80);
    \draw[>=latex,->,color=cyan] (-3.50,+1.03) -- (-3.00,+1.13);
  \end{tikzpicture}
  \caption{Warped HR\,3549 image. This figure shows the bottom half of the data
    shown in Fig.~\ref{fig:LSS_data_HR3549}, corresponding to the side where
    lies the companion, warped so as to align the dispersed speckles of the
    stellar leakages. The warping is defined by the coordinates $\rho$ and
    $\lambda$ of the pixels given by the \emph{complex} calibration model of
    the spectral and angular dispersion laws described in
    Appendix~\ref{sec:calibration_details}. The companion signal can be seen as a faint
    curved track indicated by the blue arrows.}
  \label{fig:warped-image}
\end{figure}

\citet{devaney_pex_2017} have shown that, except in the vicinity of the
coronagraphic mask, the chromatic PSF can be written in the form of a series
expansion. Applying their model to the star and taking into account that our
data have one angular dimension instead of two yields:
\begin{equation}
  \label{eq:multi-mode-psf}
  h_\star(\rho,\lambda) = \sum\nolimits_{k \ge 1} \gamma(\lambda)^{k}\,
  h_{\star,k}\Paren[\big]{\gamma(\lambda)\,(\rho - \rho_\star)}
\end{equation}
with $\gamma(\lambda) = \lambda^\Tag{ref}/\lambda$ a chromatic magnification
factor relative to some arbitrary reference wavelength $\lambda^\Tag{ref}$,
$\rho_\star$ the angular position of the star along the slit to account for a
possible pointing error of the instrument, and
$\Brace{h_{\star,k}}_{k=1,\ldots}$ a family of spatial PSF modes at the
reference wavelength.

Since the stellar leakages dominate the signal in the LSS image $\V d$,
Eq.~\eqref{eq:multi-mode-psf} suggests to apply specific image warping so as to
form a 2-D image $\V d^\Tag{warp}$ whose, say, first dimension varies along the
wavelength while its second dimension varies along coordinate
$s = \gamma(\lambda)\,(\rho - \rho_\star)$ (see Fig.~\ref{fig:warped-image}).
According to Eqs.~\eqref{eq:data-model} and \eqref{eq:multi-mode-psf}, the
warped image is modeled and then approximated by:
\begin{align}
  d^\Tag{warp}_{i,j} 
  \label{eq:warped-image}
  &= m\Paren*{\rho_\star + s^\Tag{warp}_{j}/\gamma\Paren*{\lambda^\Tag{warp}_{i}},
    \lambda^\Tag{warp}_{i}}
    + \varepsilon^\Tag{warp}_{i,j}\\
  \label{eq:warped-image-approx}
  &\approx \sum_{k \ge 1}
    \gamma\Paren*{\lambda^\Tag{warp}_{i}}^{k}\,
    f_\star\Paren*{\lambda^\Tag{warp}_{i}}\,
    h_{\star,k}\Paren*{s^\Tag{warp}_{j}}
\end{align}
where $\V\varepsilon^\Tag{warp}$ in Eq.~\eqref{eq:warped-image} denotes the
contribution of the noise in the warped image while the $\approx$ symbol in
Eq.~\eqref{eq:warped-image-approx} is to account for the contributions of the
potential companion and for the noise which have been neglected. In words, the
stellar leakages appear to have a simple separable decomposition in the warped
image.

\begin{figure}
  \centering
  \begin{tikzpicture}
    \draw (0,0) node{\includegraphics[scale=0.58]{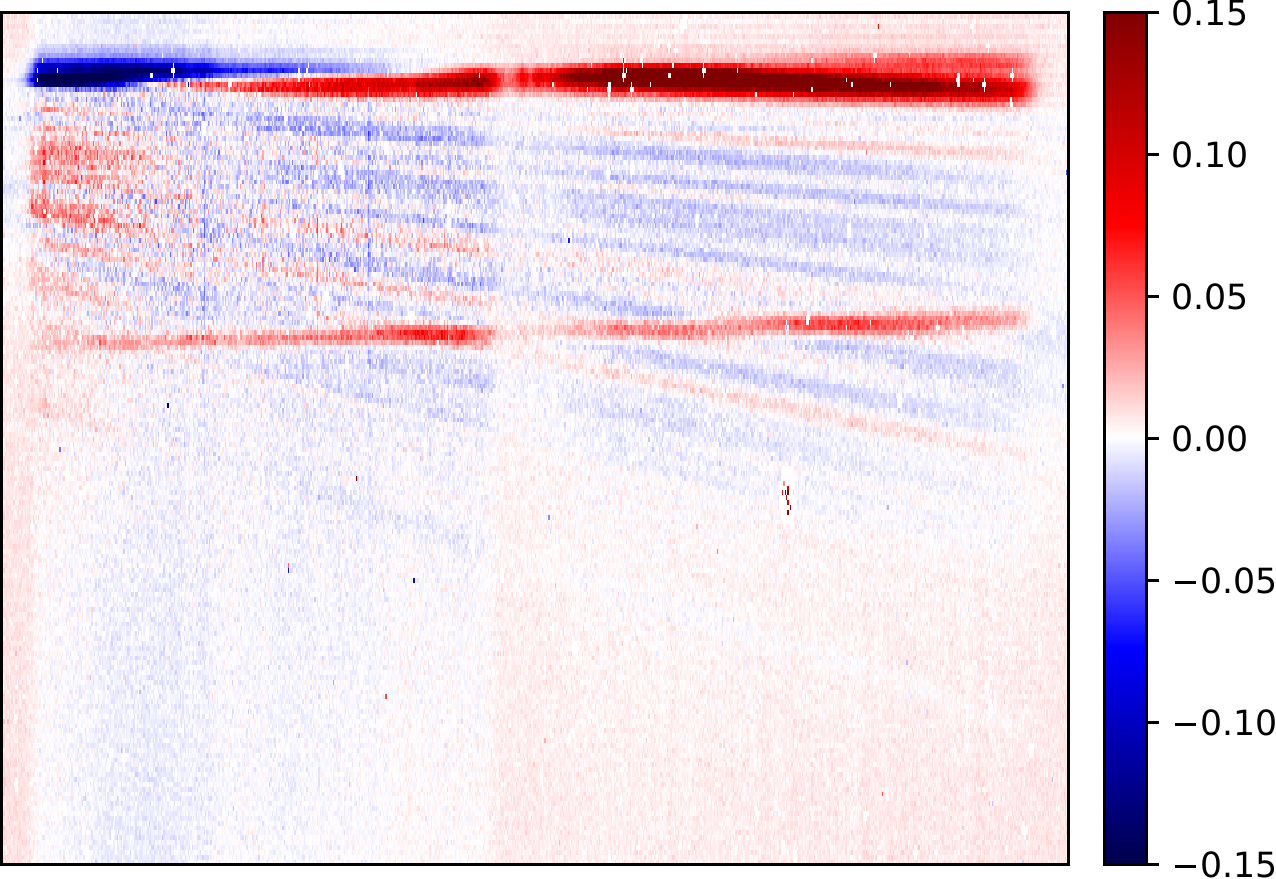}};
    \draw[>=latex,<-,color=black] (+2.05,+0.70) -- ++(+0.40,+0.04);
    \draw[>=latex,<-,color=black] (-3.30,+0.55) -- ++(-0.40,-0.04);
  \end{tikzpicture}
  \caption{Bottom half of the residual image $\V r_\planet$ for the HR\,3549
    data with stellar leakages estimated by the TSVD method as defined in
    Eq.~\ref{eq:TSVD-residuals} and with the warped image shown in
    Fig.~\ref{fig:warped-image}. Compared to the original data shown in
    Fig.~\ref{fig:LSS_data_HR3549}, the companion signal appears more
    distinctly (indicated by the 2 arrows).}
  \label{fig:TSVD-residuals}
\end{figure}

The Singular Value Decomposition (SVD) of the warped image\footnote{considered
  as a simple matrix} writes:
\begin{align}
  \label{eq:warped-SVD}
  \V d^\Tag{warp}
  &= \sum_{k = 1}^{\min(N_{1},N_{2})} \V u_{k}\,\sigma_{k}\,\V v_{k}\T
\end{align}
where $N_{1}$ and $N_{2}$ are the dimensions of the warped image,
$\V u_{k} \in \Reals^{N_{1}}$ is the $k$-th left singular vector of the
decomposition, $\sigma_{k} \ge 0$ is the $k$-th singular value, and
$\V v_{k} \in \Reals^{N_{2}}$ is the $k$-th right singular vector. Comparing
Eq.~\eqref{eq:warped-image-approx} and Eq.~\eqref{eq:warped-SVD}, the SVD of
$\V d^\Tag{warp}$ readily provides a decomposition similar to the contribution
of the stellar leakages with, for each index $k$, the left singular vector
$\V u_{k}$ sampling $\gamma(\lambda)^{k}\,f_\star(\lambda)$ as a function of
$\lambda$ and the right singular vector $\V v_{k}$ sampling $h_{\star,k}(s)$ as
a function of $s$ (both up to a normalization factor that depends on $k$). The
\emph{truncated singular value decomposition} (TSVD) of the warped image is
obtained by limiting the sum in the right-hand side of
Eq.~\eqref{eq:warped-SVD} to the $k_\Tag{max} \le \min(N_{1},N_{2})$ first
terms and, by Eckart--Young--Mirsky \citep{Eckart_Young-1936-approximation,
  Mirsky-1960-symmetric_gauge} theorem, it is the best possible approximation
of $\V d^\Tag{warp}$ of rank $k_\Tag{max}$ in the least squares sense. Hence,
it may be assumed that, for a suitable choice of $k_\Tag{max}$, the TSVD of
$\V d^\Tag{warp}$ provides a good approximation of the stellar leakages without
being too much affected by the companion signal (if the companion is not too
bright) and by the noise. A residual image that mostly depends on the companion
can then be formed by subtracting the un-warped TSVD of the warped image
$\V d^\Tag{warp}$ from the LSS image $\V d$:
\begin{align}
  \label{eq:TSVD-residuals}
  \V r_\planet = \V d -
  \mathcal{U}\Paren*{\sum_{k=1}^{k_\Tag{max}} \V u_{k}\,\sigma_{k}\,\V v_{k}\T}
\end{align}
where $\mathcal{U}$ denotes the un-warping operation\footnote{\eg, a simple
  separable 2-D interpolation}. As illustrated by
Fig.~\ref{fig:TSVD-residuals}, the signal of interest, the companion SED, is
then
easier to extract from the residual image $\V r_\planet$. This can be done by
standard aperture photometry tools.

As pointed by \citet{devaney_pex_2017}, there are a number of issues in using
the TSVD to get rid of the stellar leakages in multi-wavelengths high-contrast
data. First, to produce a rectangular warped image (that can be interpreted as
a matrix to perform the SVD), quite substantial regions of the original data
$\V d$ have to be discarded (near the coronagraphic mask and the edges of the
formed image). This limits the range of admissible angular positions for the
companion and gets rid of data that might be valuable to improve the estimation
of the stellar leakages. Second, the presence of a companion in the original
data $\V d$ yields a positive bias in the approximation of the stellar leakages
by the TSVD. This results in a negative bias in the residual image and, hence,
in the estimated companion SED. This artifact is known in the literature as
\emph{"self-subtraction"}. Third, the least squares fit performed by the TSVD
of $\V d^\Tag{warp}$ is sub-optimal regarding the distribution of the noise in
the warped image. Indeed, least squares are only optimal for independent
identically distributed (i.i.d.) noise which is certainly not the case for
$\varepsilon^\Tag{warp}_{i,j}$: at least, the shot noise in the image $\V d$
has a non-uniform distribution and a side effect of the transform of $\V d$ to
yield the warped image $\V d^\Tag{warp}$ is to introduce correlations.
Moreover, defective pixels, which are quite numerous for the kind of detectors
used by NIR instruments such as LSS, must be corrected, usually by
averaging their neighbors values, before warping the image. This correction can
only introduce additional correlations.

In spite of these drawbacks, proposed processing methods \citep{IRDIS_LSS,
  MesaHR3549B} are similar to the TSVD approximation of warped LSS images
(optimal linear combination of a set of images). Some refinements have been
proposed to limit the \emph{self-subtraction} bias \citep{MesaHR3549B} but the
other issues have been largely left unaddressed. In the remaining of this
paper, we propose a new inverse problems approach to solve all aforementioned
limitations.

\section{Inverse problems approach}
\label{sec:inverse-problems}

To avoid the issues resulting from warping the LSS image, we propose to solve
an \emph{inverse problem} which consists in jointly estimating the parameters
of the direct model of the data given in Eqs.~\eqref{eq:data-model} and
\eqref{eq:light-distribution} without transforming the data themselves. For an
optimal information extraction, we model the likelihood of the data to consider
the uneven quality of the data and, therefore, account for defective pixels or
missing data in a consistent way. Besides, our approach relies on a precise
calibration of the spectro-spatial instrumental dispersion as seen by the
detector. The proposed method includes auto-calibration stages to refine the
calibration parameters and thus accounts for a possible misalignment of the
science exposures.

\subsection{Assumed continuous model}
\label{sec:continuous-model}

\begin{figure*}[!t]
	\includegraphics[width=\textwidth]{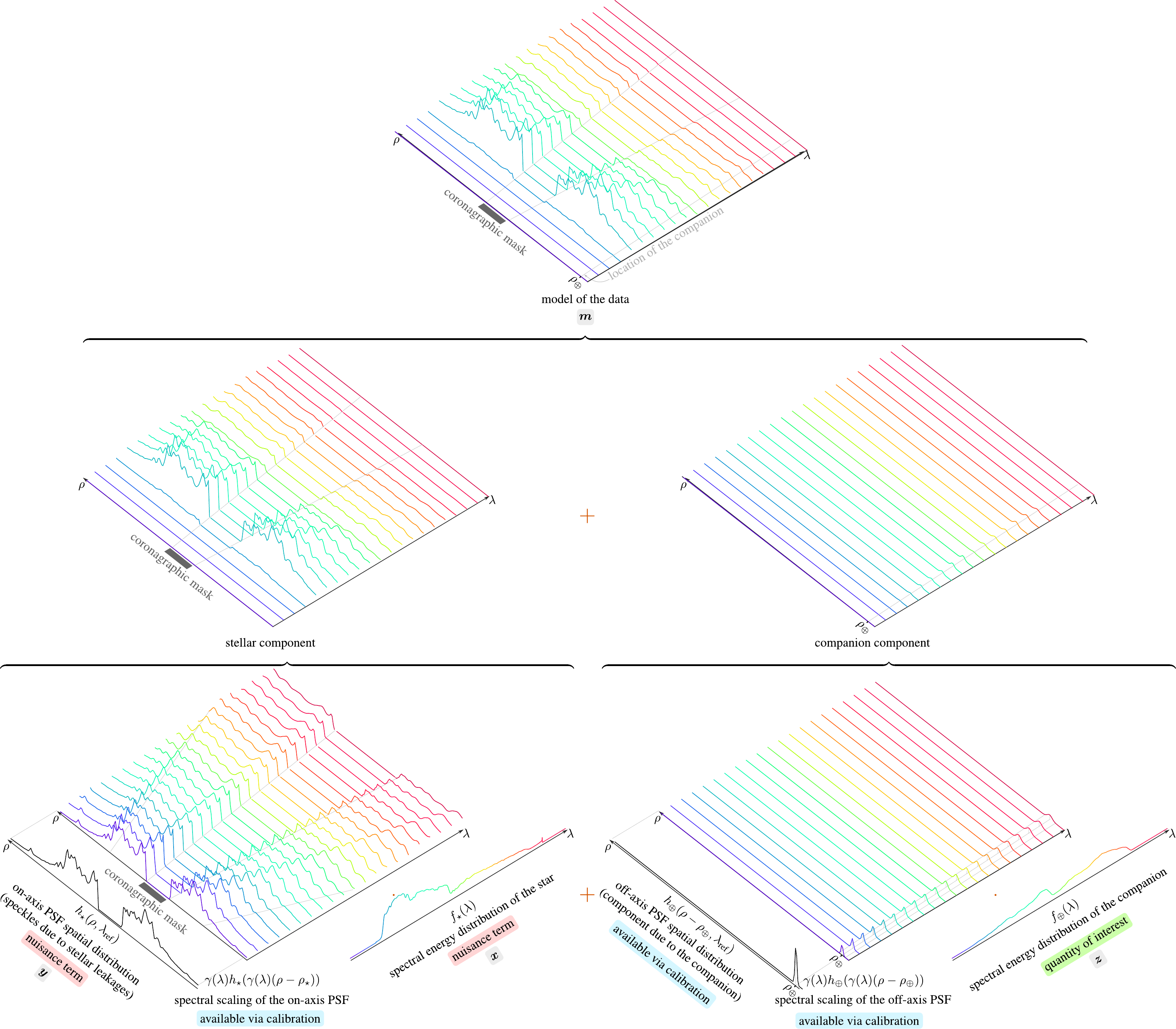}
	\caption{Illustration of the direct model for high-contrast long-slit
    spectroscopy given in Eq.(\ref{eq:light-distribution-simplified}): the data
    are modeled as the sum of two components: a stellar component and a
    companion component. Extracting the SED of the companion also requires the
    estimation of the on-axis PSF and the SED of the host star. The
      labels ``\emph{available via calibration}'' denote components that may be
      self-calibrated by \Exospeco directly from the science data (see
      Section~\ref{sec:The_EXOSPECO_algorithm} for details).}
	\label{fig:direct_model}
\end{figure*}

To simplify the on-axis PSF model in Eq.~\eqref{eq:multi-mode-psf}, we keep
only the first and most significant of these modes and thus assume
that:
\begin{equation}
  \label{eq:on-axis-psf}
  h_\star(\rho,\lambda) = \gamma(\lambda)\,
  h_\star\Paren[\big]{\gamma(\lambda)\,(\rho - \rho_\star)}
\end{equation}
with $h_\star(\rho) = h_{\star,1}(\rho)$ the first spatial mode of the on-axis
PSF. As shown in Section \ref{sec:validation}, this simple model of the
  stellar leakages already gives excellent results. Likewise, the chromatic
off-axis PSF $h_\planet$ can also be written as:
\begin{equation}
  \label{eq:off-axis-psf}
  h_\planet(\rho, \lambda) \approx \gamma(\lambda)\,
  h_\planet\Paren[\big]{\gamma(\lambda)\,(\rho - \rho_\planet)}
\end{equation}
with $h_\planet(\rho) = h_\planet(\rho, \lambda^\Tag{ref})$ the off-axis PSF at
the reference wavelength $\lambda^\Tag{ref}$ and $\rho_\planet$ the angular
position of the companion along the slit. Note that the $\gamma(\lambda)$
factor ensures that the on-axis and off-axis PSFs be normalized at all
wavelengths provided the PSF at the reference wavelength be also normalized, \ie
$\int h(\rho,\lambda)\,\mathrm{d}\rho = 1$ ($\forall\lambda$). These
approximations for the on-axis and off-axis PSFs yield the following simplified
model
that we consider in the rest of the paper:
\begin{align}
  m(\rho,\lambda) = \gamma(\lambda)\,\bigl[
  &f_\star(\lambda)\,h_\star\Paren[\big]{\gamma(\lambda)\,(\rho - \rho_\star)}
    \notag\\
  &+ f_\planet(\lambda)\,
    h_\planet\Paren[\big]{\gamma(\lambda)\,(\rho - \rho_\planet)}
    \bigr].
    \label{eq:light-distribution-simplified}
\end{align}

\subsection{Discretized distribution}
\label{sec:discrete-model}

In order to fit the data, the model $m(\rho,\lambda)$ in
Eq.~\eqref{eq:light-distribution-simplified} has to be estimated at each
angular and spectral coordinates $(\rho_n,\lambda_n)$ of the $N$ pixels of the
detector. These pixel coordinates can be identified by fitting angular and
spectral dispersion laws to calibration data, as explained in
Appendix~\ref{sec:calibration_details}. Because of these angular and spectral
dispersion laws, continuous functions are needed to model the SEDs of the star
and the companion ($f_\star$ and $f_\planet$) and their respective PSFs
($h_\star$ and $h_\planet$) on the sensor pixel grid. We explain next how we
parameterize these functions.

Our models of the star SED $f_\star(\lambda)$, of the on-axis PSF
$h_\star(\rho)$, and of the companion SED $f_\planet(\lambda)$ are given by the
following linear interpolations:
\begin{subequations}
  \begin{align}
    \label{eq:star-SED-continuous}
    & f_\star(\lambda) = \sum_{j=1}^{N_{\V x}}
      \varphi_\star\Paren*{\lambda - \lambda^\Tag{grd}_{\star,j}}\,
      \underbrace{f_\star\Paren*{\lambda^\Tag{grd}_{\star,j}}}_{\displaystyle x_j},\\
    \label{eq:star-PSF-continuous}
    & h_\star(\rho) = \sum_{j=1}^{N_{\V y}}
      \psi_\star\Paren*{\rho - \rho^\Tag{grd}_{\star,j}}\,
      \underbrace{h_\star\Paren*{\rho^\Tag{grd}_{\star,j}}}_{\displaystyle y_j},\\
    \label{eq:planet-SED-continuous}
    & f_\planet(\lambda) = \sum_{j=1}^{N_{\V z}}
      \varphi_\planet\Paren*{\lambda - \lambda^\Tag{grd}_{\planet,j}}\,
      \underbrace{f_\planet\Paren*{\lambda^\Tag{grd}_{\planet,j}}}_{\displaystyle z_j},
  \end{align}
\end{subequations}
with $\varphi_\star\from\Reals\to\Reals$, $\psi_\star\from\Reals\to\Reals$, and
$\varphi_\planet\from\Reals\to\Reals$ chosen interpolation functions, and where
$\V\lambda^\Tag{grd}_\star \in \Reals^{N_{\V x}}$ is an evenly spaced grid of
wavelengths to sample the star SED $f_\star$,
$\V\rho^\Tag{grd}_\star \in \Reals^{N_{\V y}}$ is an evenly spaced grid of
angles to sample the on-axis PSF $h_\star$, and
$\V\lambda^\Tag{grd}_\planet \in \Reals^{N_{\V z}}$ is an evenly spaced grid of
wavelengths to sample the companion SED $f_\planet$. At the coordinates
$(\rho_n,\lambda_n)$ of any pixel $n \in \IntRange{1,N}$ of the detector, our
linear interpolation yields:
\begin{subequations}
  \begin{align}
    \label{eq:star-SED-discrete}
    f_{\star,n} &= f_\star(\lambda_n)
      = \sum\nolimits_{j = 1}^{N_{\V x}} \underbrace{
      \varphi_\star\Paren*{\lambda_n - \lambda^\Tag{grd}_{\star,j}}
      }_{\displaystyle F_{\star,n,j}}\,x_j
      = \Paren*{\M F_\star\,\V x}_n,\\
    \label{eq:star-PSF-discrete}
      h_{\star,n} &= h_\star\bigl(\gamma_n\,(\rho_n-\rho_\star)\bigr)
      = \sum\nolimits_{j=1}^{N_{\V y}} \underbrace{
      \psi_\star\Paren*{\gamma_n\,\Paren*{\rho_n - \rho_\star - \rho^\Tag{grd}_{\star,j}}}
      }_{\displaystyle H_{\star,n,j}}\,y_j
      \nonumber\\
      &= \Paren*{\M H_\star\,\V y}_n,\\
    \label{eq:planet-SED-discrete}
     f_{\planet,n} &= f_\planet(\lambda_n)
      = \sum\nolimits_{i=1}^{N_{\V z}} \underbrace{
      \varphi_\planet\Paren*{\lambda_n - \lambda^\Tag{grd}_{\planet,j}}
      }_{\displaystyle F_{\planet,n,i}}\,z_i
      = \Paren*{\M F_\planet\,\V z}_n,
  \end{align}
\end{subequations}
with $\gamma_n = \gamma(\lambda_n)$, and the matrices $\M
F_\star\in\Reals^{N\times N_{\V x}}$,
$\M H_\star\in\Reals^{N\times N_{\V y}}$, and
$\M F_\planet\in\Reals^{N\times N_{\V z}}$ defined in
Eqs.~\eqref{eq:star-SED-discrete}--\eqref{eq:planet-SED-discrete} represent
interpolation operators\footnote{In practice, the interpolation operators are
  very sparse and only their non-zero entries need to be stored or computed on
  the fly.}. These operators are applied to the \emph{vectors}\footnote{We use
  boldface lowercase letters to denotes \emph{vectors}, that is quantities that
  depend on a single index, and boldface uppercase letters to denotes
  \emph{linear operators}, that is quantities that depend on two indices.}
$\V x\in\Reals^{N_{\V x}}$, $\V y\in\Reals^{N_{\V y}}$, and
$\V z\in\Reals^{N_{\V z}}$ defined in
Eqs.~\eqref{eq:star-SED-continuous}--\eqref{eq:planet-SED-continuous}. They
form the unknown parameters of our models of the star SED $f_\star$, of the
on-axis PSF
$h_\star$, and of the companion SED $f_\planet$.

The interpolation functions ($\varphi_\star$, $\psi_\star$, and
$\varphi_\planet$) and the sampling lists ($\V\lambda^\Tag{grd}_\star$,
$\V\rho^\Tag{grd}_\star$, and $\V\lambda^\Tag{grd}_\planet$) may be chosen
differently for each component of the model. If the spectral sampling lists and
spectral interpolation functions are the same (as we chose for our
experiments), then the two spectral interpolation operators $\M F_\planet$ and
$\M F_\star$ are the same. In our implementation of the method, we selected the
\citet{CATMULL1974317} cardinal cubic spline $\varphi$ as the interpolation
function:
$\varphi_\star(\lambda) = \varphi(\lambda/\Delta\lambda^\Tag{grd}_\star)$,
$\psi_\star(\rho) = \varphi(\rho/\Delta\rho^\Tag{grd}_\star)$, and
$\varphi_\planet(\lambda) = \varphi(\lambda/\Delta\lambda^\Tag{grd}_\planet)$
with $\Delta\lambda^\Tag{grd}_\star$, $\Delta\rho^\Tag{grd}_\star$, and
$\Delta\lambda^\Tag{grd}_\planet$ the sampling steps of
$\V\lambda^\Tag{grd}_\star$, $\V\rho^\Tag{grd}_\star$, and
$\V\lambda^\Tag{grd}_\planet$.

For the off-axis PSF $h_\planet(\rho)$ at the reference wavelength, we consider
a simple parametric model. Since the principal lobe of the off-axis PSF
represents most of the energy received from the companion, we assume a Gaussian
approximation:
\begin{equation}
  h_\planet(\rho) = \frac{1}{\sqrt{2\pi}\sigma_\planet}\,
  \exp\Paren[\Big]{-\frac{\rho^2}{2\sigma_\planet^2}}.
  \label{eq:off-axis-psf-at-lambda-ref}
\end{equation}
Hence $\V h_\planet$, the sampled off-axis PSF at the reference wavelength for
the companion, depends on the angular position of the companion $\rho_\planet$
and on $\sigma_\planet$ the standard deviation of the PSF at the reference
wavelength. Other parametric models of the off-axis PSF could be considered
with a simple adaptation of the algorithm proposed in
Section~\ref{sec:minimization-strategy}.

Finally, we introduce the $N$-vectors $\V m\in\Reals^{N}$, $\V
\gamma\in\Reals^{N}$ and $\V h_{\planet}\in\Reals^{N}$ defined by:
\begin{subequations}
  \begin{align}
    m_n &= m(\rho_n,\lambda_n),\\
    \gamma_n
        &=  \gamma(\lambda_n) = \lambda^\Tag{ref}/\lambda_n,\\
    h_{\planet,n}
        &= h_\planet\Paren[\big]{\gamma_n\,(\rho_n - \rho_\planet)},
          \label{eq:off_axis_psf_discrete}
  \end{align}
\end{subequations}
for $n \in \IntRange{1,N}$. The discretized model of the light distribution in
Eq.~\eqref{eq:light-distribution-simplified} then writes:
\begin{align}
  \label{eq:direct_model}
  \V m(\V x,\V y,\V z,\V\nu)
  &= \V\gamma\HadamardProduct(\M H_\star(\V\nu)\,\V y)\HadamardProduct(\M F_\star\,\V x)
    + \V\gamma\HadamardProduct\V h_\planet(\V\nu)\HadamardProduct(\M F_\planet\,\V z)
\end{align}
with $\HadamardProduct$ the Hadamard product (entry-wise multiplication) and
$\V\nu = (\V\nu_\star,\V\nu_\planet)$ the \emph{calibration parameters} of the
model which are the other unknown parameters than $\V x$, $\V y$, or $\V z$.
With our Gaussian approximation of the off-axis PSF at the reference
wavelength, the calibration parameters for the companion are
$\V\nu_\planet=(\rho_\planet,\sigma_\planet)$. To account for a possible
misalignment between the coronagraphic mask and the star, the calibration
parameters for the star are just $\V\nu_\star = (\rho_\star)$, with
$\rho_\star$ the angular position of the star along the slit.

The signal-processing problem then amounts to estimating the companion's SED
$\V z$ as well as the other nuisance parameters of the model, $\V x$, $\V y$,
and $\V \nu $. A method to perform this task is
proposed in the next section.

\subsection{Objective function and regularization}
\label{sec:Criterion_and_regularization}

After proper calibration of the detector, raw images are pre-processed to
compensate for bias and gain non-uniformity and to identify defective pixels
(\ie pixels with a non-linear response). This pre-processing produces the
long-slit spectroscopy data $\V d\in\Reals^N$ considered here and modeled by
$\V m(\V x, \V y, \V z, \V\nu)$ in Eq.~\eqref{eq:direct_model}. Due to photon
and detector noises as well as modeling inaccuracies, some discrepancies are
expected between the data $\V d$ and our model $\V m(\V x, \V y, \V z, \V\nu)$.
Due to the observed flux level, there are enough photons detected per pixel for
the data $\V d$ to approximately follow a Gaussian distribution of mean the
model $\V m(\V x, \V y, \V z, \V\nu)$ and of precision matrix\footnote{the
  precision matrix is the inverse of the covariance matrix} $\M W$. Since
  we directly consider the data without any pixel interpolation (i.e.\ no image
  warping to align the dispersed speckles and no attempt to fix defective
  pixels), no correlations are introduced in the data and the pixels can be
  considered as mutually independent. The precision matrix is thus diagonal,
$\M W = \Diag(\V w)$ where $\V w \in\Reals^N$ collects the diagonal entries of
$\M W$ and is given by:
\begin{equation}
  \label{eq:weights}
  w_{n} = \begin{cases}
    0 & \text{if $n$-th pixel is invalid,}\\
    1/\Var{d_n} & \text{otherwise.}
  \end{cases}
\end{equation}
where $\Var{d_n}$ can be estimated by different pre-processing methods
  \citep{mugnier2004_Mistral,berdeuPIC}. We consider as \emph{invalid} all
pixels for which the model is incorrect, this includes defective pixels, pixels
too much impacted by the coronagraphic mask, and pixels located outside of the
field of view (see Fig.~\ref{fig:HR3549_data_with_mask_valid_pixel}). We assume
that the estimation of the variances and the identification of defective pixels
are part of the pre-processing stage. The definition of the precision matrix in
Eq.~\eqref{eq:weights} amounts to assuming that the variance of invalid pixels
is infinite. In other words, this expresses that the values of invalid pixels
should not be considered at all. Given the large number of unknowns, the
estimation of the stellar and companion components $\V x$, $\V y$, and $\V z$
cannot be performed solely by fitting the data: regularity constraints are
necessary to prevent noise amplification and cope with missing data
\citep{Titterington-1985-regularization}. We consider regularized estimators
obtained by minimizing the following criterion:
\begin{equation}
  \label{eq:criterion}
  \Criterion(\V x,\V y,\V z,\V\nu,\V \mu) =
  \Norm*{\V d - \V m(\V x,\V y,\V z,\V\nu)}^2_{\M W}
  + \RegulXYZ(\V x,\V y,\V z, \V\mu)\,,
\end{equation}
where the first term is a statistical distance between the model and the data
(the co-log-likelihood)
while $\RegulXYZ(\V x,\V y,\V z, \V\mu)$ is a regularization term parameterized
by the vector $\V \mu$ of so-called \emph{hyper-parameters}. In the above
equation, $\Norm{\V u}^2_{\M W} = \V u\T\,\M W\,\V u$ denotes the squared
\citet{Mahalanobis-1936} norm. Our estimators $\estim{\V x}$, $\estim{\V y}$,
$\estim{\V z}$, and $\estim{\V\nu}$ of the parameters of interest are the ones
that jointly minimize the criterion in Eq.~\eqref{eq:criterion}:
\begin{equation}
  \label{eq:joint_estimation_problem}
  \Paren*{
    \estim{\V x}(\V\mu), \estim{\V y}(\V\mu),
    \estim{\V z}(\V\mu), \estim{\V\nu}(\V\mu)
  } = \argmin_{\substack{\V x\ge0,\: \V y\ge0,\\\V z\ge0,\: \V\nu\in\Omega}}
  \Criterion(\V x,\V y,\V z,\V\nu,\V\mu).
\end{equation}
These estimators depend on the hyper-parameters $\V\mu$, as made explicit by
the notation. As the parameters $\V x$, $\V y$, and $\V z$ represent
nonnegative quantities, their estimators are improved by enforcing
nonnegativity as indicated by the inequality constraints in
Eq.~\eqref{eq:joint_estimation_problem} such as $\V x\ge0$ which hold
element-wise. The calibration parameters
  $\V\nu = (\V\nu_\star,\V\nu_\planet)$ are constrained to belong to a set
  $\Omega = \Omega_{\star}\times\Omega_{\planet}$ where
  $\Omega_{\star}$ and $\Omega_{\planet}$ are the respective feasible
  sets for the stellar and companion calibration parameters defined based on
  physical considerations.

The SEDs and the on-axis PSF at the reference wavelength being mutually
independent, the regularization function can be decomposed as:
\begin{equation}
  \label{eq:composite-regul}
  \RegulXYZ(\V x,\V y,\V z,\V\mu)
  = \mu_{\V x}\,\RegulX(\V x)
  + \mu_{\V y}\,\RegulY(\V y)
  + \mu_{\V z}\,\RegulZ(\V z) \,.
\end{equation}
The complete set of hyper-parameters is then:
\begin{equation}
  \V\mu = (\mu_{\V x}, \mu_{\V y}, \mu_{\V z})
\end{equation}
where $\mu_{\V x} > 0$, $\mu_{\V y} > 0$, and $\mu_{\V z} > 0$ tune the weights
of the different regularization terms.

There are many regularizations that are suitable for our problem.
Regularization terms should enforce some kind of continuity or smoothness of
the sought uni-dimensional distributions. In the following and for the
  sake of simplicity, we consider simple smoothness regularizations imposed by
  the quadratic penalty \citep{Tikhonov-1977}:
\begin{equation}
  \label{eq:l2-regul}
  \Regul(\V u) = \sum\nolimits_{j=1}^{N_{\V u}-1}
  \Paren[\big]{\underbrace{u_{j+1} - u_j}_{(\M D\,\V u)_j}}^2
  = \Norm{\M D\,\V u}^2_2,
\end{equation}
with $N_{\V u}$ the size of $\V u = \V x$, $\V y$, or $\V z$, and
$\M D \in \Reals^{(N_{\V u} - 1)\times N_{\V u}}$ a finite difference operator.

\subsection{Alternating minimization strategy}
\label{sec:minimization-strategy}

The joint minimization of the criterion defined in Eq.~\eqref{eq:criterion}
requires to cope with a highly non-linear function whose conditioning may be
very bad and depends on the scaling of the parameters. We propose to solve the
problem by an alternated minimization strategy, that is estimating each set of
parameters given the others. Such a strategy consists in sequentially solving
the following sub-problems:
\begin{subequations}
  \begin{align}
    \proxy{\V x}(\V y,\V r_\star,\V\nu_\star,\mu_{\V x})
    &= \argmin_{\V x \ge 0}
      \Criterion\Paren{\V x, \V y, \V z, \V\nu, \V\mu} \notag\\
    &= \argmin_{\V x \ge 0} \Brace*{
      \Norm{\M A_\star\,\V x - \V r_\star}^2_{\M W}
      + \mu_{\V x}\,\Regul_{\V x}(\V x)
      },
      \label{eq:estim_x}\\
    \proxy{\V y}(\V x,\V r_\star,\V\nu_\star,\mu_{\V y})
    &= \argmin_{\V y \ge 0}
      \Criterion\Paren{\V x, \V y, \V z, \V\nu, \V\mu} \notag\\
    &= \argmin_{\V y \ge 0} \Brace*{
      \Norm{\M B_\star\,\V y - \V r_\star}^2_{\M W}
      + \mu_{\V y}\,\Regul_{\V y}(\V y)
      },
      \label{eq:estim_y}\\
    \proxy{\V\nu}_\star(\V x,\V y,\V r_\star)
    &= \argmin_{\V\nu_\star \in \Omega_\star}
      \Criterion\Paren{\V x, \V y, \V z, \V\nu, \V\mu} \notag\\
    &= \argmin_{\V\nu_\star \in \Omega_\star}
      \Norm{\V m_\star(\V x,\V y,\V\nu_\star) - \V r_\star}^2_{\M W},
      \label{eq:estim_others_star}\\
    \proxy{\V z}(\V r_\planet,\V\nu_\planet,\mu_{\V z})
    &= \argmin_{\V z \ge 0}
      \Criterion\Paren{\V x, \V y, \V z, \V\nu, \V\mu} \notag\\
    &= \argmin_{\V z \ge 0} \Brace*{
      \Norm{\M A_\planet\,\V z - \V r_\planet}^2_{\M W}
      + \mu_{\V z}\,\Regul_{\V z}(\V z)
      },
    \label{eq:estim_z}\\
    \proxy{\V\nu}_\planet(\V z,\V r_\planet)
    &= \argmin_{\V\nu_\planet \in \Omega_\planet}
      \Criterion\Paren{\V x, \V y, \V z, \V\nu, \V\mu} \notag\\
    &= \argmin_{\V\nu_\planet  \in \Omega_\planet}
      \Norm{\V m_\planet(\V z,\V\nu_\planet) - \V r_\planet}^2_{\M W},
      \label{eq:estim_others_planet}
  \end{align}
\end{subequations}
with:
\begin{subequations}
  \begin{align}
    \forall\V x,\,\M A_\star\V x
    &= \V\gamma\HadamardProduct\Paren[\big]{\M H_\star(\V\nu)\,\V y}
      \HadamardProduct(\M F_\star\V x),\\
    \forall\V y,\,\M B_\star\V y
    &= \V\gamma\HadamardProduct\Paren[\big]{\M F_\star\,\V x}
      \HadamardProduct(\M H_\star(\V\nu)\V y),\\
     \forall\V z,\,\M A_\planet\V z
    &= \V\gamma\HadamardProduct\Paren[\big]{\V h_\planet(\V\nu)}
      \HadamardProduct(\M F_\planet\V z),\\
    \label{eq:residuals-for-star}
    \V r_\star &= \V d - \V m_\planet(\V z, \V\nu_\planet),\\
    \label{eq:residuals-for-planet}
    \V r_\planet &= \V d - \V m_\star(\V x,\V y,\V\nu_\star),\\
    \label{eq:star-contrib}
    \V m_\star(\V x,\V y,\V\nu_\star)
    &= \V\gamma\HadamardProduct
      (\M H_\star(\V\nu)\,\V y)\HadamardProduct(\M F_\star\,\V x)
      = \M A_\star\,\V x = \M B_\star\,\V y,\\
    \label{eq:planet-contrib}
    \V m_\planet(\V z, \V\nu_\planet)
    &= \V\gamma\HadamardProduct
      \V h_\planet(\V\nu)\HadamardProduct(\M F_\planet\,\V z)
      = \M A_\planet\,\V z.
  \end{align}
\end{subequations}
We enforce positivity constraints for the variables $\V x$, $\V y$, and $\V z$,
while $\Omega_\star$ and $\Omega_\planet$ respectively denote the feasible set
of parameters $\V\nu_\star$ and $\V\nu_\planet$. Note that
$\V m_\star(\V x,\V y,\V\nu_\star)$ and $\V m_\planet(\V z,\V\nu_\planet)$
defined in Eqs.~\eqref{eq:star-contrib} and \eqref{eq:planet-contrib} are the
respective contributions of the star and companion.

When the convex regularization defined in \eqref{eq:l2-regul} is chosen and
$\M A_\star\T\M W \M A_\star$, $\M B_\star\T\M W \M B_\star$,
and $\M A_\planet\T\M W \M A_\planet$ are invertible, each of the
Problems~\eqref{eq:estim_x}, \eqref{eq:estim_y}, and \eqref{eq:estim_z} is
strictly convex and thus has a unique solution which can be found by using
existing algorithms\footnote{For example, in the unconstrained case and with
  quadratic regularizations, the solution of one of these sub-problem has a
  closed-form expression. Otherwise, each of these sub-problems can be solved
  by optimization algorithms such as quasi-Newton methods with bound
  constraints \citep[\eg, ][]{Thiebaut-2002-optim_bdec}.}. This is another
advantage of the alternated strategy. Since the original minimization
problem~\eqref{eq:joint_estimation_problem} is not jointly convex with respect
to all unknowns, only a local minimum is reached by the alternating
minimization scheme, though.

\begin{algorithm}[t]
  \caption{\textsc{FitStar} --- fit stellar parameters.}
  \label{alg:star}
  \KwIn{
    $\V r_\star \in \mathbb{R}^N$, $\M W \in \Reals^{N \times N}$,
    $\V x\Iter0 \in \Reals^{N_{\V x}}$, $\V\nu_\star\Iter0$,
    $\mu_{\V x} > 0$, and $\alpha_0 > 0$.
  }
  \KwOut{$\estim{\V x}$, $\estim{\V y}$, and $\estim{\V\nu}_\star$ a local minimum of $\Criterion$ in
    $\V x$, $\V y$, and $\V\nu_\star$.}
  $k = 0$\;
  $\mu_{\V y} = 1$\;
  \While{not converged}{
      \While{\True}{
          $\blacktriangleright$ Update on-axis PSF\;
          $\V y\Iter{k+1} = \proxy{\V y}\Paren[\big]{
            \V x\Iter{k}, \V r_\star, \V\nu_\star\Iter{k}, \alpha_{k}^{-2}\,\mu_{\V y}}$%
          \Comment*{Eq.~\eqref{eq:estim_y}}
          $\alpha_{k+1/2} = \proxy{\alpha}\Paren[\big]{
            \V x\Iter{k}, \V y\Iter{k+1}, \mu_{\V x}, \mu_{\V y}}$%
          \Comment*{Eq.~\eqref{eq:best-scaling_factor}}
          \lIf{$k \ge 1$ \Or $\alpha_{k+1/2} \approx \alpha_{k}$}{\Break}
          $\alpha_{k} = \alpha_{k+1/2}$\;
        }
        $\blacktriangleright$ Update star SED\;
        $\V x\Iter{k+1} = \proxy{\V x}\Paren[\big]{
          \V y\Iter{k+1}, \V r_\star, \V\nu_\star\Iter{k}, \alpha_{k+1/2}^{2}\,\mu_{\V
            x}}$%
        \Comment*{Eq.~\eqref{eq:estim_x}}
        $\alpha_{k+1} = \proxy{\alpha}\Paren[\big]{
          \V x\Iter{k+1}, \V y\Iter{k+1}, \mu_{\V x}, \mu_{\V y}}$%
        \Comment*{Eq.~\eqref{eq:best-scaling_factor}}
        $\blacktriangleright$ Auto-calibration (optional)\;
        $\V \nu_\star\Iter{k+1} \leftarrow \proxy{\V\nu}_\star\Paren[\big]{\V x\Iter{k+1},\V y\Iter{k+1},\V r_\star}$%
        \Comment*{Eq.~\eqref{eq:estim_others_star}}
        $k \gets k + 1$\;
      }
      $\estim{\V x} \gets \alpha_{k}\,\V x\Iter{k}$\;
      $\estim{\V y} \gets \V y\Iter{k}/\alpha_{k}$\;
      $\estim{\V\nu}_\star \gets \V \nu_\star\Iter{k}$\;
\end{algorithm}

We solve for the two stellar components $\V x$ and $\V y$ following the
alternated method proposed by \citet{TheTDS20} to exploit the \emph{scaling
  indetermination} of this problem (see
Appendix~\ref{sec:Scaling_indetermination} for details). This method is
implemented by Algorithm~\ref{alg:star} and takes as inputs the residuals
$\V r_\star =\V d - \V m_\planet(\V z, \V\nu_\planet)$ (\ie the data without
the contribution of the companion), the precision matrix $\M W$, initial
calibration parameters $\V\nu_\star\Iter0$, the hyper-parameters
$\mu_{\V x} > 0$ (hyper-parameter $\mu_{\V y}$ is set to the arbitrary value 1
in Algorithm \ref{alg:star}), initial estimates $\V x\Iter0$ of the stellar
SED, and initial estimate $\alpha_0 > 0$ of the scaling parameter.
Algorithm~\ref{alg:star} deserves some remarks:
\begin{enumerate}
\item The initial stellar SED $\V x\Iter0$ must be such that
  $\Regul_{\V x}(\V x\Iter0) > 0$ to be able to apply
  formula~\eqref{eq:best-scaling_factor} to compute the optimal scaling factor
  (i.e., a non-flat SED). The initial stellar SED can be provided by
  calibration data (see
  Appendix~\ref{sec:star_SED_calibration});
  otherwise it can be computed from the science data $\V d$ by the following
  weighted mean:
  \begin{equation}
    \label{eq:initial-x}
    \forall j \in \IntRange{1,N_{\V x}}:\quad
    x_{\star,j}\Iter0 = \frac{
      \sum_{n \in \mathcal{X}_j} w_n\,d_n
    }{
      \sum_{n \in \mathcal{X}_j} w_n
    }
  \end{equation}
  with $w_n = W_{n,n}$ the $n$-th diagonal term of the precision matrix and:
  \begin{equation}
    \label{eq:model-wavelength-subset}
    \mathcal{X}_j = \Brace[\Big]{
      n \in \IntRange{1,N} \SuchThat
      \Abs*{\lambda^\Tag{grd}_{\star,j} - \lambda_n} = \min_{j'\in\IntRange{1,N_{\V x}}}
      \Abs*{\lambda^\Tag{grd}_{\star,j'} - \lambda_n}
    }
  \end{equation}
      the set of pixels whose nearest wavelength in the model grid is the
      $j$-th one. Since Algorithm~\ref{alg:star} scales the final
      components $\V x\Iter{k}$ and $\V y\Iter{k}$ by the corresponding optimal
      scaling factor, $\alpha_0 = 1$ is a natural choice for the initial
      scaling factor in subsequent calls to Algorithm~\ref{alg:star} (to refine
      the solution or after having improved the other parameters).

\item The inner loop of Algorithm~\ref{alg:star} avoids sensitivity to the
  initial scaling of the parameters \citep{TheTDS20}.

\item The convergence criterion of Algorithm~\ref{alg:star} is left
      unspecified. In our implementation, we chose to stop the algorithm
      when the relative change, in norm, between two consecutive iterates is
      smaller than $10^{-3}$.

\item Although they represent very different physical quantities, the problem
  is quite symmetric in variables $\V x$ and $\V y$. Thus a variant of
  Algorithm~\ref{alg:star} can be easily implemented to start with an initial
  estimate $\V y\Iter0$ of the stellar on-axis PSF at the reference wavelength
  instead of an initial estimate $\V x\Iter0$ of the stellar SED. For the very
  first run, this variant of Algorithm~\ref{alg:star} is started with the
  weighted average of the on-axis PSF defined by:
  \begin{equation}
    \label{eq:initial-y}
    \forall j \in \IntRange{1,N_{\V y}}:\quad
    y_{\star,j}\Iter0 = \frac{
      \sum_{n \in \mathcal{Y}_j} w_n\,d_n
    }{
      \sum_{n \in \mathcal{Y}_j} w_n
    }
  \end{equation}
  with:
  \begin{equation}
    \label{eq:model-distance-subset}
    \mathcal{Y}_j = \Brace[\Big]{
      n \in \IntRange{1,N} \SuchThat
      \Abs*{\rho^\Tag{grd}_{\star,j} - \rho_n} = \min_{j'\in\IntRange{1,N_{\V y}}}
      \Abs*{\rho^\Tag{grd}_{\star,j'} - \rho_n}
    }
  \end{equation}
  the set of pixels whose nearest angular position in the model grid is the
  $j$-th one.

\item When the SED $\V z$ of the companion is not yet known, it is sufficient
  to call Algorithm~\ref{alg:star} with the weights of the pixels the most
  impacted by the companion set to zero (we write the corresponding precision
  matrix $\M W_\star$) to estimate the components $\V x$ and $\V y$ of the
  stellar leakages without introducing a significant bias due to the
  contribution of the companion.
\end{enumerate}

\begin{algorithm}[t]
    \caption{\textsc{FitCompanion} --- fit companion parameters.}
    \label{alg:companion}
    \KwIn{residuals $\V r_\planet \in \Reals^N$, precision matrix $\M W$,
      initial off-axis PSF parameters $\V\nu_\planet\Iter0$, $\mu_{\V z} > 0$.}
    \KwOut{$\estim{\V z}$ and $\estim{\V\nu}_\planet$, a local minimum of
      $\Criterion$ in $\V z$ and $\V\nu_\planet$ given $\V m_\star(\V x, \V
      y, \V\nu_\star)$ the model of the stellar contribution.}
    $k = 0$\;
    \While{not converged}{
        $\blacktriangleright$ Update companion SED\;
        $\V z\Iter{k+1} = \proxy{\V z}\Paren[\big]{
            \V r_\planet, \V\nu_\planet\Iter{k}, \mu_{\V z}}$%
        \Comment*{Eq.~\eqref{eq:estim_z}}
        $\blacktriangleright$ Update off-axis PSF (optional)\;
        $\V\nu_\planet\Iter{k+1} = \proxy{\V\nu}_\planet\Paren[\big]{
            \V z\Iter{k+1}, \V r_\planet}$%
        \Comment*{Eq.~\eqref{eq:estim_others_planet}}
        $k \gets k + 1$\;
    }
    $\estim{\V z} \gets \V z\Iter{k}$\;
    $\estim{\V\nu}_\planet \gets \V\nu_\planet\Iter{k}$\;
\end{algorithm}

Algorithm~\ref{alg:companion} (\textsc{FitCompanion}) implements an alternated
strategy to estimate the parameters $\V z$ and $\V\nu$ of the companion SED and
its off-axis PSF at the reference wavelength. It takes as inputs the residuals
$\V r_\planet = \V d - \V m_\star(\V x, \V y, \V\nu_\star)$ (\ie the data
without the contribution of the star) and their respective weights $\M W$, the
hyper-parameter $\mu_{\V z} > 0$ and an initial estimate
$\V\nu\Iter0_\planet \in \Omega_\planet$ of the parameters of the off-axis PSF
at the reference wavelength. These latter parameters can be given by the
calibration described in Appendix~\ref{sec:star_SED_calibration}.
Algorithm~\ref{alg:companion} also deserves some remarks:
\begin{enumerate}
\item The outputs of the algorithm only depend on the residual data
  $\V r_\planet$ defined in Eq.~\eqref{eq:residuals-for-planet} that need
  to be computed only once (on entry of the algorithm and not at each
  iterations).

\item Like for Algorithm~\ref{alg:star}, various stopping criteria may be
  implemented to break the loop.

\item Like for Algorithm~\ref{alg:star}, we can use the VMLM-B algorithm
  \citep{Thiebaut-2002-optim_bdec} to solve Problem~\eqref{eq:estim_z} to
  estimate $\V z$ under a non-negativity constraint.
\end{enumerate}

In both algorithms, there are optional self-calibration steps performed by
solving Problem~\eqref{eq:estim_others_star} in Algorithm~\ref{alg:star}
(\textsc{FitStar}) and Problem~\eqref{eq:estim_others_planet} in
Algorithm~\ref{alg:companion} (\textsc{FitCompanion}) to estimate the
parameters of the on-axis and off-axis PSFs. These minimizations can be carried
out by a derivative-free minimization algorithm. When there is a single
calibration parameter, we use \citet{brent2013algorithms} \textsc{Fmin}
algorithm; if there are several parameters, we use one of Powell's
derivative-free methods \textsc{Newuoa} or \textsc{Bobyqa}
\citep{Powell-2006-NEWUOA,powell_bobyqa} depending on the constraints defined
by $\Omega$.

\subsection{The \Exospeco algorithm}
\label{sec:The_EXOSPECO_algorithm}

\begin{algorithm}[t]
  \caption{\Exospeco algorithm}
  \label{alg:EXOPSEC}
  \KwIn{The data $\V d$ and its precision matrix $\M W$, the masked precision
    matrix $\M W_\star$, initial estimates $\V x\Iter0$, $\V\nu_\star\Iter0$,
    and $\V\nu_\planet\Iter0$, and hyper-parameters $\mu_{\V x} > 0$ and
    $\mu_{\V z} > 0$.  } \KwOut{$\estim{\V x}$, $\estim{\V y}$, $\estim{\V
      z}$, $\estim{\V\nu}_\star$, and $\estim{\V\nu}_\planet$ a local minimum
    of $\Criterion$.}
  $\mu_y = 1$\;
  $\V z\Iter{0} = \V 0$\;
  $\estim{\alpha} = 1$\;
  $k = 0$\;
  \While{not converged}{
      \uIf{$k = 0$}{
        $\blacktriangleright$ Hide companion\;
        $\M W' = \M W_\star$\Comment*{Eq.~\eqref{eq:W-star}}
      }\Else{
        $\blacktriangleright$ Account for companion\;
        $\M W' = \M W$\;
      }
      $\blacktriangleright$ Update star leakage model\;
      $\V r_\star\Iter{k} = \V d - \V m_\planet\Paren[\big]{
        \V z\Iter{k}, \V\nu_\planet\Iter{k}
      }$\Comment*{Eq.~\eqref{eq:residuals-for-star}}
      $\Paren[\big]{
        \V x\Iter{k+1},
        \V y\Iter{k+1},
        \V\nu_\star\Iter{k+1}
      } =$\;
      \hspace*{24mm}%
      $\textsc{FitStar}\bigl(\V r_\star\Iter{k}, \M W', \V x\Iter{k},
      \V\nu_\star\Iter{k}, \mu_{\V x},\estim{\alpha}\bigr)$\;
      $\blacktriangleright$ Update companion model\;
      $\V r_\planet\Iter{k+1} = \V d - \V m_\star\Paren[\big]{
        \V x\Iter{k+1}, \V y\Iter{k+1}, \V\nu_\star\Iter{k+1}
      }$\Comment*{Eq.~\eqref{eq:residuals-for-planet}}
      $\Paren[\big]{
        \V z\Iter{k+1},
        \V\nu_\planet\Iter{k+1}
      } = \textsc{FitCompanion}\Paren[\big]{
        \V r_\planet\Iter{k+1}, \M W,
        \V\nu_\planet\Iter{k}, \mu_{\V z}}$\;
      $k \gets k + 1$\;
    }
    $(\estim{\V x}, \estim{\V y}, \estim{\V z}, \estim{\V\nu}_\star,
    \estim{\V\nu}_\planet) \gets
    \Paren[\big]{\V x\Iter{k}, \V y\Iter{k},
      \V z\Iter{k}, \V\nu_\star\Iter{k}, \V\nu_\planet\Iter{k}}$\;
\end{algorithm}

\textsc{FitStar} (Algorithm~\ref{alg:star}) and \textsc{FitCompanion}
(Algorithm~\ref{alg:companion}) are the building blocks of the \Exospeco method
given in Algorithm~\ref{alg:EXOPSEC} for estimating all unknowns. Only a few
additional remarks are worth being made:

\begin{enumerate}
\item For the first estimation of the stellar leakage parameters, it is
beneficial to
  define a masked version $\M W_\star$ of the precision matrix of the data to
  avoid a significant bias of the first estimates due to the signal from the
  companion, which would slow down the convergence of Algorithm
  \ref{alg:EXOPSEC}. The masked
  precision matrix is simply given by:
  \begin{equation}
    \label{eq:W-star}
    \M W_\star = \Diag(\V w_\star),
  \end{equation}
  where the weights $\V w_\star$ are those of the precision matrix $\M W$ of
  the data except that they are set to zero for the pixels that are the most
  impacted by the companion:
  \begin{equation}
  \forall n \in \IntRange{1,N}:\quad
  w_{\star,n} = \begin{cases}
                  0 & \text{if $\gamma_n\,\Abs{\rho_n - \rho_\planet} \le \tau$}\\
                  w_n & \text{otherwise}
                \end{cases}
  \end{equation}
  with $\tau > 0$ the angular half-width at the reference wavelength of the
  impacted region. In practice $\tau$ is taken to be 2-3 times
  $\sigma_\planet\Iter0$ the initial angular standard deviation of the off-axis
  PSF at the reference wavelength.

\item The model of the stellar leakage only depends on either $\mu_{\V x}$ or
  $\mu_{\V y}$, the other being arbitrarily chosen. For this reason
  Algorithm~\ref{alg:EXOPSEC} takes as inputs only 2 hyper-parameters
  $\mu_{\V x}$ and $\mu_{\V z}$, the remaining hyper-parameter being set to
  $\mu_{\V y} = 1$.

\item After extracting the companion's spectrum by \Exospeco
  Algorithm~\ref{alg:EXOPSEC}, it is possible to express it as a contrast
  relative to the host star which can be multiplied by a reference spectrum of
  the star to get rid of the atmospheric absorption (see
  Appendix~\ref{sec:star_SED_calibration}).

\item The auto-calibration steps in \textsc{FitStar} (Algorithm~\ref{alg:star})
  and \textsc{FitCompanion} (Algorithm~\ref{alg:companion}) are optional and
  consist in the resolution of Problems~\eqref{eq:estim_others_star} and
  \eqref{eq:estim_others_planet}. As these problems are non-convex, activating
  the auto-calibration at the beginning of the method can lead to a local
  minimum. To avoid such a behavior, it is possible to start the
  self-calibration of $\V\nu_\star$ and $\V\nu_\planet$ only after a few
      iterations of \Exospeco (Algorithm~\ref{alg:EXOPSEC}).

\item Controlling the number of \emph{inner iterations} to solve each
      sub-problem could be done by changing the value of the stopping parameter
      $\epsilon$ (cf.\ remark 3 in Section~\ref{sec:minimization-strategy}):
      the smaller $\epsilon$ the more \emph{inner iterations} are needed and
      conversely. But this is expected to also impact the number of \emph{outer
      iterations}. Owing to the modest amount of time (2-3 min.) taken by our
      implementation of \Exospeco to solve the entire problem, we did not
      investigate whether the algorithm can be effectively accelerated by
      changing $\epsilon$ and keep the value $\epsilon = 10^{-3}$ suggested
      before.

\end{enumerate}

\section{Calibration}
\label{sec:calibration}

The direct model in Eq.~\eqref{eq:direct_model} assumes known the physical
coordinates $(\rho_n,\lambda_n)$ of each pixel $n$ of the detector. We describe
in this section a consistent approach to derive the spectro-angular dispersion
laws of the instrument from calibration data.

\subsection{Calibration data}
\label{sec:calibration_data}

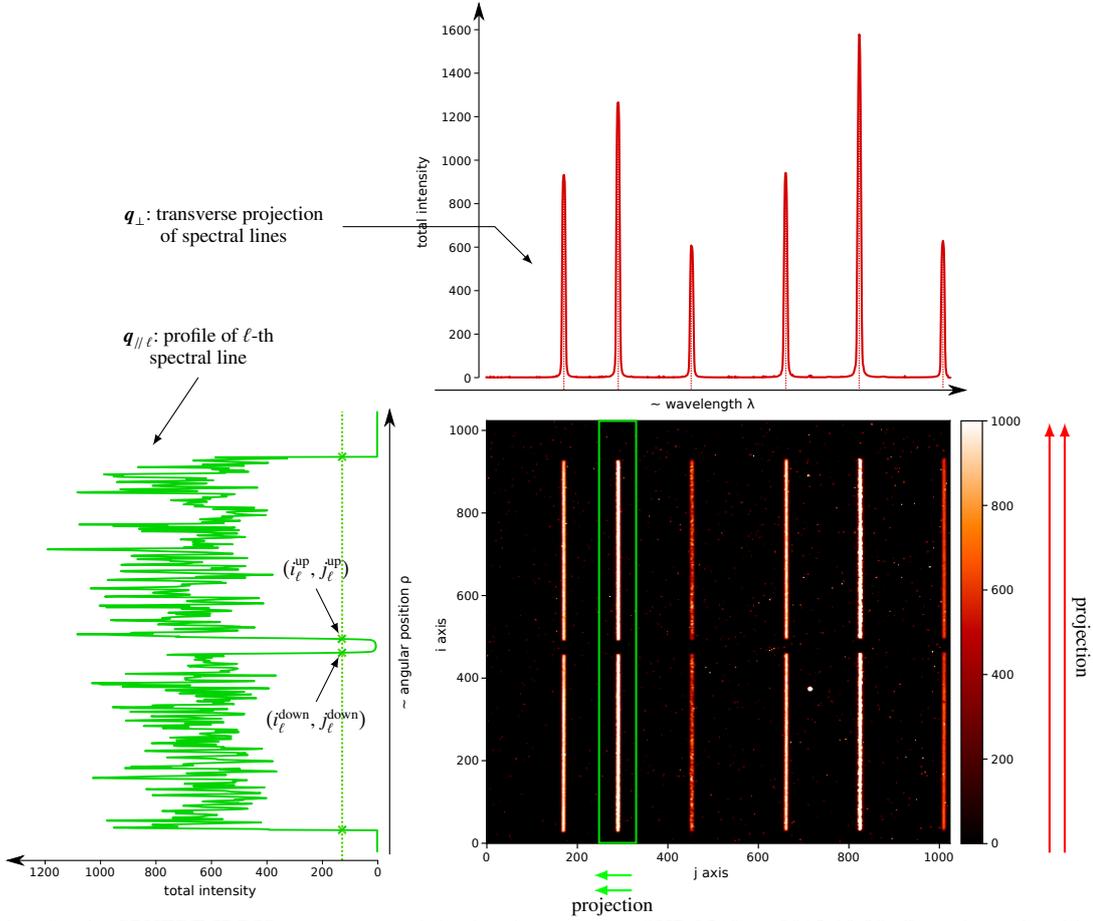
\begin{figure*}
  \centering
  \includegraphics[scale=.55]{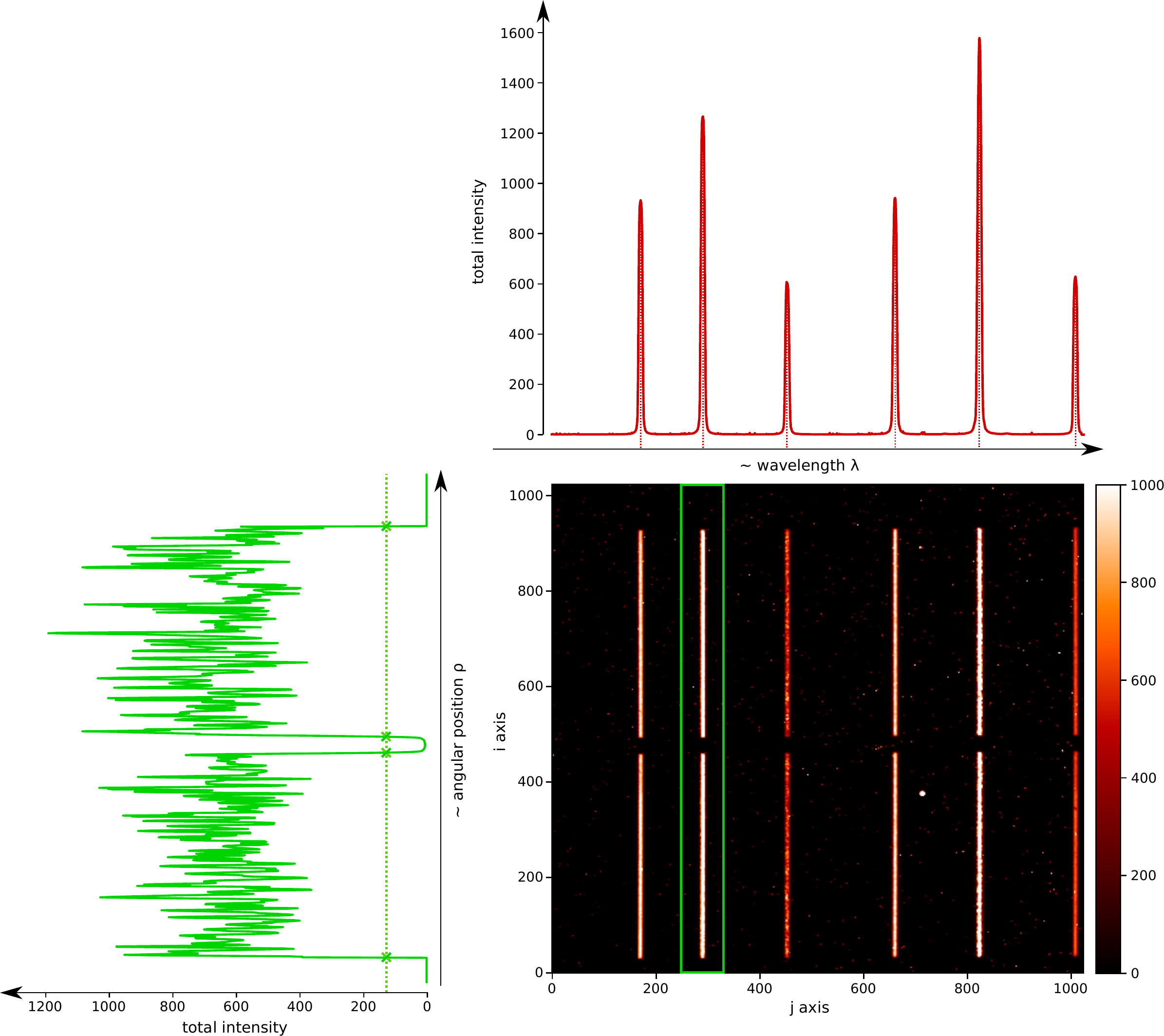}
  \begin{tikzpicture}[overlay, remember picture]
    \draw[>=latex,->,thick,color=red] (0.3,0.6) -- (0.3,6.3);
    \draw[>=latex,->,thick,color=red] (0.5,0.6) -- (0.5,6.3) node[midway,above,color=black,scale=.75,rotate=-90]{projection};

    \draw[>=latex,->,thick,color=green] (-5.2,0.3) -- (-5.7,0.3);
    \draw[>=latex,->,thick,color=green] (-5.2,0.1) -- (-5.7,0.1) node[midway,below,color=black,scale=.75]{projection};

    \draw[>=latex,<-,color=black] (-6.5,8.4) -- ++(-.5,.5) -- ++(-2,0) node[left,scale=.75]{%
      \begin{tabular}{c}
        $\V q_\bot$: transverse projection\\
        of spectral lines
      \end{tabular}
    };

    \draw[>=latex,<-,color=black] (-11.5,6) -- ++(.6,.9)
    node[above,scale=.75]{%
      \begin{tabular}{c}
        $\V q_{/\!/\,\ell}$: profile of $\ell\nth$\\
        spectral line
      \end{tabular}
    };

    \draw[>=latex,<-,color=black] (-9.05,3.2) -- ++(-0.3,-0.6)
    node[below,scale=.75]{$\Paren[\big]{i^\Tag{down}_\ell,j^\Tag{down}_\ell}$};

    \draw[>=latex,<-,color=black] (-9.05,3.5) -- ++(-0.3,0.6)
    node[above,scale=.75]{$\Paren[\big]{i^\Tag{up}_\ell,j^\Tag{up}_\ell}$};

  \end{tikzpicture}
  \caption{Calibration data for the SPHERE/IRDIS instrument and for the observations of HR\,3549 on 2015/12/28. Central panel: calibration image. Left and top panels: projections of the calibration data along the 2nd spectral line (in green) and across all spectral lines (in red).}
  \label{fig:LSS_calibration}
\end{figure*}

Calibration data takes the form of an image such as the one shown in
Fig.~\ref{fig:LSS_calibration} which is obtained by illuminating the
spectrograph slit with $N_\lambda$ laser sources\footnote{$N_\lambda=6$ at
  wavelengths $0.9877\,\micron$, $1.1237\,\micron$, $1.3094\,\micron$,
  $1.5451\,\micron$, $1.73\,\micron$, and $2.015\,\micron$ for SPHERE/IRDIS}.
This produces $N_\lambda$ mono-chromatic lines on the detector, each being
interrupted by the coronagraphic mask. The calibration image $\V d_\Tag{cal}$
is of size $I\times J$ and, for the calibration procedure, we denote by
$n \sim (i,j)$ the one to one mapping between the pixel number $n$ and its
indices $i\in\IntRange{1,I}$ and $j\in\IntRange{1,J}$ along the first and
second dimensions of the detector.

\begin{figure}[!t]
  \centering\includegraphics[scale=0.58]{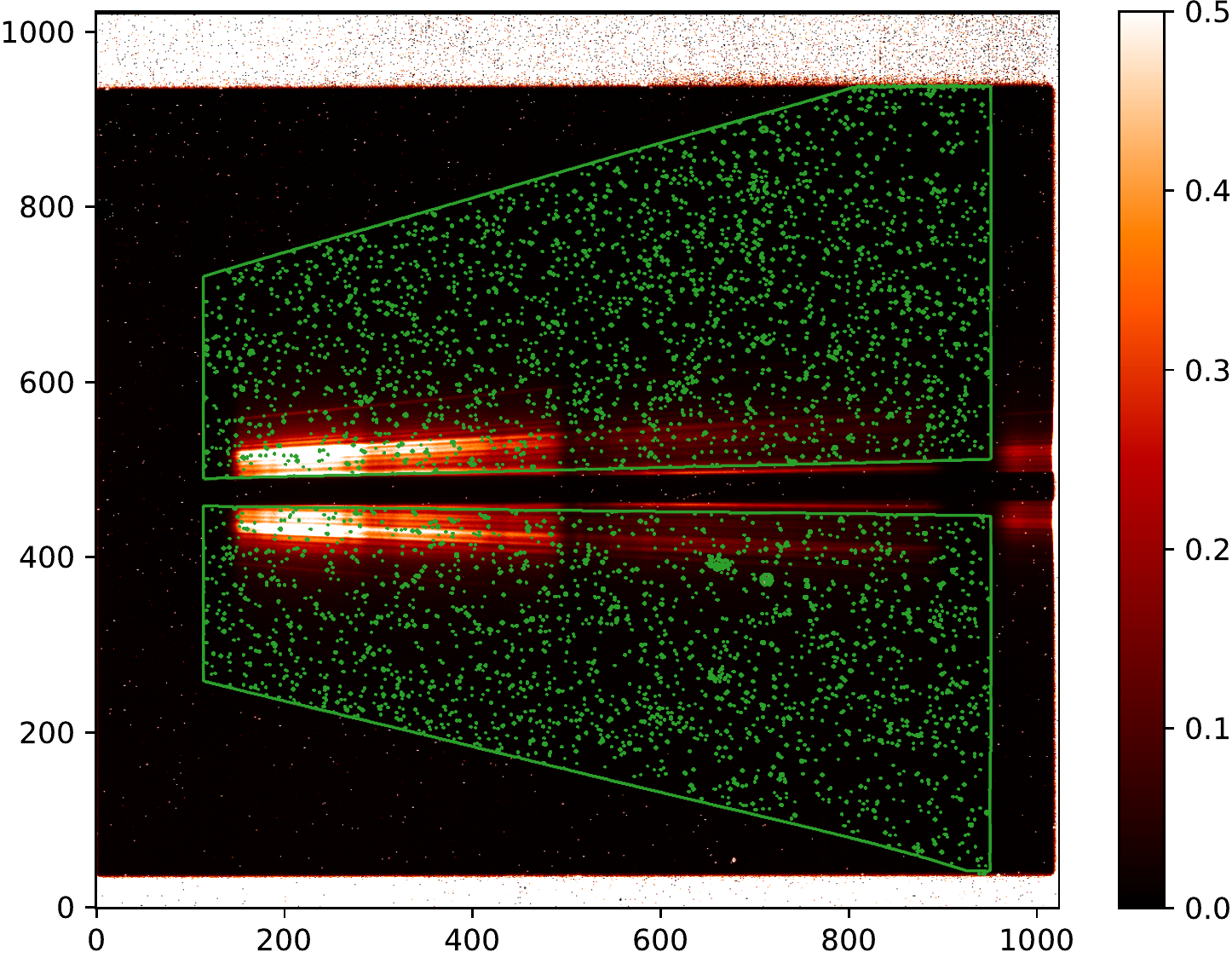}
  \caption{Valid pixel mask for the HR\,3549 data observed on 2015-12-28
      in MRS mode.}
  \label{fig:HR3549_data_with_mask_valid_pixel}
\end{figure}

The calibration image shall have been pre-processed to compensate for bias and
non-uniform response of the detector. Furthermore, we assume known a \emph{mask
  of valid pixels}:
\begin{equation}
  \label{eq:valid-pixel-mask}
  w_{\mask,i,j} = \begin{cases}
    1 & \text{if the pixel $(i,j)$ is valid,}\\
    0 & \text{else}.
  \end{cases}
\end{equation}
We consider a pixel as being \emph{invalid} if its value cannot follow the
assumed direct model given in Eq.~\eqref{eq:direct_model}. Invalid pixels
include pixels outside the field of view, pixels under the coronagraphic mask
or close to this mask, and defective pixels whose level does not linearly
depend on the illumination. Figure~\ref{fig:HR3549_data_with_mask_valid_pixel}
shows the mask of valid pixels for the HR\,3549 data: the field of view and the
coronagraphic mask are outlined by the two green trapezes while the defective
pixels are marked by green dots.

\subsection{Dispersion laws}
\label{sec:dispersion_laws}

There are two dispersion laws to calibrate: $\Lambda(i,j)$ for the wavelength
and $\varrho(i,j)$ for the separation angle along the slit. To determine the
best approximation of these spectro-angular dispersion laws, we compared three
models:
\begin{itemize}
\item The \emph{standard model} which assumes that the dispersion laws are
      uni-dimensional polynomials with spectral and angular directions aligned
      with the detector axes:
      \begin{subequations}
      \begin{align}
        \label{eq:standard_spectral_law}
        \Lambda_\Tag{sta}(i,j) &= \sum\nolimits_{p = 0}^{P_\lambda} a_p\, j^p,\\
        \label{eq:standard_angular_law}
        \varrho_\Tag{sta}(i,j) &= \sum\nolimits_{p = 0}^{P_\rho} s_p\,i^p,
      \end{align}
      \end{subequations}
      with $P_\lambda$ and $P_\rho$ the degrees of the polynomials and
      $\Brace{a_p}_{p\in\IntRange{0,P_\lambda}}$ and
      $\Brace{s_p}_{p\in\IntRange{0,P_\rho}}$ their coefficients. For $P_\rho=1$
      and $P_\lambda=3-5$, the \emph{standard model} reproduces
      what is done in the software by \citet{IRDIS_LSS} usually used to process
      SPHERE/LSS data.

\item A \emph{model of medium complexity}, also assuming 1D polynomials for the
      dispersion laws but accounting for misalignment angles $\phi_\lambda$ and
      $\phi_\rho \approx \phi_\lambda + 90^\circ$ respectively between the
      spectral and angular directions and the detector axes:
      \begin{subequations}
      \begin{align}
        \label{eq:medium_spectral_law}
        \Lambda_\Tag{med}(i,j)
        &= \sum\nolimits_{p = 0}^{P_\lambda} a_p\,(i\,\sin\phi_\lambda + j\,\cos\phi_\lambda)^p,\\
        \label{eq:medium_angular_law}
        \varrho_\Tag{med}(i,j)
        &= \sum\nolimits_{p = 0}^{P_\rho} s_p\,(i\,\sin\phi_\rho + j\,\cos\phi_\rho)^p.
      \end{align}
      \end{subequations}
      Note that taking $\phi_\lambda = 0^\circ$ and $\phi_\varrho = 90^\circ$
      yields the standard model.

\item A \emph{more complex model} which assumes 2D polynomials for the
      dispersion laws and, depending on the degree of these polynomials, can
      account for more complex image distortions than a simple rotation:
      \begin{subequations}
      \begin{align}
        \label{eq:complex_spectral_law}
        \Lambda_\Tag{\Exospeco}(i,j)
        &= \sum\nolimits_{p_{1} = 0}^{P_\lambda} \sum\nolimits_{p_{2} = 0}^{P_\lambda - p_1}
          a_{p_1,p_2}\,i^{p_1}\,j^{p_2},\\
        \label{eq:complex_angular_law}
        \varrho_\Tag{\Exospeco}(i,j)
        &= \sum\nolimits_{p_{1} = 0}^{P_\rho} \sum\nolimits_{p_{2} = 0}^{P_\rho - p_1}
          s_{p_1,p_2}\,i^{p_1}\,j^{p_2}.
      \end{align}
      \end{subequations}

\end{itemize}
To summarize, the considered dispersion laws are polynomials of respective
degree $P_\lambda$ and $P_\rho$. Their calibration amounts to fitting their
coefficients $\V a$ and $\V s$ given the calibration image $\V d_\Tag{cal}$ as
explained in the next sub-sections.

\subsection{Calibration of the spectral dispersion law $\Lambda$}

To calibrate the spectral dispersion law $\Lambda$, we extract from the
calibration image $\V d_\Tag{cal}$ (see Fig.~\ref{fig:LSS_calibration})
$N_\lambda$ lists of pixel coordinates following the path of each spectral
line on the detector and estimate the coefficients $\V a$ by a least squares
fit:
\begin{equation}
  \label{eq:fit_spectral_law}
    \estim{\V a} = \argmin_{\V a} \sum_{\ell=1}^{N_\lambda}
    \sum_{(i,j)\in\mathcal{C}_\ell(\phi_\lambda)}
    \Paren[\big]{\lambda_\ell - \Lambda(i,j)}^2
\end{equation}
where $\mathcal{C}_\ell(\phi_\lambda)$ denotes the list of, possibly fractional, pixel
coordinates $(i,j)$ along the $\ell$-th spectral line on the detector. Since
$\Lambda(i,j)$ linearly depends on the coefficients $\V a$, the solution
$\estim{\V a}$ of the above problem has a closed form expression
\citep{Lawson_Hanson-1974-solving_least_squares_problems} that is easy to
compute.

To extract the paths $\mathcal{C}_\ell(\phi_\lambda)$ of the spectral lines, we first compute
a \emph{transverse} projection $\V q_\bot(\phi_\lambda)$ of the calibration
image $\V d_\Tag{cal}$ tuning the projection angle $\phi_\lambda$ so as to
maximize the peak values in of the resulting projection. This transverse
projection is plotted in red in the top panel of Fig.~\ref{fig:LSS_calibration}
and corresponds to $\phi_\lambda \approx 0^{\circ}$ for the considered
calibration data. Equations~\eqref{eq:transverse_projection} in
Appendix~\ref{sec:transverse_projection} formally define how we carefully
compute the projection avoiding invalid pixels. We then use the procedure
described in Appendix~\ref{sec:spectral_peaks_detection} to locate the position
of the $N_\lambda$ most significant peaks in the transverse projection
$\V q_\bot(\phi_\lambda)$ which can be seen as a mean cross section of the
spectral lines. Finally, we use the method described in
Appendix~\ref{sec:spectral_paths_extraction} to extract the coordinates of the
points defining the $N_\lambda$ paths $\mathcal{C}_\ell(\phi_\lambda)$. These coordinates
are given by the centers of gravity (again accounting for invalid pixels thanks
to the mask) of the calibration data in small sliding rectangular windows along
each spectral lines (see Appendix~\ref{sec:spectral_paths_extraction} for
details).

\subsection{Calibration of the angular dispersion law $\varrho$}
\label{sec:angular_law_calibration}

To calibrate the angular dispersion law $\varrho$, we extract from the
calibration image $\V d_\Tag{cal}$ (see Fig.~\ref{fig:LSS_calibration}) the
positions of the edges of the coronagraphic mask for each of the $N_\lambda$
spectral lines and estimate the coefficients $\V s$ of the polynomial and the
width $\Delta\rho$ of the mask by a least squares fit:
\begin{align}
    \label{eq:fit_angular_law}
    \Paren{\estim{\phi}_\rho,\estim{\Delta\rho},\estim{\V s}} =
    \argmin_{\phi_{\rho},\Delta\rho,\V s}\Bigg\{
    &\sum_{\ell=1}^{N_\lambda}
    \Paren*{
    \varrho\Paren[\big]{i^\Tag{down}_\ell,j^\Tag{down}_\ell} + \Delta\rho/2
    }^2 \notag\\
    &+\sum_{\ell=1}^{N_\lambda}
    \Paren*{
    \varrho\Paren[\big]{i^\Tag{up}_\ell,j^\Tag{up}_\ell} - \Delta\rho/2
    }^2
    \Bigg\}
\end{align}
where $\Paren[\big]{i^\Tag{down}_\ell,j^\Tag{down}_\ell}$ and
$\Paren[\big]{i^\Tag{up}_\ell,j^\Tag{up}_\ell}$ denote the coordinates of the
edges of the coronagraphic mask respectively on the downhill and uphill sides
along the profile of the $\ell$-th spectral line. To solve this problem, we
exploit that the criterion is quadratic in the unknowns $\V s$ and $\Delta\rho$
which thus have a closed-form solution
\citep{Lawson_Hanson-1974-solving_least_squares_problems} which depends on
$\phi_\rho$. Replacing this closed-form solution in the criterion yields an
uni-variate objective function that only depends on $\phi_\rho$ and which we
minimize by \citet{brent2013algorithms} \textsc{Fmin} method starting at
$\phi_\rho = \phi_\lambda + 90^\circ$.

As explained in Appendix~\ref{sec:edges_extraction}, the coordinates of the
edges of the coronagraphic mask for the $\ell$-th spectral line are obtained
from the \emph{longitudinal} profile $\V q_{{/\!/}\,\ell}(\phi_\lambda)$ of the
line which is the weighted projection, in a direction perpendicular to the
transversal projection $\V q_{\bot}(\phi_\lambda)$, of the calibration data
in a window encompassing the line. The longitudinal profile
$\V q_{{/\!/}\,\ell}(\phi_\lambda)$ for the second ($\ell=2$) line is plotted
in green in the left panel of Fig.~\ref{fig:LSS_calibration} and the
corresponding window is outlined in green in the central panel of
Fig.~\ref{fig:LSS_calibration}.

\subsection{Comparison of the calibration models}
\label{sec:comparing_calibration_models}

\begin{figure}[t!]
  \centering
  \includegraphics[scale=0.58]{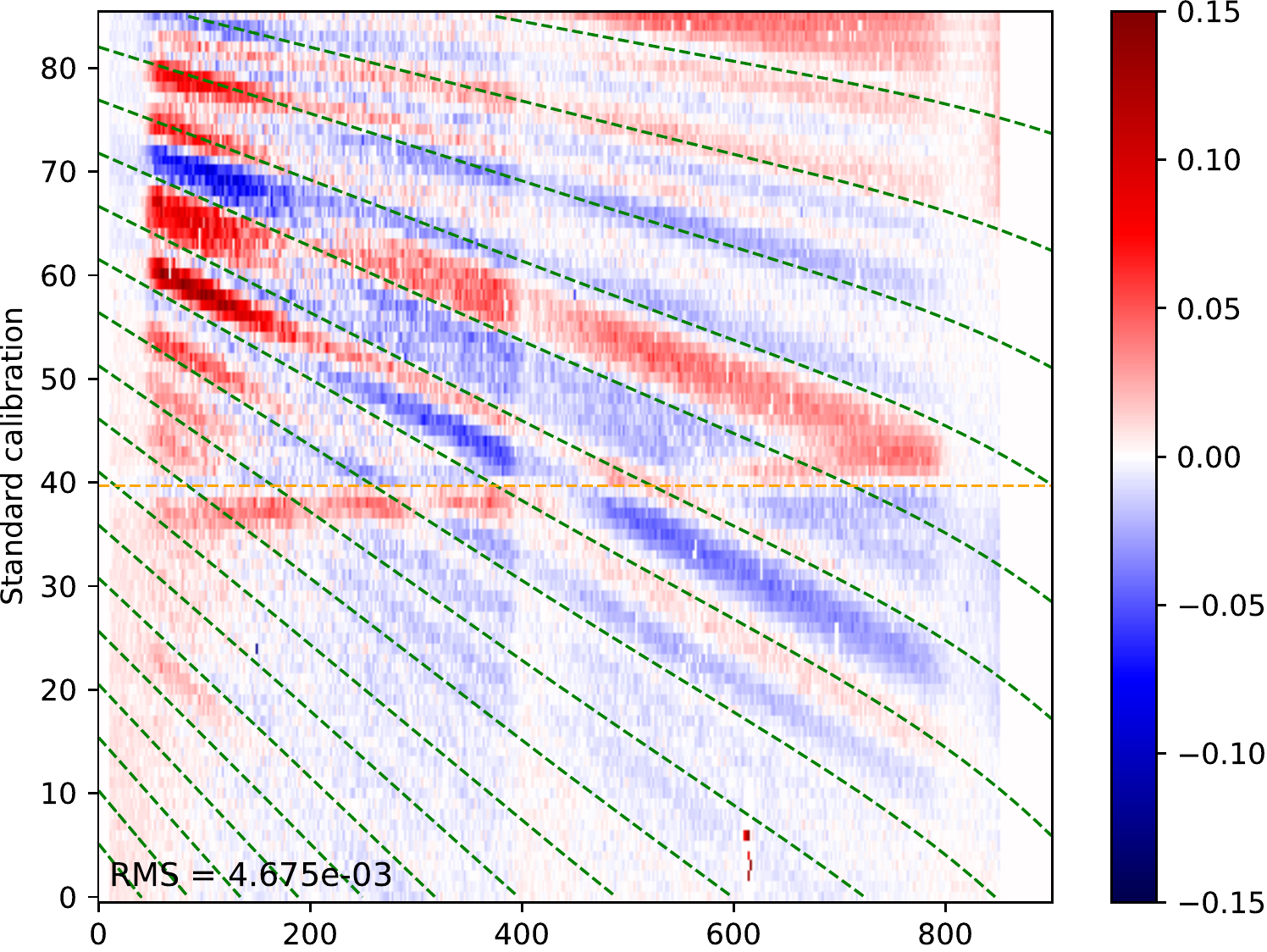}
  \includegraphics[scale=0.58]{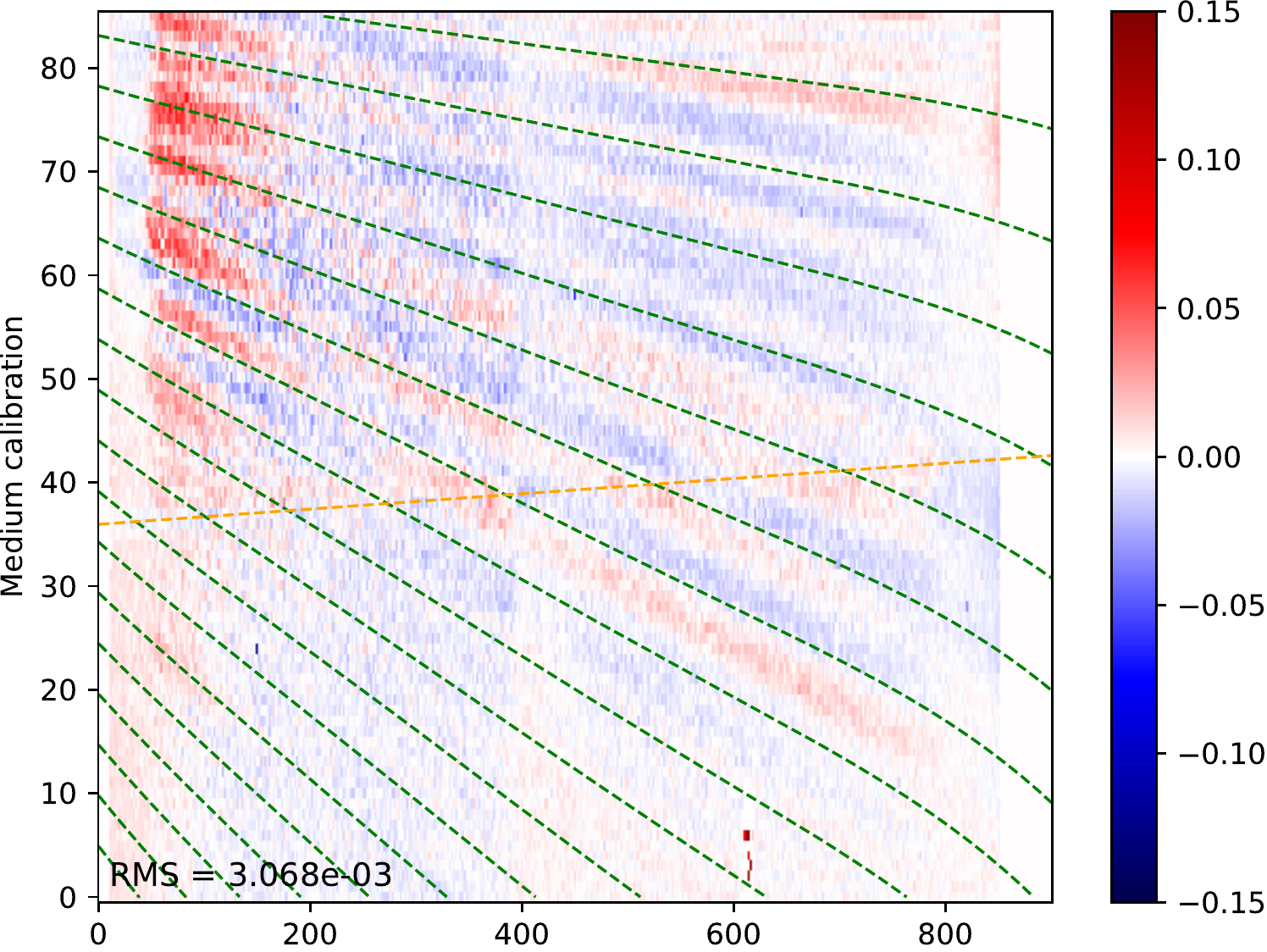}
  \includegraphics[scale=0.58]{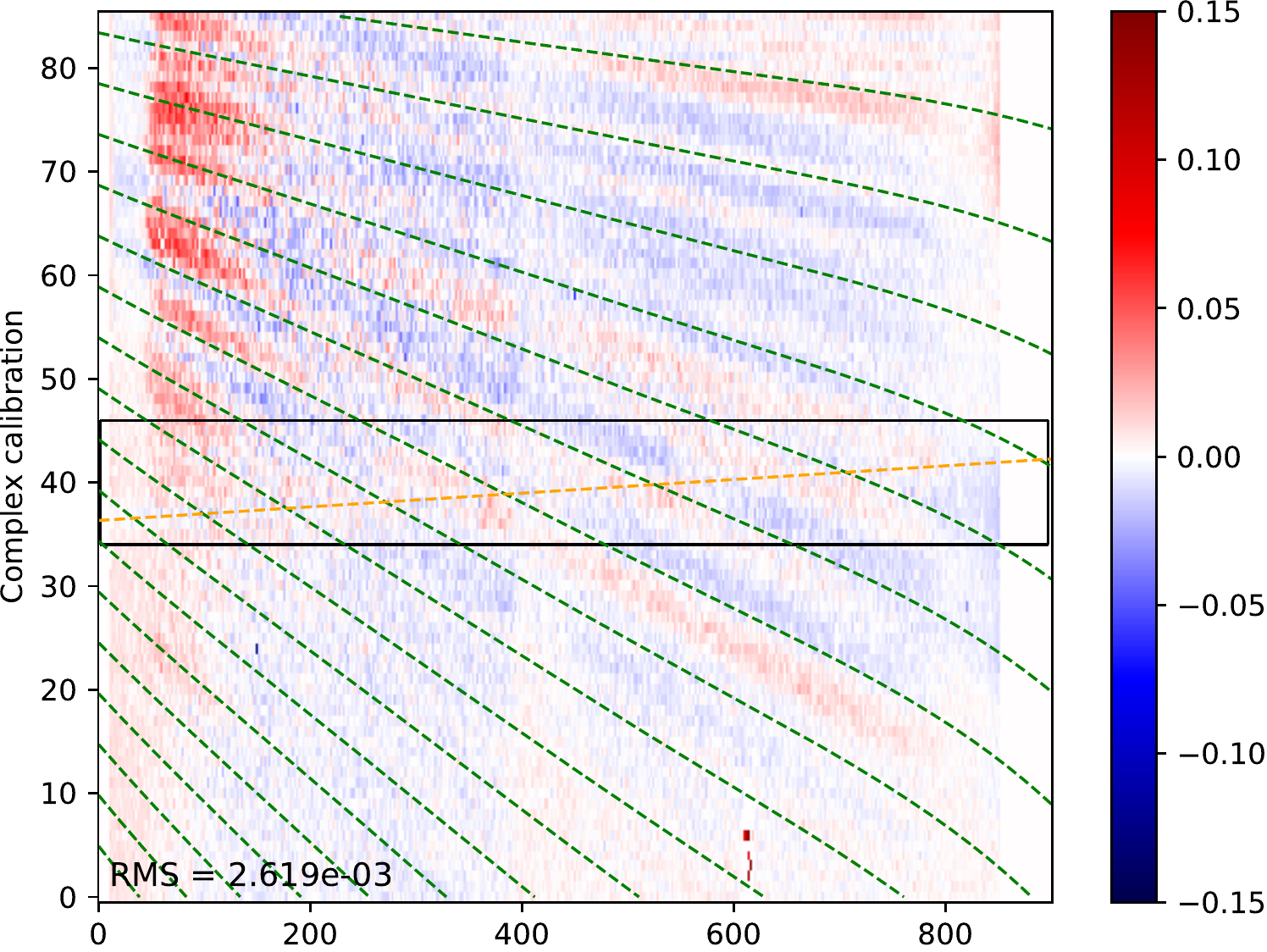}
  \caption{Residuals between the HR\,3549 scientific data and the model of the
    stellar leakages, assuming a \emph{standard} (top figure), \emph{medium}
    (center), and \emph{complex} (bottom) calibration models.  The RMS values
    of the residuals are given for each model. The iso-levels of
    $\rho/\lambda$ which are approximately followed by the dispersed stellar
    speckles are plotted as green dashed lines.  The position of the companion
    at the different wavelengths (\ie, at $\rho = \rho_\planet$) is plotted as
    an orange dashed line.}
  \label{fig:HR3549_residuals}
\end{figure}

To compare the calibration models considered in
Section~\ref{sec:dispersion_laws}, we apply \Exospeco
(Algorithm~\ref{alg:EXOPSEC}) on a scientific dataset of the star HR\,3549
observed in MRS mode of SPHERE/IRDIS on 2015/12/28.
Figure~\ref{fig:HR3549_residuals} shows the residuals $\V r = \V d - \V m$,
that is the difference between the data and their model, computed for the
different spectro-angular dispersion laws. This figure shows that the Root Mean
Square (RMS) of the residuals are significantly reduced when using more
flexible calibration models than the \emph{standard} one. The improvement
brought by the \emph{medium} complexity model compared to the \emph{standard}
model proves that accounting for a slight angular misalignment between the
spatial and spectral directions and the detector axes is important. Compared to
the \emph{medium} model, the \emph{complex} model is able to account other
image distortions than a simple rotation and thus achieves a better suppression
of the stellar leakages. These results motivate the choice of the
\emph{complex} dispersion model in \Exospeco to reduce the RMS level of the
residual stellar leakages by a factor of $\sim 2$ compared to the
\emph{standard} model and should therefore result in a better extraction of the
companion contribution. Remember that the simple \emph{standard} calibration
model is similar to what is usually done by others for these data.

\section{Validation and tuning of the method}
\label{sec:validation}

To fully validate the method, we propose in this section a study of the method on both real data and data where a synthetic companion was injected. This study allows us to both evaluate the modeling of the stellar leakages, and so its subtraction in the residuals, and the extraction of the companion, by comparing with a ground truth spectrum.

\subsection{Reduction of the self-subtraction}

\begin{figure}[t!]
  \centering
  \includegraphics[scale=0.58]{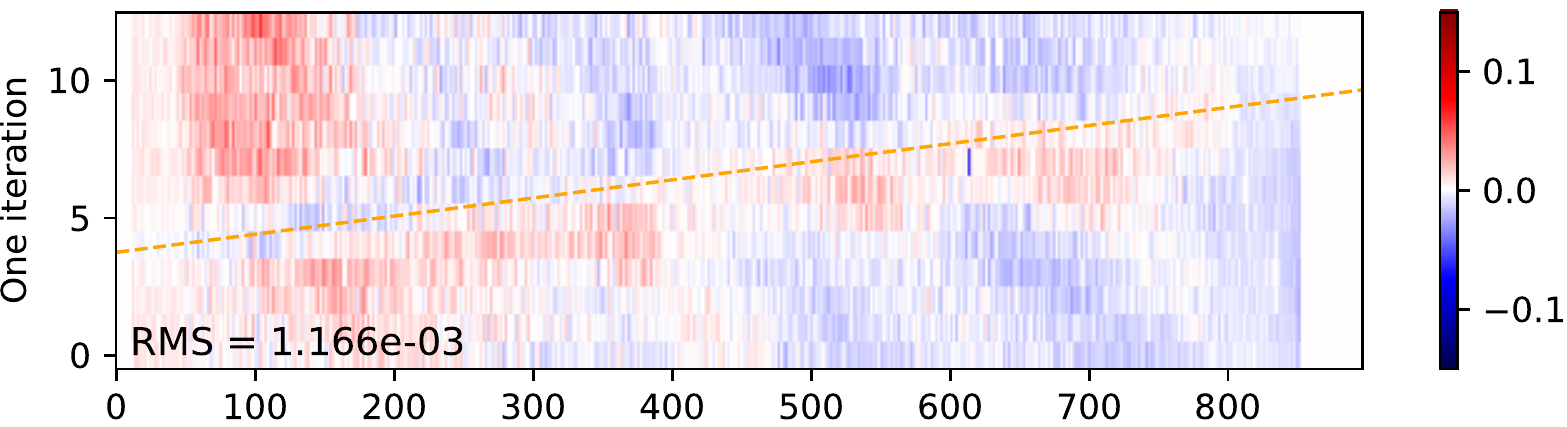}
  \includegraphics[scale=0.58]{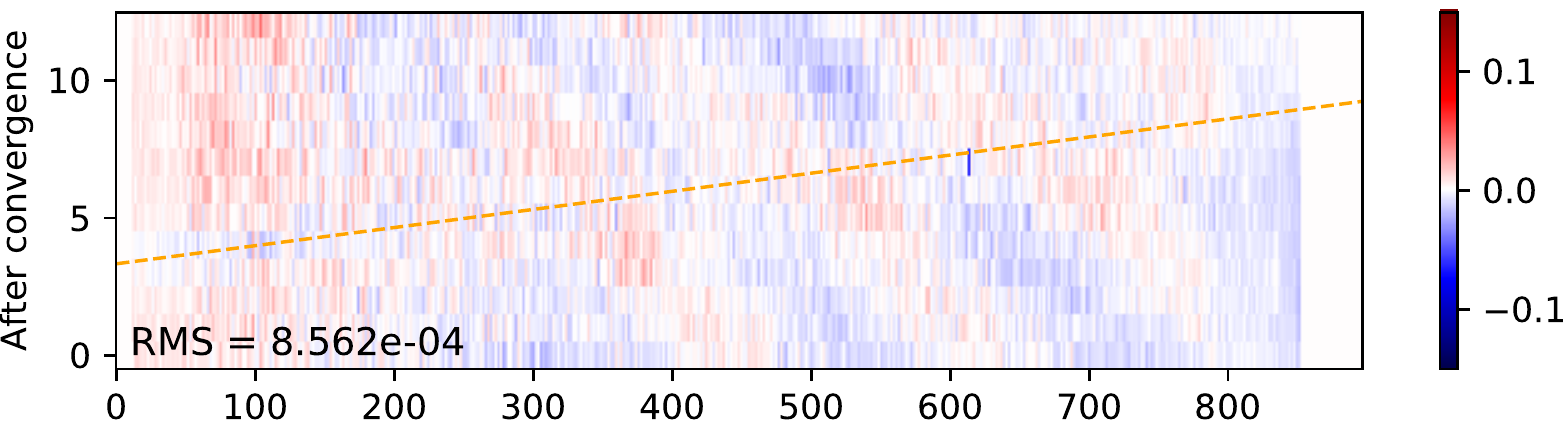}
  \includegraphics[scale=0.58]{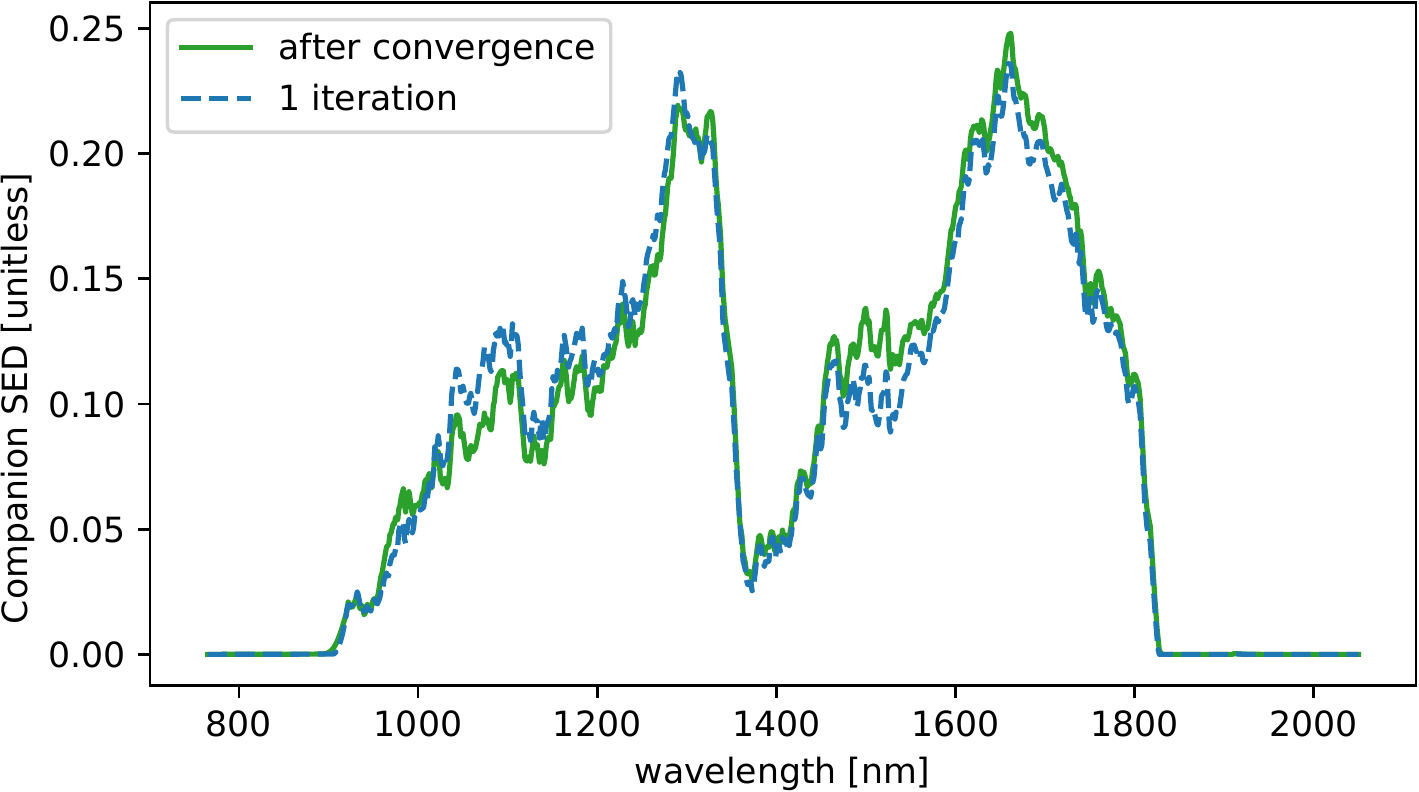}
  \caption{Residuals between the HR\,3549 scientific data and the model of the
    stellar leakages near the companion (defined by the black rectangle in the
    bottom of Fig.~\ref{fig:HR3549_residuals}) and for the \emph{complex model}
    of the spatial and spectral dispersion laws after one iteration (top) and
    after convergence (middle) of \textsc{Expospec}. The RMS of the residuals
    in this region are significantly reduced after convergence. The orange
    dashed line indicates the position of the companion at the different
    wavelengths. The SEDs of the companion (including atmospheric absorption)
    extracted from these residuals are plotted in the bottom-most panel.}
  \label{fig:HR3549_residuals_comp_cvg}
\end{figure}

One of the feature of \Exospeco is that it jointly estimates the contributions
of the star and of the companion whose parameters are iteratively refined until
convergence. Figure~\ref{fig:HR3549_residuals_comp_cvg} shows the residuals
close to the companion (in the region outlined by the black rectangle in the
bottom of Fig.~\ref{fig:HR3549_residuals}) with the \emph{complex} model of the
spatial and spectral dispersion laws in two cases: after the first outer
iteration of \Exospeco and after convergence of the algorithm. In the first
outer iteration of \Exospeco, the model of the stellar leakages is estimated by
masking the region most impacted by the companion, see Eq.~\eqref{eq:W-star},
which is similar to what is done by conventional methods. In all other outer
iterations of \Exospeco, the contribution of the other component is taken into
account when fitting a given component (star or companion). As shown by the
bottom panel of Fig.~\ref{fig:HR3549_residuals_comp_cvg}, there is a noticeable
bias in the estimated companion's SED after the first outer iteration. This
so-called \emph{self-subtraction} bias is mostly avoided by the proposed
alternating strategy.

\subsection{Tuning of the regularization parameters}
\label{sec:tuning-regularization}

As described in Appendix~\ref{sec:Scaling_indetermination}, the fact that the
model of the star leakages is bi-linear makes it possible to tune the
regularization of this component by a single hyper-parameter, the other
hyper-parameter being held fixed. Thanks to this, the solution found by
\Exospeco only depends on 2 hyper-parameters, one for the star, say,
$\mu_{\V x}$ (while $\mu_{\V y} = 1$ is imposed) and one for the companion,
$\mu_{\V z}$. In this section, we highlight the incidence on the companion SED
extracted by \Exospeco of these remaining hyper-parameters using the same
scientific data set as in Section~\ref{sec:comparing_calibration_models}.

\begin{figure}[t]
  \includegraphics[scale=0.58]{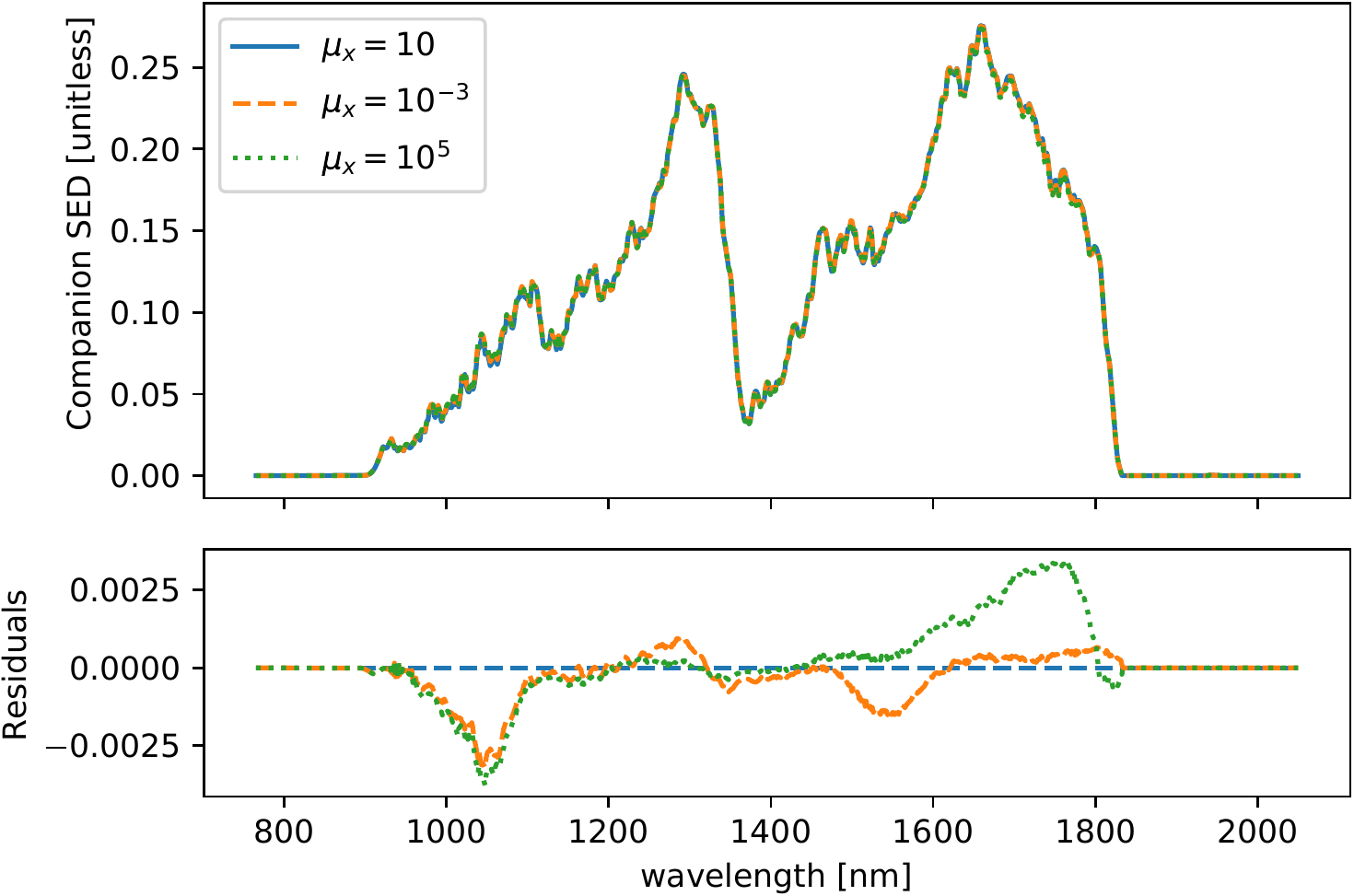}
  \caption{Top: Profiles of the companion SED $\V z$, for different levels of
    the stellar hyper-parameter $\mu_{\V x}$ and with $\mu_{\V y} = 1$ and $\mu_{\V z} = 10^5$. Bottom:
    differences between the profile for $\mu_{\V x} = 10$ and the profiles for
    $\mu_{\V x} = 10^{-3}$ (dashed orange) and $\mu_{\V x} = 10^5$ (dotted
    green)}
  \label{fig:HR3549_profiles_diff_star_regul}
\end{figure}

Figure~\ref{fig:HR3549_profiles_diff_star_regul} shows the SEDs of the
companion estimated by \Exospeco for different values of the star regularization
hyper-parameters ($\mu_{\V y} = 1$ and $\mu_{\V x} = 10^{-3}$, $10$, and
$10^5$). For such a large range of values, the differences between the
extracted companion SEDs are smaller than 1\%. The stellar regularization
hyper-parameters have thus a limited impact on the resulting companion SED. The
tuning of $\mu_{\V x}$ can thus reasonably be done by visual inspection.

\begin{figure}[t]
  \includegraphics[scale=0.58]{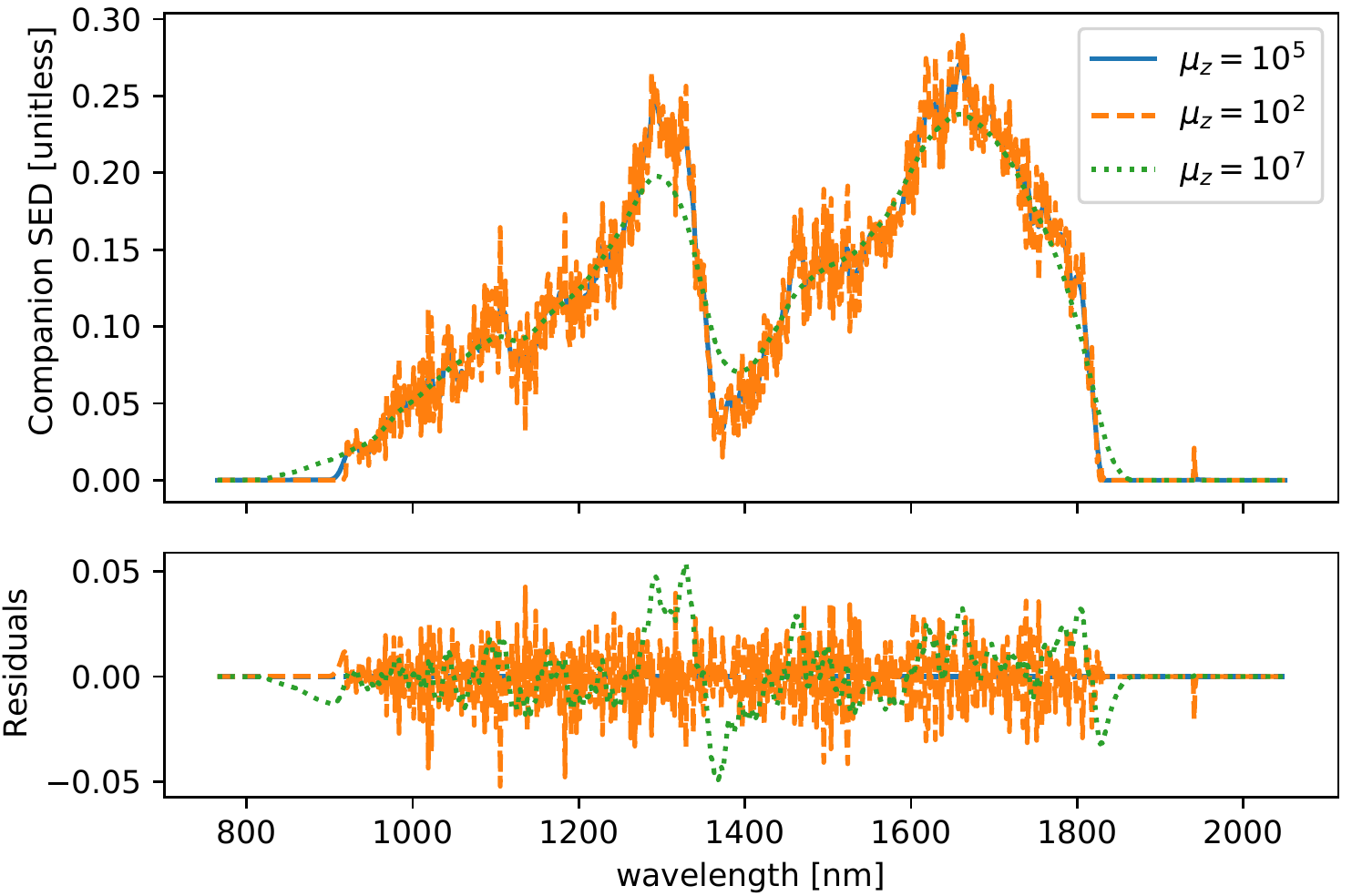}
  \caption{Top: Profiles of the companion SED $\V z$ for different levels of
    the companion regularization ($\mu_{\V z} = 10^2$ in dashed orange,
    $\mu_{\V z} = 10^5$ in blue, and $\mu_{\V z} = 10^7$ in dotted green).
    Bottom: differences between the profile for $\mu_{\V z} = 10^5$ and the
    profiles for $\mu_{\V z} = 10^2$ (dashed orange) and $\mu_{\V z} = 10^7$
    (dotted green). For all these results, the stellar hyper-parameters are
    $\mu_{\V x} = 10$ and $\mu_{\V y} = 1$.}
  \label{fig:HR3549_profiles_diff_comp_regul}
\end{figure}

On the contrary, as Fig.~\ref{fig:HR3549_profiles_diff_comp_regul} shows, the
hyper-parameter $\mu_{\V z}$ has a strong impact on the resulting companion
SED. This is expected as $\mu_{\V z}$ directly tunes the strength of the
smoothness constraint for the companion SED $\V z$. This hyper-parameter has
thus to be carefully chosen to find the best compromise between a solution that
is too smooth (\eg for $\mu_{\V z} = 10^{7}$ in
Fig.~\ref{fig:HR3549_profiles_diff_comp_regul}) or too noisy (\eg for
$\mu_{\V z} = 10^{2}$ in Fig.~\ref{fig:HR3549_profiles_diff_comp_regul}). It is
worth noticing that the correct value of $\mu_{\V z}$ strongly depends on the
considered data, so $\mu_{\V z} = 10^{5}$, which seems to be a good choice for
the HR\,3549 data (see Fig.~\ref{fig:HR3549_profiles_diff_comp_regul}), should
not be considered as a universal value.

Many methods have been proposed to automatically tune the hyper-parameter(s) of
an inverse problem: the Generalized Cross Validation \citep[CGV,
][]{Golub_Heath_Wahba-1979-GCV}, Stein's Unbiased Risk Estimate
\citep[SURE,][]{Stein-1981-SURE}, the hierachical Bayesian method
\citep{Molina-1994-hierarchical_Bayesian}, or the L-curve
\citep{Hansen_OLeary-1993-L_curve} to mention a few that could be used with our
extraction algorithm. Implementing and testing these methods for \Exospeco is
out of the scope of this paper. However, since the companion SED found by
\Exospeco does not strongly depend on the tuning of the stellar regularization,
our method is mostly driven by a single hyper-parameter, $\mu_{\V z}$, the
level of the regularization for the companion SED. This greatly reduces the
complexity of tuning the \Exospeco algorithm.

\subsection{Extraction of simulated spectrum in real data}
\label{sec:synthetic-data}

\begin{figure}[t]
  \centering
  \includegraphics[scale=0.58]{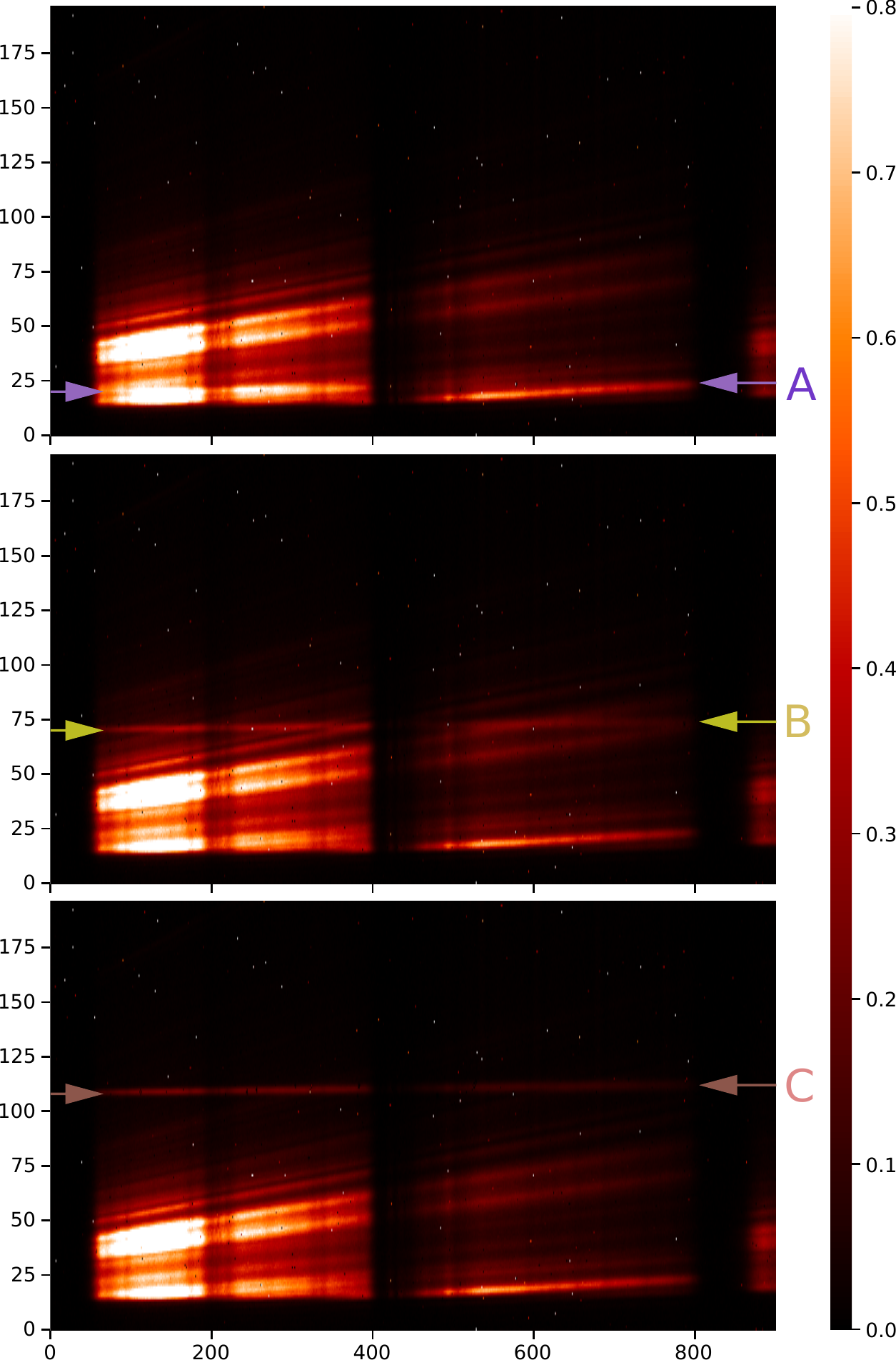}
  \caption{Scientific data of HIP\,65426 with a synthetic companion whose
    contrast is $\chi = 2\cdot10^{-4}$ relative to the star and injected at
    angular separations $\rho_\planet - \rho_\star = 273\,\mas$
    (\VioletBlue{A}), $890\,\mas$ (\DarkYellow{B}), and $1353\,\mas$
    (\DarkPink{C}) indicated by the arrows.}
  \label{fig:HIP65426_data_injections}
\end{figure}

To validate the \Exospeco method, we injected the contribution of a synthetic
companion in existing SPHERE/IRDIS MRS data $\V d$ of the star HIP\,65426
observed on 2019-05-20.
Although HIP\,65426 star hosts a planet \citep{chauvin-2017-HIP65426,Carter-2022-HIP65426b},
the frame was selected for the derotation angles hiding the planet outside the
slit. The off-axis PSF $h_\planet$ of the synthetic companion follows the model
in Eqs.~\eqref{eq:off-axis-psf-at-lambda-ref} and
\eqref{eq:off_axis_psf_discrete} with $\sigma_\planet$ set to match the
diffraction limit of the telescope at the reference wavelength
$\lambda^\Tag{ref}$ and with different angular positions $\rho_\planet$ on the
side of the coronagraphic mask where no companion was detected. The
\emph{ground truth} SED of the synthetic companion is
$\V z_\Tag{gt} = \chi\,\V x_\Tag{flux}$ where $\chi > 0$ is the mean contrast
of the companion relative to the star (without a coronagraph) and
$\V x_\Tag{flux}$ is the SED of the star HIP\,65426 calibrated as explained in
Appendix~\ref{sec:star_SED_calibration}. We use a constant
contrast for all wavelengths (i.e., the SED of the star and of the companion
are the same, up to the contrast $\chi$).
Figure~\ref{fig:HIP65426_data_injections} shows examples of generated data with
a synthetic companion whose contrast with respect to the star is
$\chi = 2\cdot10^{-4}$ and which is injected at different angular separations
$\rho_\planet - \rho_\star$.

\begin{figure}[t!]
  \centering
  \includegraphics[width=\columnwidth]{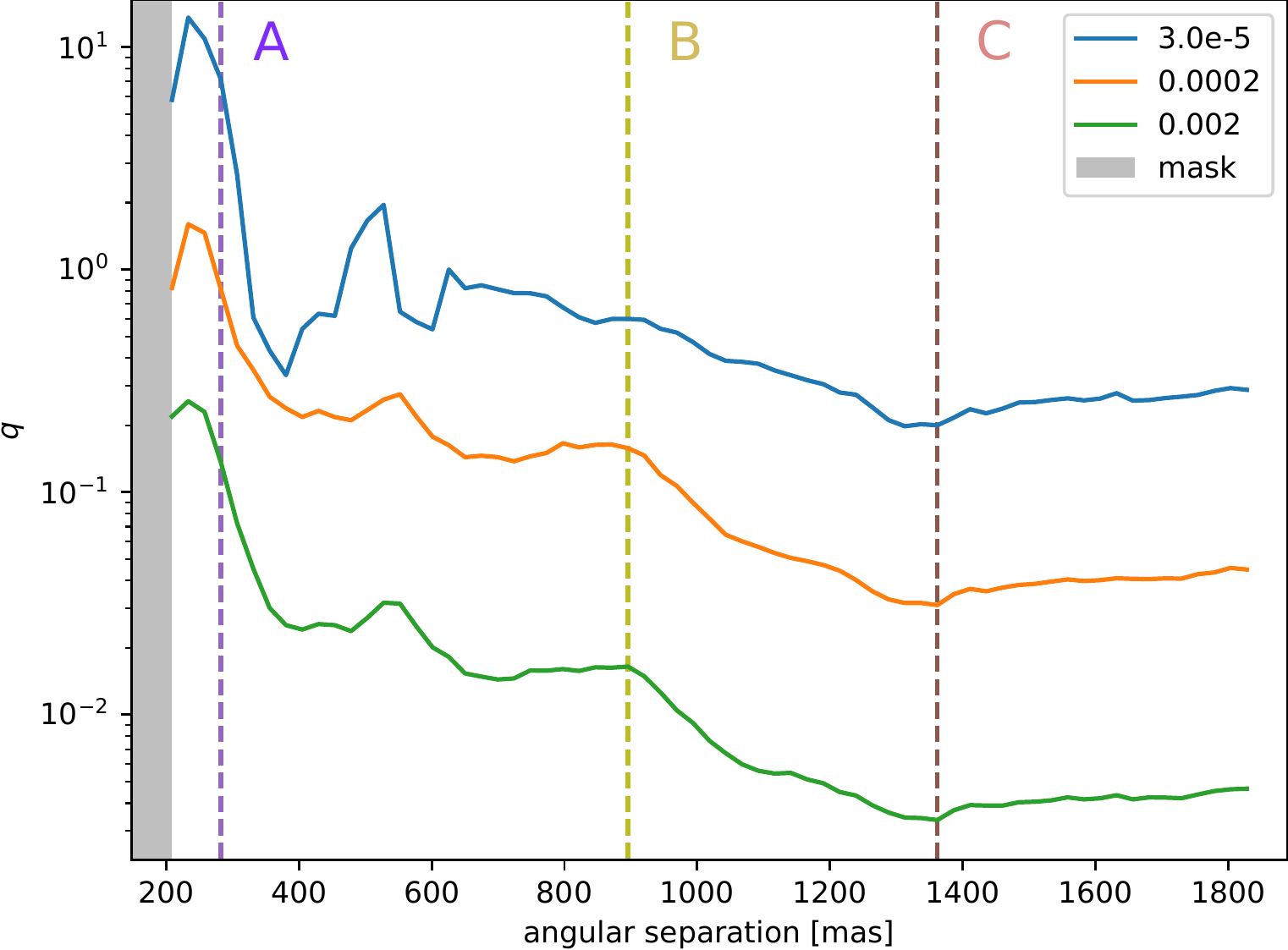}
  \caption{Relative error $q$ defined in Eq.~\eqref{eq:rel-abs-err} for
    synthetic companions injected in the scientific data of HIP\,65426 (with
    the same spectra as the host star) as a function of the angular separation
    $\rho_\planet - \rho_\star$ and for contrasts $\chi = 3\cdot10^{-5}$
    (blue), $2\cdot10^{-4}$ (orange), and $2\cdot10^{-2}$ (green). The grayed
    area represents the region invalidated by the coronagraphic mask. The
    angular separations of the three cases presented in
    Fig.~\ref{fig:HIP65426_data_injections} are highlighted by the dashed lines
    labeled \VioletBlue{A}, \DarkYellow{B}, and \DarkPink{C}.}
  \label{fig:HIP65426_injections_quality_factor}
\end{figure}

To assess the quality of the extracted companion's SED $\estim{\V z}$, we
compute the following relative error:
\begin{equation}
  q = \frac{\sum_{j=1}^{N_{\V z}}\Abs*{z_{\Tag{gt},j} - \estim{z}_j}}
  {\sum_{j=1}^{N_{\V z}}\Abs*{z_{\Tag{gt},j}}}\,.
  \label{eq:rel-abs-err}
\end{equation}
In the following tests, the value of $\mu_{\V z}$, the regularization level of
the companion's SED, has been tuned so as to minimize the relative error $q$.
Figure~\ref{fig:HIP65426_injections_quality_factor} plots the relative error
$q$ for synthetic companions injected at angular separations
$\rho_\planet - \rho_\star$ ranging from $200\,\mas$ to $1850\,\mas$ and with
contrasts $\chi = 3\cdot10^{-5}$, $2\cdot10^{-4}$, and $2\cdot10^{-3}$.
Clearly, the quality of the recovered SEDs degrades as the companion gets
closer to the mask. This is expected because, when getting closer to the mask,
not only are the stellar leakages brighter (hence causing more photon noise in
the residuals) but the approximation by the assumed off-axis PSF model also
worsens. For angular separations larger than $\sim600\,\mas$ and for all
considered contrasts, the quality of the recovered SEDs improves as the
separation increases until a plateau is reached at
$\rho_\planet - \rho_\star \sim 1400\,\mas$ where the dominant source of
nuisance is the readout noise.

\begin{figure}[t!]
  \centering
  \includegraphics[scale=0.58]{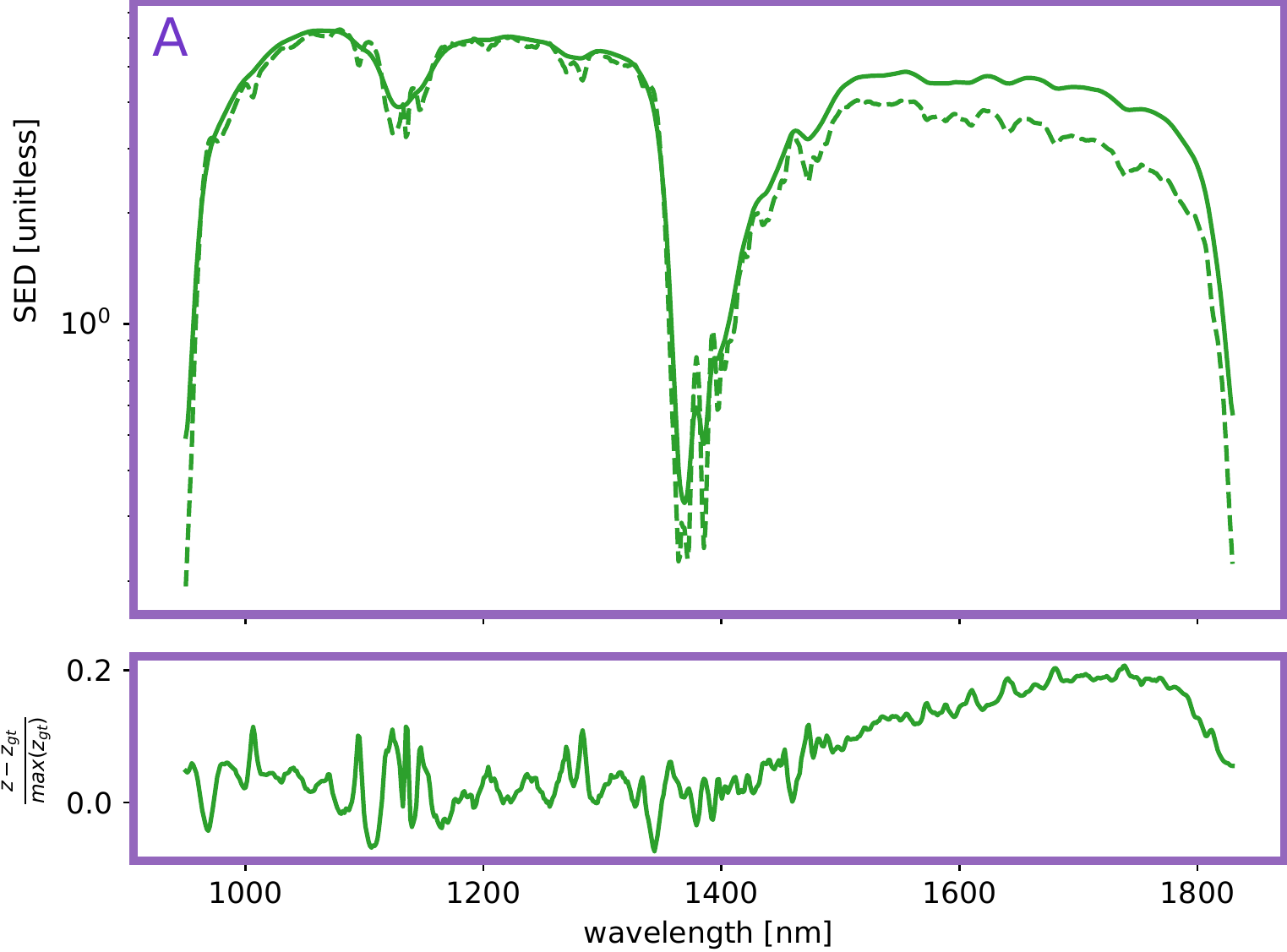}\\[2ex]
  \includegraphics[scale=0.58]{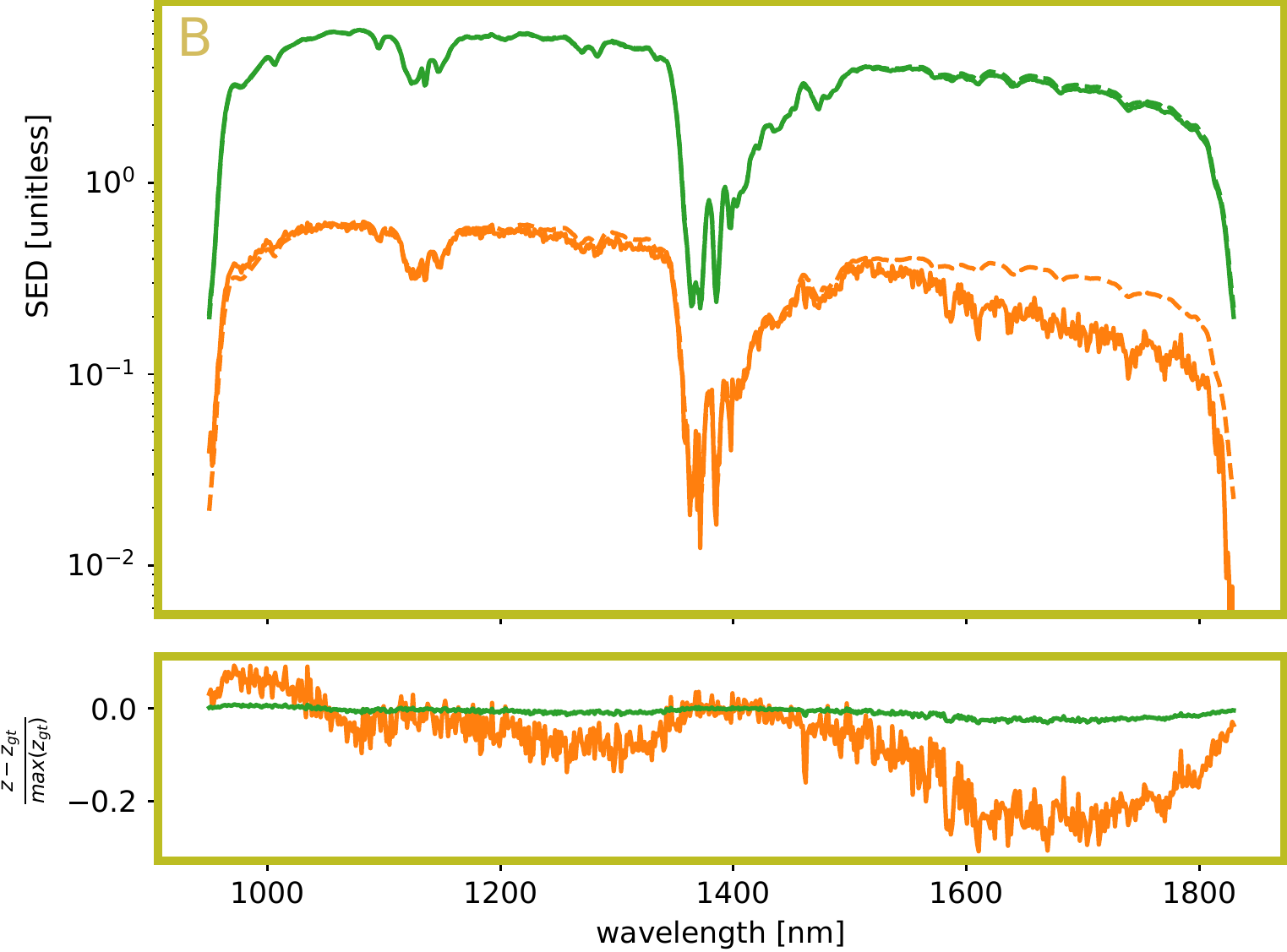}\\[2ex]
  \includegraphics[scale=0.58]{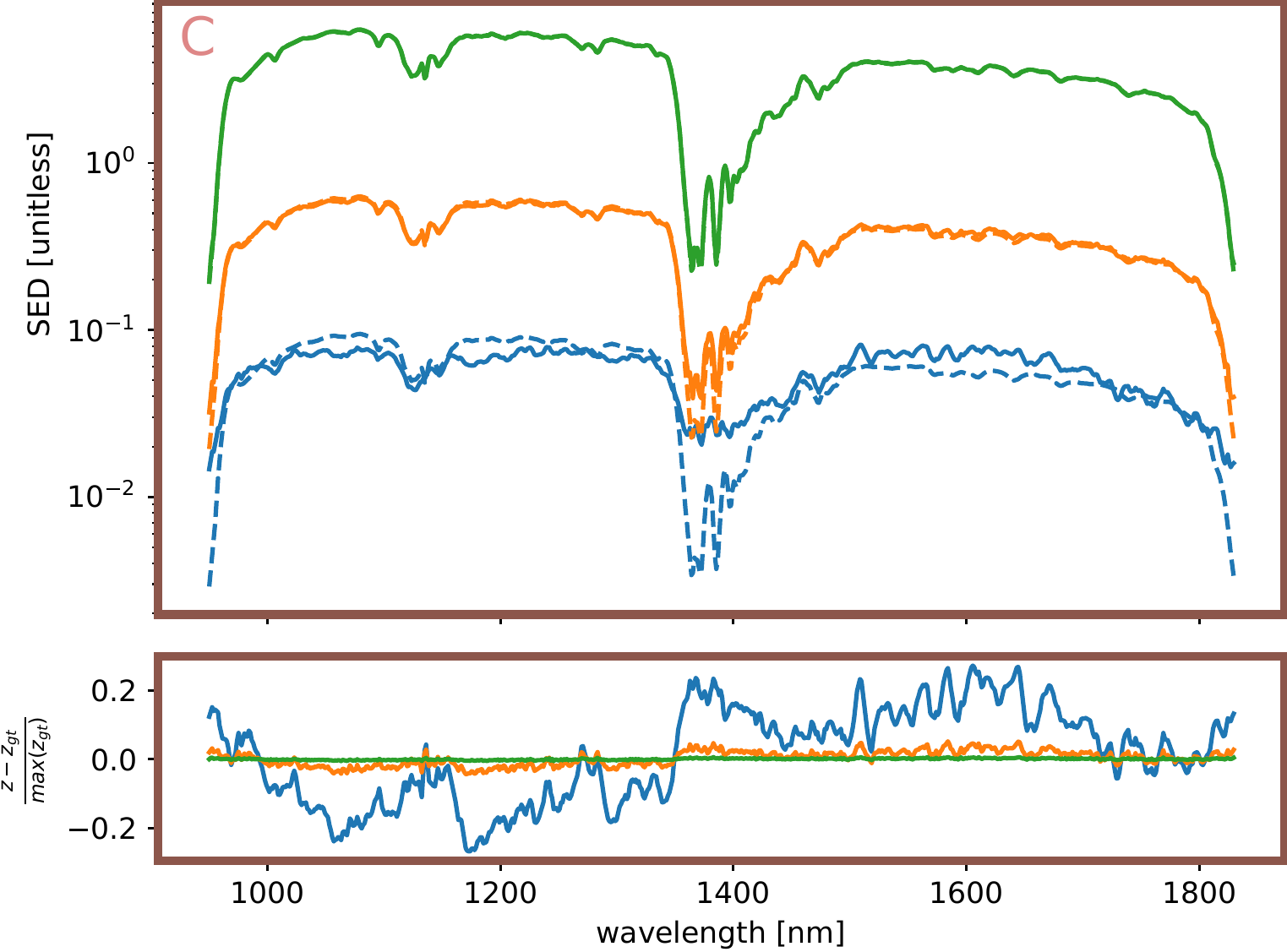}
  \caption{Examples of recovered companion SEDs $\estim{\V z}$ in the same
    conditions as in Fig.~\ref{fig:HIP65426_data_injections} and
    Fig.~\ref{fig:HIP65426_injections_quality_factor} for synthetic companions
    injected at angular separations $\rho_\planet - \rho_\star = 273\,\mas$
    (\VioletBlue{A}), $890\,\mas$ (\DarkYellow{B}), and $1353\,\mas$
    (\DarkPink{C}) with contrasts $\chi = 3\cdot10^{-5}$ (blue curves),
    $2\cdot10^{-4}$ (orange curves), and $2\cdot10^{-2}$ (green curves). The
    ground truth SED $\V z_\Tag{gt}$ is plotted in dashed lines. The normalized
    residuals are plotted below each panel.}
  \label{fig:HIP65426_injections_profiles}
\end{figure}

Figure~\ref{fig:HIP65426_injections_profiles} shows examples of recovered
companion SEDs $\estim{\V z}$ at angular separations
$\rho_\planet - \rho_\star = 273\,\mas$ (A), $890\,\mas$ (B), and $1353\,\mas$
(C) for the same contrasts $\chi$ as in
Fig.~\ref{fig:HIP65426_injections_quality_factor}.
Figure~\ref{fig:HIP65426_injections_profiles} confirms that the relative error
$q$ does reflect the ability of our method to reliably recover the companion
SED. When $q \le 0.1$ (the green curves for angular separations B and C and the
orange curve for angular separation C), all the features of the SED are
correctly recovered. For $q \sim 0.2$ (the green curve for case A, the orange
curve for case B, and the blue curve for case C), the global shape of the SED
is restored but with small spectral features smoothed out and some photometric
biases. These cases prove that it is possible to extract a coarse but still
exploitable SED for bright companions quite close to the mask, typically
$\chi \ge 10^{-3}$ for $\rho_\planet - \rho_\star \sim 250\,\mas$, from a
single MRS exposure. The angular separation must be larger for fainter
companions; for example, $\rho_\planet - \rho_\star \ge 1200\,\mas$ for
$\chi \sim 2\cdot10^{-5}$. The photometric biases in the most difficult cases
(the green curve in case A, the orange one in case B, and the blue one in case
C) clearly indicates that the removal of the modeled stellar contribution
leaves non-negligible residuals compared to the companion. A possible
improvement could be to use a more complex model of the on-axis PSF and
consider more than one mode in the series expansion of
Eq.~\eqref{eq:multi-mode-psf}.

\begin{figure}
  \includegraphics[scale=0.58]{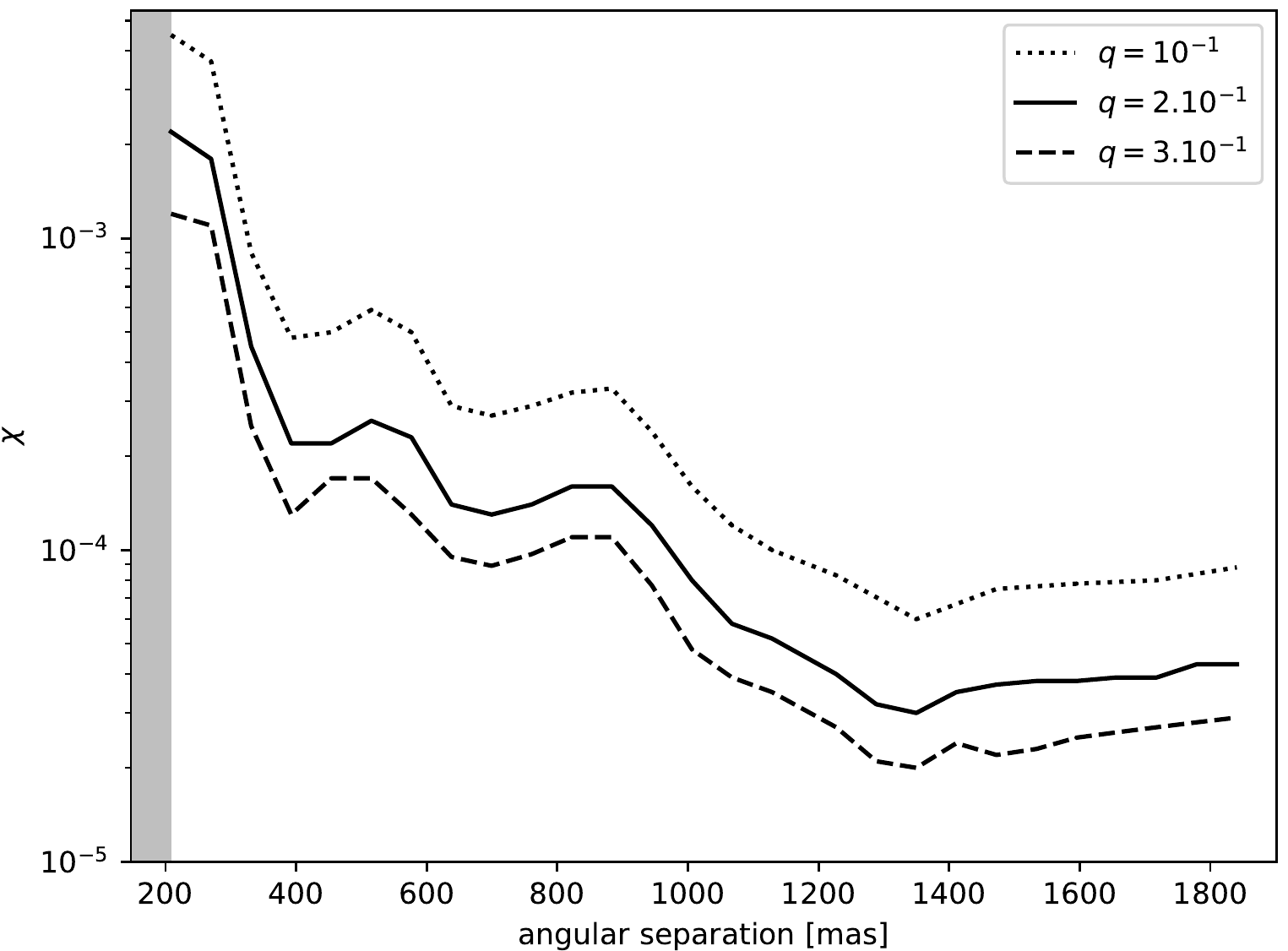}
  \caption{Minimal contrast $\chi$ required to achieve a given relative error
    $q$ as a function of the angular separation. The conditions are the same as
    in Fig.~\ref{fig:HIP65426_data_injections} and
    Fig.~\ref{fig:HIP65426_injections_quality_factor}.}
  \label{fig:HIP65426_injections_contrast_curve}
\end{figure}

To summarize the performances of the current version of \Exospeco for a single
data frame of the HIP\,65426 observations,
Fig.~\ref{fig:HIP65426_injections_contrast_curve} plots the minimal contrast
needed to achieve a given relative error $q$ as a function of the angular
separation. The figure shows that by tolerating a relative error as high as
$q = 0.3$, a companion with a contrast up to $\chi \sim 2\cdot10^{-5}$ can be
characterized. In our conclusions, we explain how to extend \Exospeco to
jointly process several data frames in order to increase the sensitivity of the
algorithm.

\subsection{Comparison with TSVD extraction}
\label{sec:Exospeco_vs_TSVD}

\begin{figure}[t]
  \centering
  \includegraphics[scale=0.58]{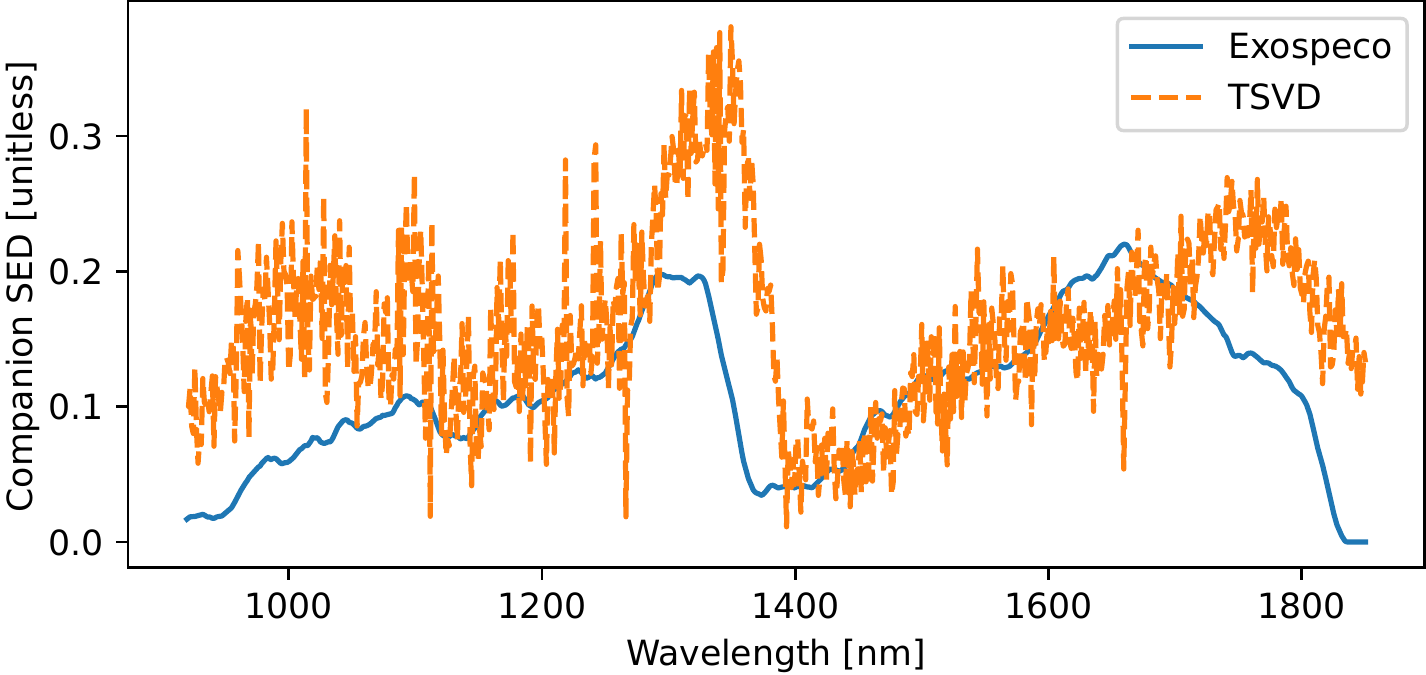}
  \caption{Comparison of the SEDs extracted from the HR\,3549 data by \Exospeco
    and by a standard TSVD method. See text for
    details.} \label{fig:TSVD_vs_Exospeco}
\end{figure}

\begin{figure}[t]
  \centering
  \includegraphics[scale=0.58]{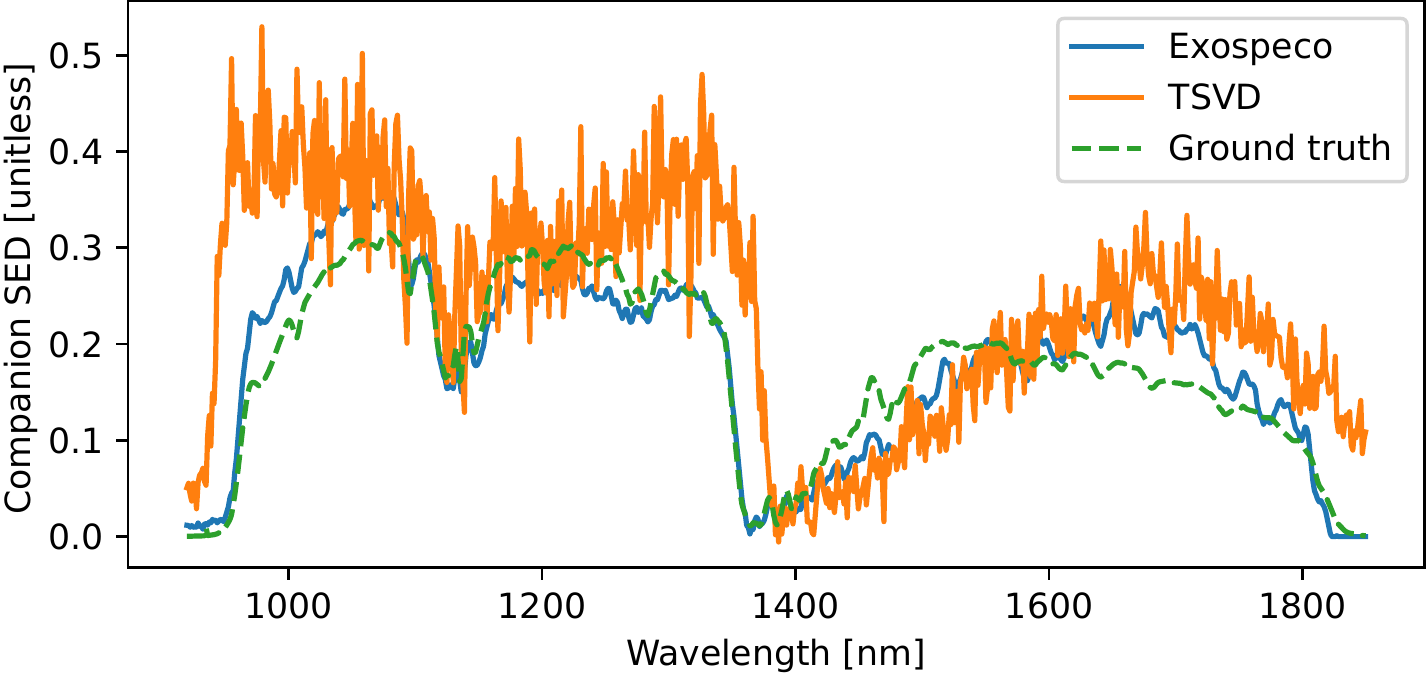}
  \caption{Comparison of the SEDs extracted from semi-synthetic data by
    \Exospeco and by a standard TSVD method. The angular separation and the
    contrast of the injected companion are respectively
    $\rho_{\planet} - \rho_{\star} = 785\,\mathrm{mas}$ and
    $\chi = 2\times10^{{-4}}$. The green dashed curve represents the ground
    truth injected spectrum which is that of HIP\,65426 multiplied by $\chi$.}
  \label{fig:TSVD_vs_Exospeco_synthetic}
\end{figure}

We compared \Exospeco to a standard approach based on the TSVD method described
in Section~\ref{sec:TSVD} to remove the stellar leakages.
Figure~\ref{fig:TSVD_vs_Exospeco} shows the companion SED extracted from the
HR\,3549 data by \Exospeco and by the TSVD approach. For the latter method, the
SED of the companion was extracted by local averaging in a 7 pixel height
sliding window along the companion signal in the residual image given by
Eq.~\ref{eq:TSVD-residuals} and shown in Fig.~\ref{fig:TSVD-residuals}. In both
cases, the same \emph{complex} calibration model of the spectral and angular
dispersion laws described in Appendix~\ref{sec:calibration_details} has been used. In
spite of this identical calibration, the two extracted SEDs are notably
different. Thanks to the optimal extraction in the maximum likelihood sense and to the spectral regularization,
the SED extracted by \Exospeco is smoother and less noisy. At a coarser
resolution, the two SEDs display quite different spectral features. However,
without a known ground truth, the two SEDs cannot be ranked. For this reason,
we also compared the results given by the two methods on a synthetic injection
done as described in Section~\ref{sec:synthetic-data}.
Figure~\ref{fig:TSVD_vs_Exospeco_synthetic} clearly demonstrates that not only
does \Exospeco produce less noisy results, but that they also better reflect
reality.

\section{Conclusion}
\label{sec:conclusion}

In this paper we presented a novel algorithm, \Exospeco, to extract the spectrum
of a companion from high-contrast long-slit spectroscopic data. The most
challenging part of such a processing is to disentangle the signal of interest
from the stellar leakages which are much brighter. Compared to existing methods,
our algorithm avoids any transform of the data, whether it is to align the
speckles of the stellar leakages at all wavelengths or to fix defective pixels.
\Exospeco has also the advantage of jointly extracting the parameters describing
the stellar leakages (the star spectrum and the on-axis PSF), the companion
spectrum, the off-axis PSF, and, optionally, some calibration parameters. By
using non-uniform statistical weights for the data pixels, our approach is
\emph{optimal} in the maximum likelihood sense, it takes into account all
available measurements and consistently treats defective pixels as missing data.
The joint optimization problem having no closed-form solution, we
proposed an alternating minimization strategy which has proven to be effective.
In spite of the numerous parameters coming into play in the algorithm, the
outputs of the method are, in practice, mostly driven by a single
hyper-parameter that tunes the level of regularization of the companion SED.

Although it is not directly part of the spectrum extraction algorithm, we have
shown that careful calibration of the instrument is critical to get rid of the
contamination by the stellar leakages. For that purpose, we described a refined
calibration method of the spectral and spatial dispersion laws from available
calibration data. In particular, \textsc{SPHERE/LSS} data present a
misalignment of the principal directions of dispersion with the detector axes
as well as a geometrical shear. If not accounted for, we show that these
distortions have a detrimental impact on the result of the processing, whether
it is by \Exospeco or by the current state-of-the-art method. A few remaining
calibration parameters that may depend on the observing conditions, such as the
off-axis PSF size and the precise locations of the star and of the companion
along the slit, can be optionally adjusted by a self-calibration procedure
built into \Exospeco. Thanks to this calibration step, our method significantly
reduces the \emph{self-subtracting} bias by better disentangling the
stellar leakages component from the companion component.

A Julia \citep{Bezanson_et_al-2017-Julia} implementation of \Exospeco is
freely available at \url{https://github.com/SJJThe/Exospeco}, while an
implementation of the calibration method described in
Section~\ref{sec:calibration} and Appendix~\ref{sec:calibration_details} is at
\url{https://github.com/SJJThe/ExospecoCalibration}.

Based on tests carried on empirical long-slit spectroscopic data and on
injections of a synthetic companion signal in these data, we demonstrated that
the proposed approach effectively avoids the \emph{self-subtraction} bias, even
very close to the coronagraphic mask. We provided curves to predict the minimal
contrast required to achieve a given quality of extraction of the companion
SED. Reliable extraction of a companion SED can be achieved from a single data
frame at contrasts as low as a few $10^{-5}$. The proposed method could boost
the characterization of known (faint) exoplanets at a spectral resolution
substantially higher than currently possible with SPHERE IFS ($R \sim 35 - 50$)
and for contrasts much better than achievable with IRDIS MRS using state of the
art methods. By capturing more efficiently the stellar contamination, the
method we propose does not require independent and thus imperfect calibration
of the speckles by rotating the slit to hide the planet signal. This will
typically gain at least 50\% telescope time while reaching, and even
surpassing, the same contrast limit. This new method also paves the way to
combining polarimetry and spectroscopic measurements with IRDIS LSS mode (R.\
Holstein private communication).

Being based on an inverse problems framework, \Exospeco is very flexible
and can be adapted to various kinds of data (such as data sequences or data from
other instruments). An example of such an extension of \Exospeco is the joint
processing of multiple frames that can be done as follows. Assuming $T$ LSS
exposures $\V d_1$ to $\V d_T$ of the same object are collected during a night,
they can be combined into a single criterion that extends
Eq.~\eqref{eq:criterion}:
\begin{align}
\Criterion(&\V x, \V y_1, \V y_2, \dots, \V y_T, \V z,\V\nu,\V \mu)\notag\\
& = \sum_{t=1}^{T}\Norm*{\V d_{t} - \V m(\V x,\V y_t,\V z,\V\nu)}^2_{\M W_{t}}
+ \mu_{\V x}\,\RegulX(\V x_t)\notag\\
& + \mu_{\V y}\,\sum_{t=1}^{T}\RegulY(\V y_t)
+ \mu_{\V z}\,\RegulZ(\V z),
\end{align}
where the statistical independence of noise between frames is considered (a
natural assumption). This criterion can be minimized in $\V x$, $\V y_{1}$,
\ldots, $\V y_{T}$, $\V z$, and $\V\nu$ following the same alternating method
as described in Section~\ref{sec:minimization-strategy}, only with more steps
in order to estimate the on-axis PSF at each of the $T$ frames. Such a joint
processing has the potential to improve the estimation of companion SEDs and
push further the achievable contrast limit.

Finally, to better disentangle the stellar leakages from the companion
spectrum, the model of the on-axis PSF could be improved by taking into
account more spatial modes of the series expansion in
Eq.~\eqref{eq:multi-mode-psf}. Indeed, as demonstrated in
\citet{devaney_pex_2017}, accounting for more such modes significantly
improves the modeling of the stellar leakages, especially near the
coronagraphic mask. Such an improvement would not call into question the
founding principles of \Exospeco, but would require to adapt the optimization
strategy.

\begin{acknowledgements}
  This work was supported by the Action Spécifique Haute Résolution Angulaire
  (ASHRA) of CNRS/INSU co-funded by CNES.

  SPHERE is an instrument designed and built by a consortium consisting of IPAG
  (Grenoble, France), MPIA (Heidelberg, Germany), LAM (Marseille, France),
  LESIA (Paris, France), Laboratoire Lagrange (Nice, France), INAF -
  Osservatorio di Padova (Italy), Observatoire de Genève (Switzerland), ETH
  Zürich (Switzerland), NOVA (Netherlands), ONERA (France) and ASTRON
  (Netherlands) in collaboration with ESO. SPHERE was funded by ESO, with
  additional contributions from CNRS (France), MPIA (Germany), INAF (Italy),
  FINES (Switzerland) and NOVA (Netherlands). SPHERE also received funding from
  the European Commission Sixth and Seventh Framework Programmes as part of the
  Optical Infrared Coordination Network for Astronomy (OPTICON) under grant
  number RII3-Ct-2004-001566 for FP6 (2004-2008), grant number 226604 for FP7
  (2009-2012) and grant number 312430 for FP7 (2013-2016).

\end{acknowledgements}

\bibliographystyle{aa} 
\bibliography{aa_LSS_bibliography}

\begin{appendix}

\section{Calibration of the spectro-angular dispersion laws}
\label{sec:calibration_details}

This appendix provides some details about the methods used for the calibration
of the spectro-angular laws described in Section~\ref{sec:calibration} and some
figures to support the results discussed in
Section~\ref{sec:comparing_calibration_models}. The considered calibration data
$\V d_\Tag{cal}$ is the image in the central panel of
Fig.~\ref{fig:LSS_calibration}.

\subsection{Transverse projection}
\label{sec:transverse_projection}

To locate the positions of the spectral lines, we compute a weighted
\emph{transverse} projection of the calibration image $\V d_\Tag{cal}$:
\begin{subequations}
  \label{eq:transverse_projection}
  \begin{equation}
    q_{\bot\,k}(\phi_\lambda) = \frac{
      \sum\nolimits_{i,j} \xi_{\bot\,i,j,n}(\phi_\lambda)\,d_{\Tag{cal},i,j}
    }{
      \sum\nolimits_{i,j} \xi_{\bot\,i,j,n}(\phi_\lambda)
    },
  \end{equation}
  with weights given by:
  \begin{equation}
    \xi_{\bot\,i,j,k}(\phi_\lambda)
    = w_{\Tag{msk},i,j}\,\varphi_\Tag{proj}\Paren[\big]{i\,\sin\phi_\lambda + j\,\cos\phi_\lambda - k}
  \end{equation}
\end{subequations}
and for a projection angle $\phi_\lambda$ chosen so as to maximize the peak
values of the resulting projection (plotted in red in the top panel of
Fig.~\ref{fig:LSS_calibration} for $\phi_\lambda \approx 0^{\circ}$). In
practice, we take $\varphi_\Tag{proj}(t) = \max\Paren{1 - \Abs{t}, 0}$, the
linear B-spline, as the interpolating function for the projection. Note that,
thanks to the weighting by the mask of valid pixels $\V w_\Tag{msk}$ defined in
Eq.~\eqref{eq:valid-pixel-mask}, invalid pixels have no incidence on the
computed projection.

\subsection{Detection of the spectral peaks}
\label{sec:spectral_peaks_detection}

We use Algorithm~\ref{alg:get_peaks} with tolerance parameter
$\delta_{\bot} = 10$ pixels to find the $N_\lambda$ most significant peaks in
the transverse projection $\V q_\bot(\phi_\lambda)\in\Reals^{N_q}$ computed
according to Eq.~\eqref{eq:transverse_projection}.

\begin{algorithm}
  \caption{Find the most significant peaks}
  \label{alg:get_peaks}
  \KwIn{$N_\lambda$, $\V q_\bot(\phi_\lambda)$, and $\delta_{\bot}$.}
  \KwOut{$\mathcal{P}(\phi_\lambda)$.}
  $\mathcal{P} \gets \varnothing$\Comment*{start with an empty list}
  $\V z = \V q_\bot(\phi_\lambda)$\Comment*{copy profile in workspace array}
  \For{$\ell = 1,\ldots,N_\lambda$}{
      $k_\ell = \argmax_{k} z_{k}$\Comment*{find position of largest value in $\V z$}
      $\mathcal{P} \gets \mathcal{P} \cup \Brace*{k_\ell}$\Comment*{update list of positions}
  \For{$k \in \IntRange{\max(1, k_\ell - \delta_{\bot}), \min(k_\ell +
    \delta_{\bot}, N_q)}$}{
      $z_{k} \gets -\infty$\Comment*{invalidate nearby pixels}
    }
  }
  \Return $\mathcal{P}$\;
\end{algorithm}

\subsection{Extraction of the paths of the spectral lines}
\label{sec:spectral_paths_extraction}

Given the projection angle $\phi_\lambda$ and the list
$\mathcal{P}(\phi_\lambda)$ of the $N_\lambda$ most significant peaks in the
transverse projection $\V q_\bot(\phi_\lambda)$, we build the $\ell$-th
spectral path $\mathcal{C}_\ell$ as a list of points along the $\ell$-th
spectral line. The coordinates
$\Paren[\big]{i_{\ell,m}^{\Tag{path}}, j_{\ell,m}^{\Tag{path}}}$ of the $m$-th
such point are given by computing the center of gravity of the calibration data
in a small rectangular window $\mathcal{W}_{\ell,m}(\phi_\lambda)$ sliding
along the considered spectral lines:
\begin{equation}
  \label{eq:center-of-gravity}
  \Paren[\big]{i_{\ell,m}^{\Tag{path}}, j_{\ell,m}^{\Tag{path}}} = \frac{
    \sum_{(i,j) \in \mathcal{W}_{\ell,m}(\phi_\lambda)} w_{\Tag{msk},i,j} \times (i,j)
  }{
    \sum_{(i,j) \in \mathcal{W}_{\ell,m}(\phi_\lambda)} w_{\Tag{msk},i,j}
  }
\end{equation}
computed for all non-empty\footnote{in the sense that it contains at least one
  valid pixel} sliding window $\mathcal{W}_{\ell,m}$ of size $\sim 1$ pixel
along the spectral line and $2\,\delta_{\bot} + 1$ pixels in the perpendicular
direction:
\begin{equation}
  \label{eq:sliding-window}
  \mathcal{W}_{\ell,m}(\phi_\lambda) = \Brace*{
    \begin{array}{l}
      \textstyle\
      (i,j) \in \IntRange{1,I}\times\IntRange{1,J} \text{ such that}\\
      \textstyle
      \Abs[\big]{i\,\sin\phi_\lambda + j\,\cos\phi_\lambda - k_\ell}
      \le \delta_{\bot} + \frac12 \\
      \textstyle\text{and }
      \Abs[\big]{i\,\cos\phi_\lambda - j\,\sin\phi_\lambda - m} \le \frac12
    \end{array}
  }
\end{equation}
where $k_\ell$ is the $\ell$-th index in the list $\mathcal{P}(\phi_\lambda)$
of the $N_\lambda$ most significant peaks in the transverse projection
$\V q_\bot(\phi_\lambda)$. Again note that, thanks to the weighting by the mask
of valid pixels, invalid pixels have no incidence on the computed coordinates.
In practice, we use the same value for the half-width of the sliding windows
and for the minimal separation between peaks in the transverse projection, that
is $\delta_{\bot} = 10$ pixels for the considered calibration data.

\subsection{Detection of the edges of the spectral bands}
\label{sec:edges_extraction}

Given the projection angle $\phi_\lambda$ and the list
$\mathcal{P}(\phi_\lambda)$ of the $N_\lambda$ most significant peaks in the
transverse projection $\V q_\bot(\phi_\lambda)$, we compute the
\emph{longitudinal} profile of each spectral line as a the following weighted
projection:
\begin{subequations}
  \label{eq:longitudinal_projection}
  \begin{equation}
  q_{{/\!/}\,\ell,k}(\phi_\lambda) = \frac{
    \sum\nolimits_{(i,j)\in\mathcal{D}_\ell(\phi_\lambda)}
    \xi_{{/\!/}\,i,j,k}(\phi_\lambda)\,d_{\Tag{cal},i,j}
  }{
    \sum\nolimits_{(i,j)\in\mathcal{D}_\ell(\phi_\lambda)} \xi_{{/\!/}\,i,j,k}(\phi_\lambda)
  }
  \end{equation}
  with weights given by:
  \begin{equation}
  \xi_{{/\!/}\,i,j,k}(\phi_\lambda) = w_{\Tag{msk},i,j}\,
  \varphi_\Tag{proj}\Paren{i\,\cos\phi_\lambda - j\,\sin\phi_\lambda - k},
  \end{equation}
\end{subequations}
and where $k$ is the index along the projection and
$\mathcal{D}_\ell(\phi_\lambda)$ is a narrow rectangular window (in green in
the central panel of Fig.~\ref{fig:LSS_calibration}) to isolate the pixels of
the calibration image $\V d_\Tag{cal}$ impacted by the considered spectral
line:
\begin{equation}
  \label{eq:spectral-line-box}
  \mathcal{D}_\ell(\phi_\lambda) = \Brace*{
    \begin{array}{l}
      \textstyle\
      (i,j) \in \IntRange{1,I}\times\IntRange{1,J} \text{ such that}\\
      \textstyle
      \Abs[\big]{i\,\sin\phi_\lambda + j\,\cos\phi_\lambda - k_\ell} \le \delta_{\bot} + \frac12
    \end{array}
  }
\end{equation}
with, as before, $\delta_{\bot} \approx 10$~pixels the half-width of the
region.

Detecting the edges of the central hole due to the coronagraphic mask in the
resulting profile (plotted in green in the left panel of
Fig.~\ref{fig:LSS_calibration}) can be done by a quite simple procedure. Each
value at index $k$ of the profile is compared with the next one up to a
threshold value $\tau$. If
$q_{/\!/\,\ell,k} \leqslant \tau \leqslant q_{/\!/\,\ell,k+1}$ and
$q_{/\!/\,\ell,k} < q_{/\!/\,\ell,k+1}$, then an ascending edge is detected. If
$q_{/\!/\,\ell,k} \geqslant \tau \geqslant q_{/\!/\,\ell,k+1}$ and
$q_{/\!/\,\ell,k} > q_{/\!/\,\ell,k+1}$, it is a descending edge. From the four
edges detected in the $\ell\nth$ spectral line (indicated by the crosses in the
left panel of Fig.~\ref{fig:LSS_calibration}), the second and third ones
correspond to the coronagraphic mask. Retrieving these edges in
$\mathcal{C}_\ell(\phi_\lambda)$ yields the coordinates
$\Paren[\big]{i^\Tag{down}_\ell,j^\Tag{down}_\ell}$ and
$\Paren[\big]{i^\Tag{up}_\ell,j^\Tag{up}_\ell}$ required in
Section~\ref{sec:angular_law_calibration} for the calibration of the angular
dispersion law.

\subsection{Results on the calibration data}

\begin{figure}
  \centering%
  \includegraphics[scale=0.58]{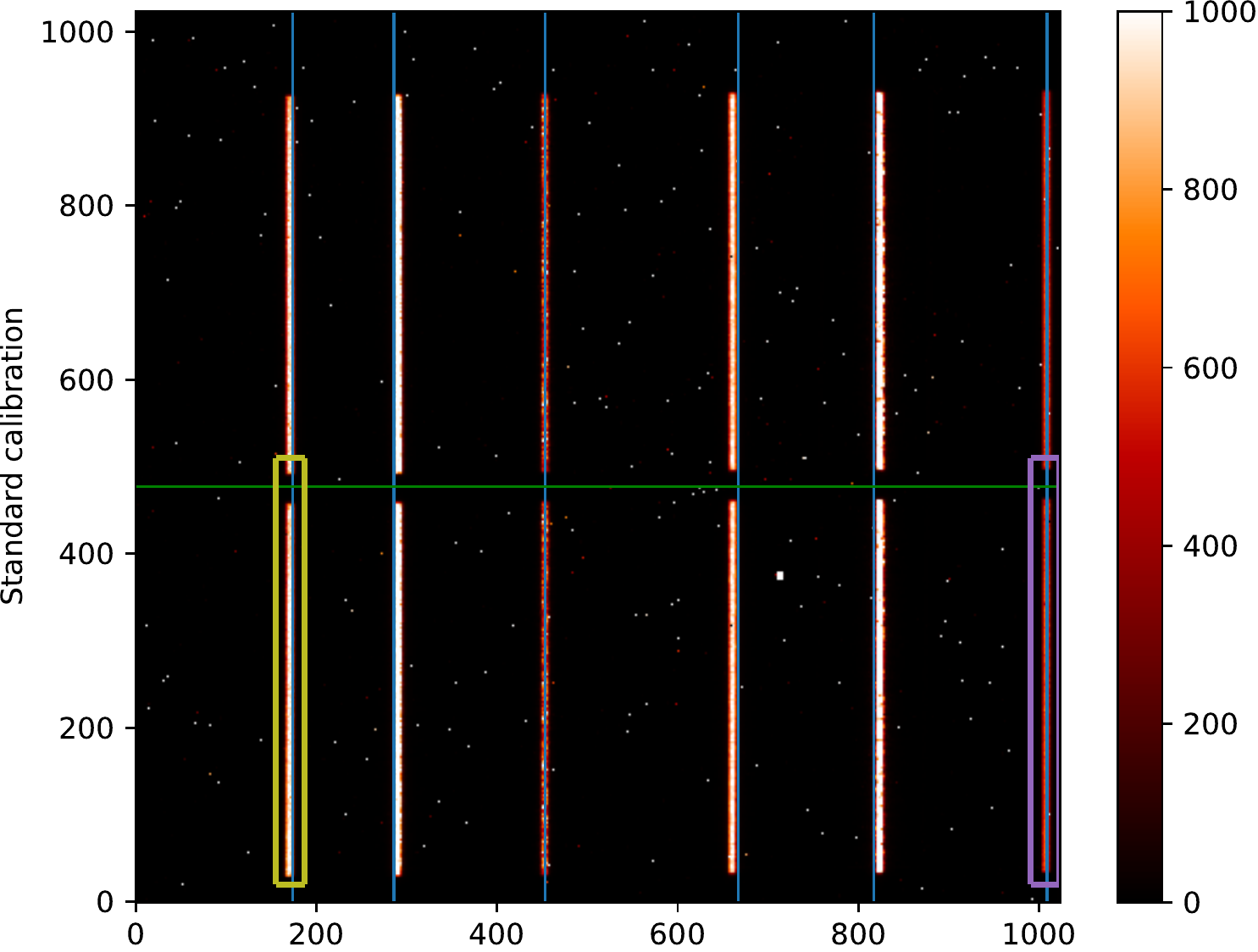}
  \includegraphics[scale=0.58]{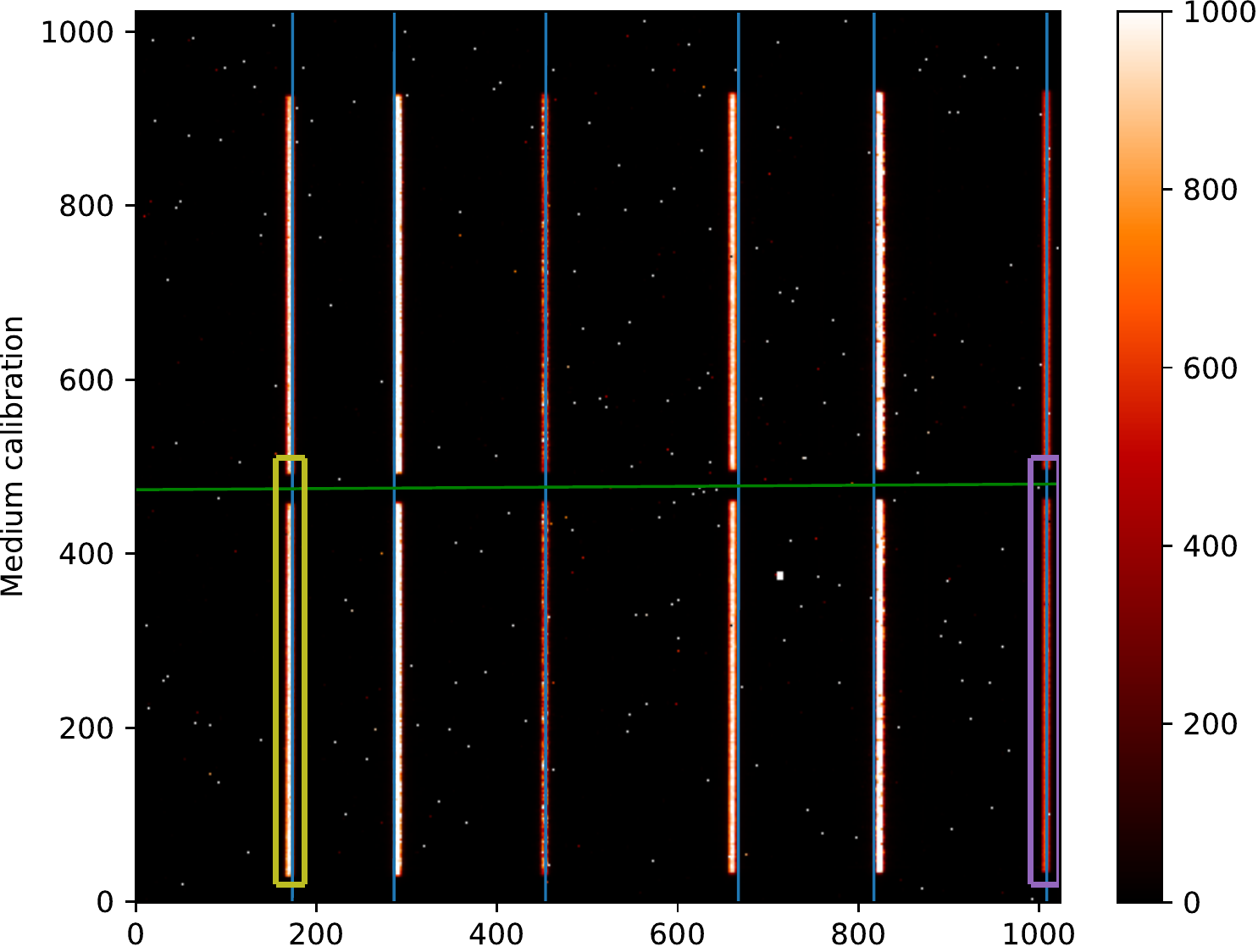}
  \includegraphics[scale=0.58]{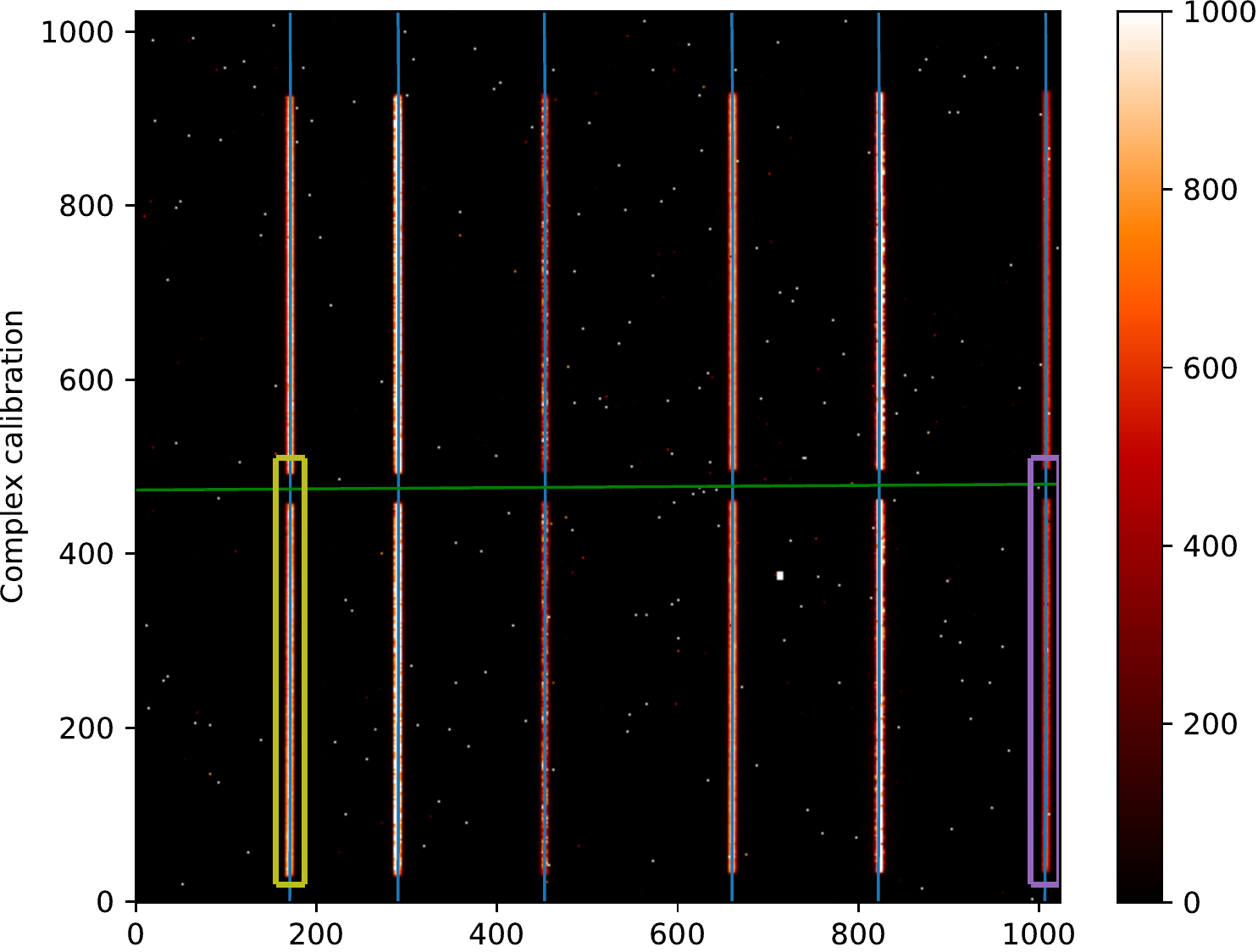}
  \caption{Iso-wavelength curves at the wavelengths of the calibration
  sources (blue lines) and iso-angular distance of the center of the
  coronagraphic mask of (green lines) presented on top of the calibration
  data $\V d_\calib$.  The upper panel presents the results for the simple
  model, the central panel shows the results for the medium model, while the
  bottom panel shows the results of using the complex model.}
  \label{fig:LSS_calibration_laws}
\end{figure}

\begin{figure}[t]
  \includegraphics[scale=0.6]{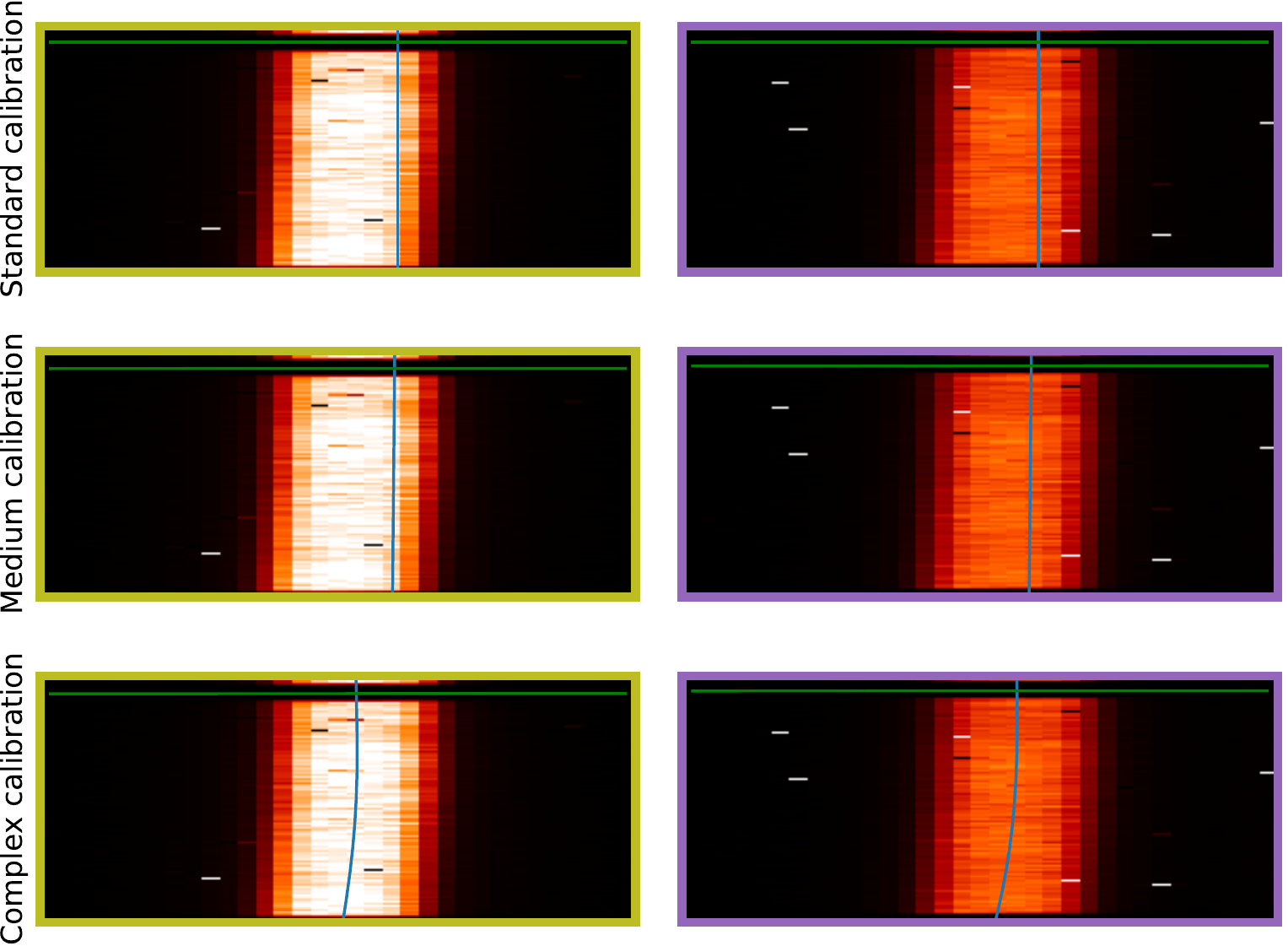}
  \caption{Magnified images of the two regions outlined by the yellow and
    purple rectangles for the three models described in Section
    \ref{sec:dispersion_laws}. To best see the
    differences between models, the magnifications are different in the two
    dimensions.}
    \label{fig:LSS_zoom_calibration_laws}
\end{figure}

\begin{figure}[t]
    \includegraphics[scale=0.58]{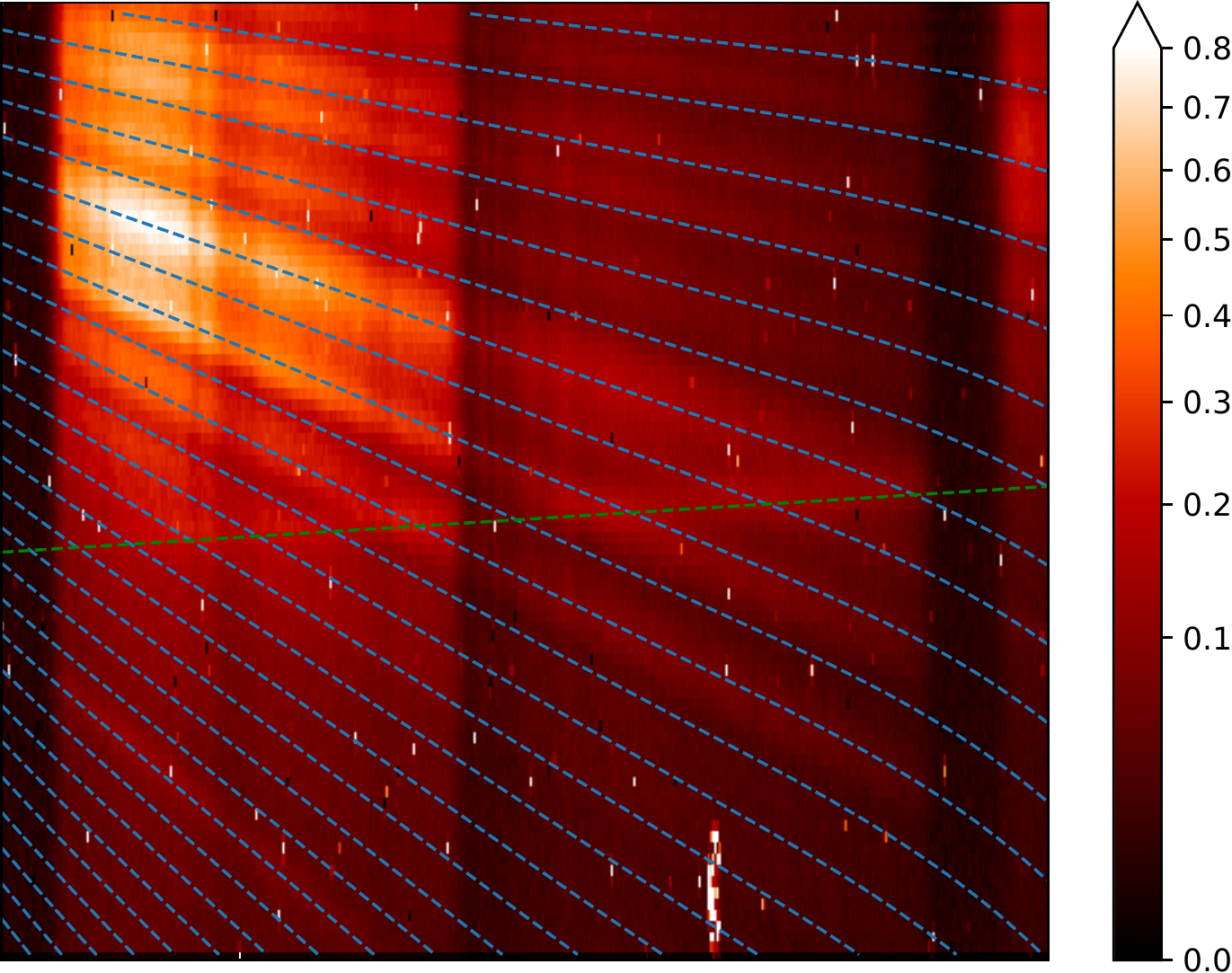}
    \caption{Iso-wavelength and iso-angular distance superposed to the HR\,3549
    data observed on 2015-12-28 with IRDIS in MRS mode.}
    \label{fig:HR3549_data_with_calibration}
\end{figure}

The results of the proposed calibration models are displayed in Fig.~\ref{fig:LSS_calibration_laws}, as blue lines for the spectral law and green lines for the spatial law. In practice, degrees $P_\lambda=5$ and $P_\rho=1$ (limited by the small number of points for $\rho$) were chosen. The three models described in Section~\ref{sec:dispersion_laws} are tested, that is a standard model, one of a medium complexity and one more complex.

As can be seen on the different zooms shown in Fig.~\ref{fig:LSS_zoom_calibration_laws}, 2D polynomials are needed to explain local distortions. Choosing this model, we plot on Fig.~\ref{fig:HR3549_data_with_calibration} some iso-wavelength (blue) and iso-angular distance (green) curves, on a zoom in of the HR3549 dataset. This figure highlights how well our proposed models for the dispersion laws are following the speckles, compared to the standard model. A strong shear effect due to the dispersive elements is visible and taken into account by our model. For all these models, we took polynomials of degrees $P_\lambda=5$ and $P_\rho=1$ for the spectral and spatial dispersion laws.

\section{Calibration of the contrast}
\label{sec:star_SED_calibration}

\begin{figure}
  \centering
  \includegraphics[scale=0.58]{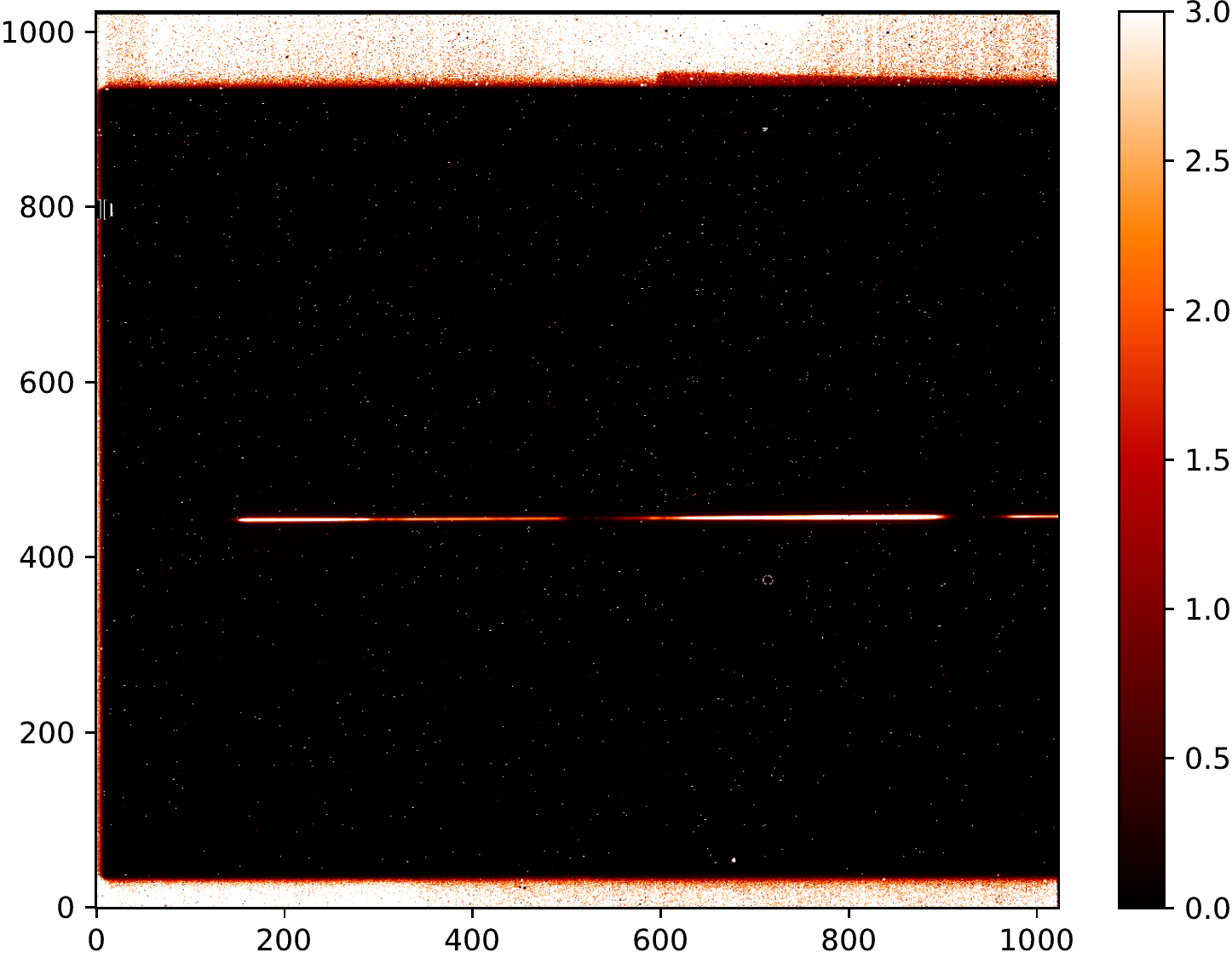}
  \includegraphics[scale=0.58]{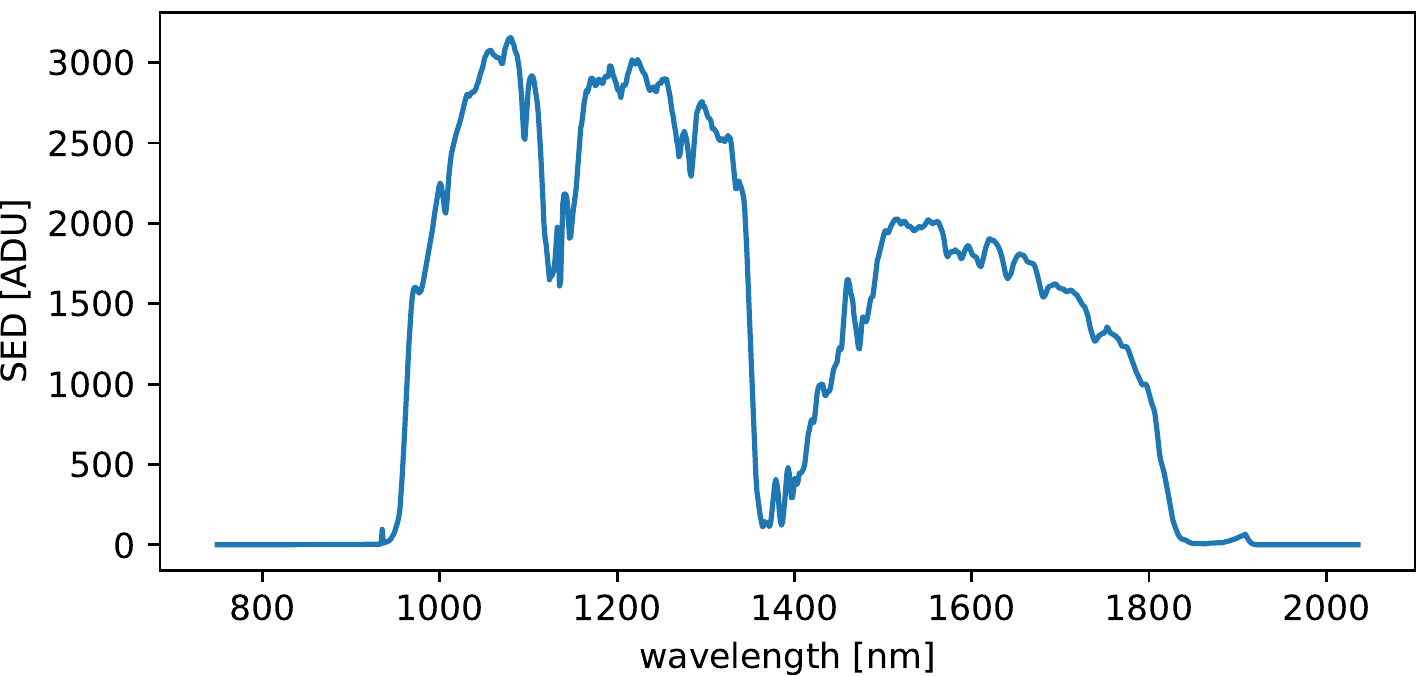}
  \caption{Calibration of the SED of the star HIP\,65426. \emph{Top:}
    Calibration image $\V d_\Tag{flux}$ observed on 2019-05-20 with the MRS
    mode of SPHERE/IRDIS. \emph{Bottom:} SED of the star extracted by
    \textsc{FitStar} (Algorithm~\ref{alg:companion}) and corrected from the
    density filter.}
  \label{fig:HIP65426_flux}
\end{figure}

In order to express the SED of the companion in terms of contrast with respect
to the star, we use specific calibration data $\V d_\Tag{flux}$ (shown in
Fig.~\ref{fig:HIP65426_flux} \emph{top}) for which the star is placed in the
spectrograph slit but shifted away from the coronagraphic mask and with a
neutral density inserted in the optical path to avoid detector saturation.
Applying \textsc{FitCompanion} (Algorithm~\ref{alg:companion}) to
$\V d_\Tag{flux}$ and dividing the resulting SED by the spectral transmission
of the neutral filter yields the SED of the star $\V x_\Tag{flux}$ (shown in
Fig.~\ref{fig:HIP65426_flux} \emph{bottom}). The so derived parameters of the
off-axis PSF can be later used in \textsc{FitCompanion}
(Algorithm~\ref{alg:companion}) to extract the companion SED.

\section{Exploiting the scaling indetermination}
\label{sec:Scaling_indetermination}

The estimated components $\estim{\V x}$, $\estim{\V y}$, $\estim{\V z}$, and
$\estim{\V\nu}$ of the direct model defined in Eq.~\eqref{eq:direct_model}
depend on hyper-parameters which include the regularization weights
$\mu_{\V x}$, $\mu_{\V y}$, and $\mu_{\V z}$. In this Appendix, we show how to
adapt the approach of \citet{TheTDS20} to reduce the effective number of
regularization parameters and also accelerate the minimization.

We first note that the regularizations considered for the stellar components
are homogeneous functions of degree 2, \ie the following property holds:
\begin{equation}
  \label{eq:homogeneous-function}
  \Regul_{\V u}(\alpha\,\V u) = \alpha^{2}\,\Regul_{\V u}(\V u),
\end{equation}
whatever the component $\V u = \V x$ or $\V y$ considered and $\alpha \ge 0$.

The contribution of the star, the first right-hand side term in
Eq.~\eqref{eq:direct_model}, is a bilinear function of the parameters $\V x$
and $\V y$. As a consequence:
\begin{align}
  \label{eq:scaling-invariance}
  \V m(\alpha\,\V x, \V y/\alpha, \V z, \V\nu) = \V m(\V x, \V y, \V z, \V\nu),
\end{align}
holds for any scaling factor $\alpha > 0$.

Combining the properties in Eqs.~\eqref{eq:homogeneous-function} and
\eqref{eq:scaling-invariance} with the definition of the objective function in
Eq.~\eqref{eq:criterion} and that of the regularization in
Eq.~\eqref{eq:composite-regul}, it can be seen that for any $\alpha > 0$:
\begin{equation}
  \label{eq:criterion_scaled}
  \Criterion(\alpha\,\V x,\V y/\alpha, \V z, \V\nu,
  \mu_{\V x},\mu_{\V y}, \mu_{\V z})
  = \Criterion(\V x,\V y,\V z,\V\nu,
  \alpha^{2}\,\mu_{\V x},\alpha^{-2}\,\mu_{\V y},\mu_{\V z}).
\end{equation}
In other words, scaling the unknowns $\V x$ and $\V y$ without changing the
model is equivalent to scaling their regularization weights. Exploiting this,
it is possible to compute an optimal scaling factor:
\begin{align}
  \proxy{\alpha}(\V x,\V y,\mu_{\V x},\mu_{\V y})
  &= \argmin\limits_{\alpha > 0}
    \Criterion(\alpha\,\V x, \V y/\alpha, \V z, \V\nu, \mu_{\V x},
    \mu_{\V y},\mu_{\V z}) \notag \\
  &= \argmin\limits_{\alpha > 0} \Criterion(\V x,\V y, \V z, \V\nu,
    \alpha^{2}\,\mu_{\V x}, \alpha^{-2}\,\mu_{\V y},\mu_{\V z})
    \notag \\
  &= \argmin\limits_{\alpha > 0} \Brace*{\alpha^{2}\,\mu_{\V x}\,
    \Regul_{\V x}(\V x) + \alpha^{-2}\,\mu_{\V y}\,\Regul_{\V y}(\V y)}
    \notag \\
  &= \Paren*{
    \frac{\mu_{\V y}\,\Regul_{\V y}(\V y)}{\mu_{\V x}\,\Regul_{\V x}(\V x)}
    }^{\frac{1}{4}}.
    \label{eq:best-scaling_factor}
\end{align}
Plugging this expression in the definition of the criterion yields:
\begin{align}
  \Criterion^+(\V x,&\,\V y, \V z, \V\nu,
                      \mu_{\V x},\mu_{\V y}, \mu_{\V z}) \notag\\
                    &= \min\limits_{\alpha > 0}
                      \Criterion(\alpha\,\V x, \V y/\alpha, \V z, \V\nu,
                      \mu_{\V x}, \mu_{\V y},\mu_{\V z}) \notag \\
                    &= \Norm*{\V d - \V m(\V x, \V y, \V z, \V\nu)}^2_{\M W}
                      + \mu_{\V x,\V y}\,\Regul_{\V x,\V y}(\V x,\V y)
                      + \mu_{\V z}\,\Regul_{\V z}(\V z)
\label{eq:criterion-entangled}
\end{align}
where:
\begin{equation}
  \label{eq:entangled-regul}
  \Regul_{\V x,\V y}(\V x,\V y) = \sqrt{\Regul_{\V x}(\V x)\,\Regul_{\V y}(\V y)},
\end{equation}
and:
\begin{equation}
  \label{eq:entangled-regul-weight}
  \mu_{\V x,\V y} = 2\,\sqrt{\mu_{\V x}\,\mu_{\V y}}.
\end{equation}
This shows that the regularizations of the stellar components $\V x$ and $\V y$
are entangled and that the effect of these regularizations on the shape of
these components is effectively controlled by a single hyper-parameter, here
$\mu_{\V x,\V y}$. As shown by Eq.~\eqref{eq:best-scaling_factor}, the ratio
$\mu_{\V x}/\mu_{\V y}$ of the hyper-parameters controls the scaling, not the
shape, of $\V x$ and $\V y$, one of the two can be fixed. This saves us the
hassle of adjusting both parameters at the same time.


\end{appendix}

\end{document}